\documentclass[sn-mathphys, Numbered]{test}%

\usepackage{times}
\usepackage{graphicx}%
\usepackage{multirow}%
\usepackage{amsmath,amssymb,amsfonts}%
\usepackage{amsthm}%
\usepackage{mathrsfs}%
\usepackage[title]{appendix}%
\usepackage{xcolor}%
\usepackage{textcomp}%
\usepackage{manyfoot}%
\usepackage{booktabs}%
\usepackage{algorithm}%
\usepackage{algorithmicx}%
\usepackage{algpseudocode}%
\usepackage{listings}%
\usepackage[utf8]{inputenc}
\usepackage[T1]{fontenc}

\begin{document}

\title[ ]{Electronic dynamics created at conical intersections and its dephasing in aqueous solution}

\author[1,2]{\fnm{Yi-Ping} \sur{Chang}}
\equalcont{These authors contributed equally to this work.}

\author[1,3]{\fnm{Tadas} \sur{Balciunas}}
\equalcont{These authors contributed equally to this work.}

\author*[3,4]{\fnm{Zhong} \sur{Yin}}\email{yinz@tohoku.ac.jp}
\equalcont{These authors contributed equally to this work.}

\author[5]{\fnm{Marin} \sur{Sapunar}}
\equalcont{These authors contributed equally to this work.}

\author[6,7]{\fnm{Bruno N. C.} \sur{Tenorio}}
\equalcont{These authors contributed equally to this work.}

\author[8]{\fnm{Alexander C.} \sur{Paul}}
\equalcont{These authors contributed equally to this work.}

\author[9,10]{\fnm{Shota} \sur{Tsuru}}

\author*[8]{\fnm{Henrik} \sur{Koch}}\email{henrik.koch@ntnu.no}

\author*[1]{\fnm{Jean--Pierre} \sur{Wolf}}\email{jean-pierre.wolf@unige.ch}

\author*[6]{\fnm{Sonia} \sur{Coriani}}\email{soco@kemi.dtu.dk}

\author*[3]{\fnm{Hans Jakob} \sur{W{\"o}rner}}\email{hwoerner@ethz.ch}

\affil[1]{\orgdiv{GAP--Biophotonics}, \orgname{Universit{\'e} de Gen{\`e}ve}, \postcode{1205} \city{Geneva}, \country{Switzerland}}

\affil[2]{\orgname{European XFEL}, \orgaddress{\postcode{22689} \city{Schenefeld}, \country{Germany}}}

\affil[3]{\orgdiv{Laboratory of Physical Chemistry}, \orgname{ETH Z{\"u}rich}, \postcode{8093} \city{Z{\"u}rich}, \country{Switzerland}}

\affil[4]{\orgdiv{International Center for Synchrotron Radiation Innovation Smart}, \orgname{Tohoku University}, \postcode{980-8577} \orgaddress{\city{Sendai}, \country{Japan}}}

\affil[5]{\orgdiv{Division of Physical Chemistry}, \orgname{Ruđer Bošković Institute},  \postcode{10000} \city{Zagreb}, \country{Croatia}}

\affil[6]{\orgdiv{Department of Chemistry}, \orgname{Technical University of Denmark}, \postcode{2800} \orgaddress{\city{Kongens Lyngby}, \country{Denmark}}}

\affil[7]{\orgdiv{Instituto Madrile\~{n}o de Estudios Avanzados en Nanociencia}, \orgname{IMDEA-Nanociencia}, \postcode{28049} \orgaddress{\city{Madrid}, \country{Spain}}}

\affil[8]{\orgdiv{Department of Chemistry}, \orgname{Norwegian University of Science and Technology}, \postcode{7034} \orgaddress{\city{Trondheim}, \country{Norway}}}

\affil[9]{\orgdiv{Lehrstuhl f\"{u}r Theoretische Chemie}, \orgname{Ruhr-Universit\"{a}t Bochum}, \postcode{44801} \orgaddress{\city{Bochum}, \country{Germany}}}

\affil[10]{\orgdiv{RIKEN Center for Computational Science}, \orgname{RIKEN}, \postcode{650-0047} \orgaddress{\city{Kobe}, \country{Japan}}}


\abstract{A dynamical rearrangement in the electronic structure of a molecule can be driven by different phenomena, including nuclear motion, electronic coherence or electron correlation. Recording such electronic dynamics and identifying their fate in aqueous solution has remained a challenge. Here, we reveal the electronic dynamics induced by electronic relaxation through conical intersections in pyrazine molecules using X-ray spectroscopy. We show that the ensuing created dynamics corresponds to a cyclic rearrangement of the electronic structure around the aromatic ring. Furthermore, we find that such electronic dynamics are entirely suppressed when pyrazine is dissolved in water. Our observations confirm that conical intersections can create electronic dynamics that are not directly excited by the pump pulse and that aqueous solvation can dephase them in less than 40 fs. These results have implications for the investigation of electronic dynamics created during light-induced molecular dynamics and shed light on their susceptibility to aqueous solvation.}


\maketitle

Electronic dynamics are the central concept underlying multiple emerging research fields that range from attosecond spectroscopy \cite{drescher02a,kluender11a} over attochemistry \cite{calegari14a, kraus15b, woerner17a, matselyukh22a} to quantum biology \cite{engel07a,cao20a}. Despite their importance, identifying electronic dynamics and distinguishing them from purely vibrational dynamics has remained a major challenge causing numerous controversies. Especially in the case of excited-state dynamics, the distinction of electronic from purely vibrational dynamics often remains out of reach because of the strong coupling between the two types of dynamics in the vicinity of conical intersection (CIs) and/or an insufficient contrast in the sensitivity of the probe to electronic vs. vibrational dynamics. 
Additionally, an important question for the viability and broader impact of the above-mentioned research areas is the lifetime of electronic dynamics under ambient conditions, which imply a rapidly fluctuating environment \cite{gustin23a}. Moreover, many technologically and fundamentally important processes, especially in chemistry and biology, take place in water such that exploiting electronic dynamics in such processes \cite{scholes17a} necessitates a method that can trace and disambiguate them in aqueous solutions. Although two-dimensional spectroscopies in the optical domain have in principle such capabilities, the assignment and interpretation of the observed signals often remain sufficiently ambiguous that it has created controversies, e.g., regarding the nature and lifetime of electronic vs. vibrational dynamics observed in light-harvesting systems \cite{engel07a,cao20a}.

Here, we report two experimental breakthroughs, namely (i) the observation of electronic and vibrational dynamics corresponding to a circular rearrangement of the electronic structure that is created by CI dynamics and (ii) their sub-40-fs dephasing induced by aqueous solvation. This has been achieved by directly comparing the dynamics of UV-excited pyrazine in the gas phase and in aqueous solution with a new experimental scheme that 
unites element specificity, site selectivity, and the transparency of water with the required time resolution. This scheme combines for the first time one-photon excitation in the ultraviolet (UV) domain with a soft-X-ray probe \cite{pertot17a,saito19a,scutelnic2021,zinchenko2021} covering the entire water window \cite{Schmidt2018}.

The interpretation of our results is supported by the latest advances in quantum-chemical calculations of X-ray absorption (XA) spectra, coupled to nonadiabatic dynamical simulations, which also include solvation effects. These calculations enable an unambiguous assignment of the transient spectral features observed in the experimental spectra.

The heteroaromatic pyrazine molecule (C$_{4}$H$_{4}$N$_{2}$), a paradigmatic system for both theoretical and experimental studies of electronically nonadiabatic dynamics, serves as a demonstration of the opportunities opened by our work. The gas-phase measurements indeed show that the electronic relaxation of the initially photoexcited $^1$B$_{\rm 2u}$($\pi\pi^*$) state (historically referred to as S$_2$) through CIs induces electronic and vibrational dynamics involving the $^1$B$_{\rm 3u}$(n$\pi^*$, known as S$_1$) and the $^1$A$_{\rm u}$(n$\pi^*$) state. 

Our gas-phase results moreover resolve a decades-old controversy between ever-evolving dynamics simulations \cite{werner2008nonadiabatic,kanno2015ab, mignolet2018ultrafast} that eventually converged on the prediction of oscillatory population flow \cite{sala2014role,tsuru2019time, huang2021ab, gelin2021ab, freibert2022femtosecond,kaczun2023pyrazineN} and experiments that have always negated them. Specifically, time-resolved photoelectron spectroscopy neither revealed the population of the $^1$A$_{\rm u}$ state, nor quasi-periodic electronic dynamics \cite{stert2000electron, suzuki2006femtosecond, horio2009probing, suzuki2010time, horio2016full}, whereas carbon K-edge transient absorption suggested a population of the $^1$A$_{\rm u}$ state in $\sim$200~fs, up to ten times slower than the theoretical predictions, and did not reveal any oscillations \cite{scutelnic2021}. As shown in the present work, the key to resolving this controversy was the present experimental advance that combined single-photon excitation with transient absorption spectroscopy at the nitrogen K-edge located at $\sim$405~eV. The unprecedented capability of nitrogen K-edge spectroscopy allowed us to not only confirm the predicted oscillatory population flow, but to show that it corresponds to a cyclic rearrangement of the electronic structure around the aromatic ring of pyrazine.

Beyond confirming the possibility of creating electronic dynamics at CIs and resolving the important controversy regarding the electronic-relaxation pathway of the isolated pyrazine molecule, our work additionally reveals the effect of aqueous solvation on the paradigmatic dynamics of this molecule. Specifically, we find that the electronic and vibrational dynamics are completely dephased by solvation within 40~fs. Comparison with dynamical calculations indeed confirms the fact that solvation strongly damps the observed dynamics, but reveals that two explicit water molecules combined with a continuum-solvation model are insufficient to fully capture the dephasing observed in the solution-phase experiments.

\begin{figure}[h!]
\centering\includegraphics[width=\textwidth]{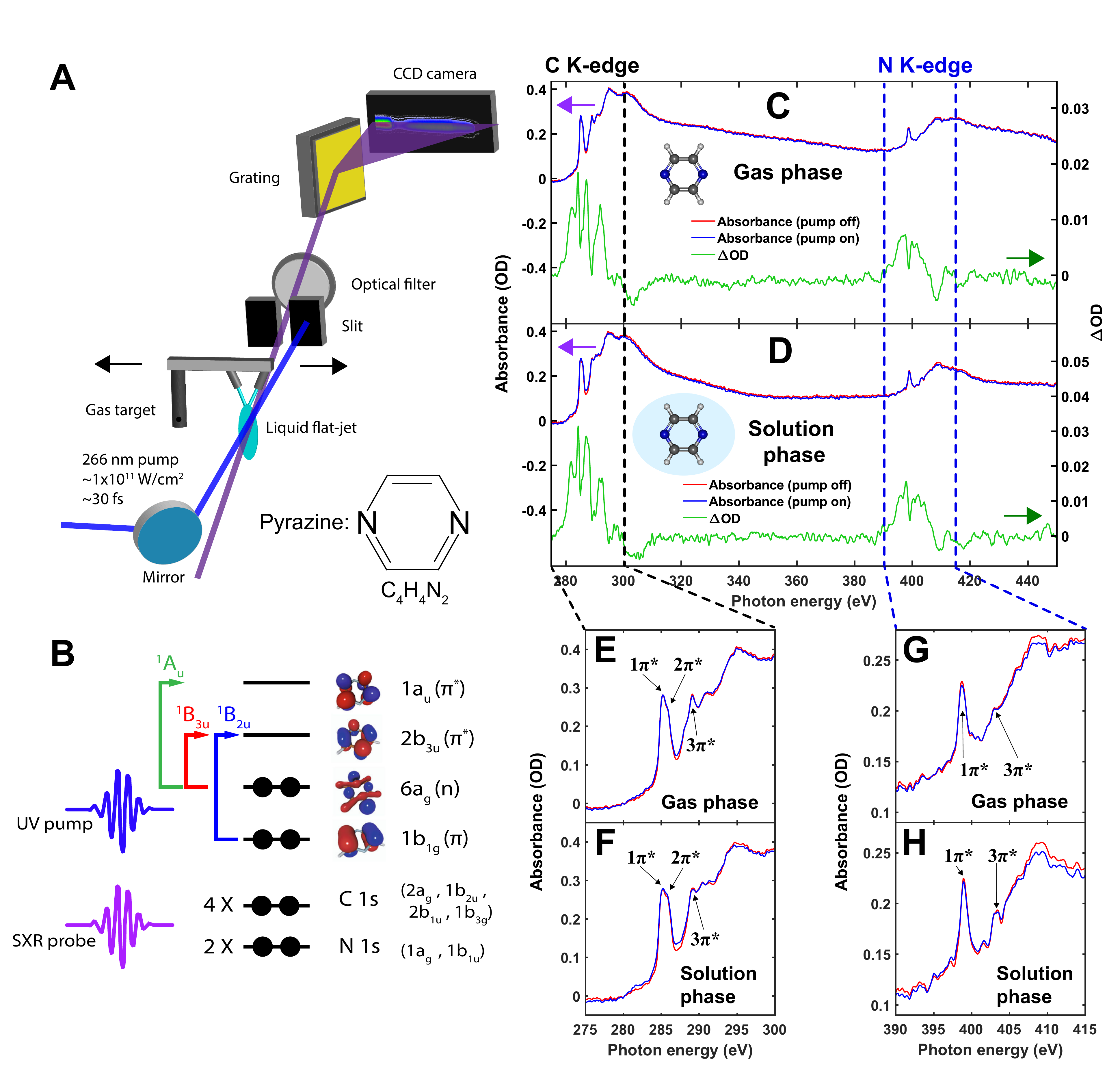}
\caption{\textbf{Overview of the experimental methods and results.}
\textbf{(A)} Scheme of the experimental setup with both gas and solution flat-jet targets. \textbf{(B)} Molecular-orbital diagram of photoexcitation of pyrazine, with arrows indicating the main transition character of the valence excited states: $^1$B$_{\mathrm{2u}}$($\pi\pi^*$), $^1$B$_{\mathrm{3u}}$(n$\pi^*$) and $^1$A$_{\mathrm{u}}$(n$\pi^*$). XA spectra and time-averaged differential absorbances (over 150 fs) covering the water window from carbon to nitrogen K-edges for: \textbf{(C)} the gas phase and \textbf{(D)} the solution phase. No energy shift in the ground-state pyrazine (static measurement) at the carbon pre-edge peak (285.3 eV and 285.8 eV) has been observed between the solution/gas phase. The nitrogen pre-edge peak has a value of 398.7 eV in the gas phase compared to 398.9 eV in solution phase, with an energy shift of about 0.2 eV. \textbf{(E,F)} Zoomed-in on the carbon K-edge. \textbf{(G,H)} Zoomed-in on the nitrogen K-edge.}
\label{fig:absorbances and scheme}
\end{figure}

Figure\,\ref{fig:absorbances and scheme} provides an overview of the experimental setup (A), the diagram of the relevant orbitals of pyrazine (B) and the XA spectra (C-H).
The experiments were performed by photoexciting pyrazine with a 30-fs pump pulse centered at 266~nm, focused to an intensity of $\sim$1$\times 10^{11}$W/cm$^2$ and probing with a soft-X-ray supercontinuum extending beyond 450~eV obtained from high-harmonic generation of a $\sim$12-fs pulse centered at 1.8~$\mu$m in helium. Here, we demonstrate the unique capability of soft-X-ray spectroscopy in the water window to directly compare the dynamics of the same molecules in the gas and solution phase. For this purpose, a dedicated target system has been constructed that allows rapid switching between a gas cell delivering an effusive beam of pyrazine vapor and a liquid flat jet running a 5M aqueous solution of pyrazine. More details on the experimental setup are given in the supporting information (SI).

Figure\,\ref{fig:absorbances and scheme}C, D shows the static XA spectra of gaseous pyrazine and a 5M aqueous pyrazine solution from the carbon to the nitrogen K-edges (red curves), respectively. The time-averaged transient absorption of the excited sample is shown by the blue curves. The differential absorbance spectrum ($\Delta$OD) is shown in green. Magnified portions of the absorption spectra at each edge for both phases are shown in Fig.\,\ref{fig:absorbances and scheme}E-H.

At the carbon K-edge, a strong pre-edge absorption feature consisting of two sub-peaks is observed. In both gaseous and aqueous samples, the two pre-edge peaks are at 285.3 eV and 285.8 eV, and can be assigned to C 1s $\to 1\pi^*$ [2a$_\textrm{g}$\,$\rightarrow$ 2b$_{\mathrm{3u}}$] and
1s $\to$ 2$\pi^*$ [1b$_\textrm{3g}$  \,$\rightarrow$ 1a$_{\mathrm{u}}$]  
transitions, see SI, Figs. S11, S14.
The two final core-excited states have B$_\textrm{3u}$ symmetry.
In the gas phase, a shoulder at $\sim$288.2 eV and a peak at $\sim$289.1 eV can be observed, corresponding to 1s $\to$ $\sigma^*$/Rydberg and 1s $\to$ 3$\pi^*$
[2b$_\textrm{1u}$\,$\rightarrow$ 2b$_\textrm{2g}$] transitions,
respectively.
In solution phase, the shoulder at 288.2 eV cannot be resolved, while the peak at $\sim$289.0 eV is still present with a small energy shift, but the same assignments as in the gas phase apply.

At the nitrogen K-edge, a single strong pre-edge absorption peak can be observed at 398.7 eV and 398.9 eV, for gaseous and aqueous samples, respectively. This corresponds to the
N~1s $\to$ 1$\pi^*$ [1a$_\textrm{g}$\,$\rightarrow$ {2}b$_{\mathrm{3u}}$] 
transition and indicates an energy shift of +0.2 eV from the gas to the solution phase. In the gas phase, a peak at 402.9 eV corresponding to the 
{N 1s $\to$ 3$\pi^*$ [1b$_{\textrm{1u}}$\,$\rightarrow$ {2}b$_{\textrm{2g}}$]} 
transition can be observed. 
In solution phase, a broad peak is seen at $\sim$403.3 eV.

The experimental spectra have been calibrated by aligning the experimental gaseous pyrazine carbon and nitrogen pre-edge peaks with synchrotron measurements \cite{vall2008c}. In both gas and solution phases, the $\Delta$OD spectra exhibit an increased absorption up to 10 eV below and above the pre-edge. 
Most of the pre-edge absorption features originate from allowed transitions into valence vacancies created by the pump pulse.
The rest of the pre-edge absorption features, as well as the above-pre-edge ones, share core-excitation characters with those of transitions from the ground state.

\begin{figure}[h!]
\centering
\includegraphics[width=\textwidth]{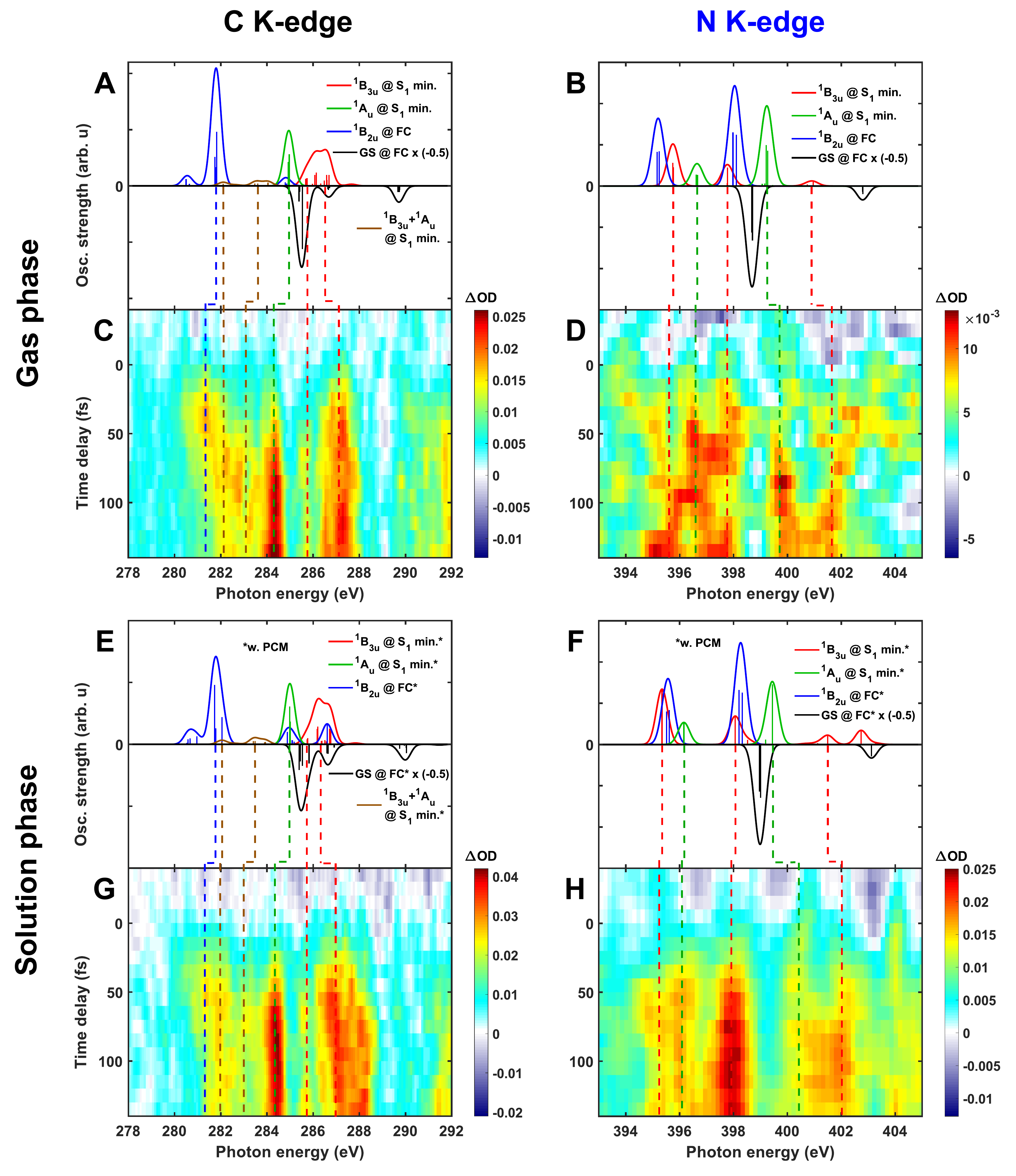} 
\caption{\textbf{Time-resolved differential absorbance spectra of pyrazine in the gas and aqueous-solution phases}. \textbf{(C,D)} Time-resolved differential absorbance spectra at the carbon and nitrogen K-edges of gaseous pyrazine, respectively. \textbf{(G,H)} Time-resolved differential absorbance spectra at the carbon and nitrogen K-edges of 5M aqueous pyrazine, respectively. \textbf{(A,E)} Carbon K-edge excited-state XA spectra calculated at the RASPT2/RAS2(10e,8o) level at both the Franck-Condon and the relaxed geometry for the first valence excited state (S$_1$), with and without PCM, respectively. \textbf{(B,F)} same as (A,E) at the nitrogen K-edge.
}
\label{fig:2}
\end{figure}

Figure\,\ref{fig:2} shows the time-resolved $\Delta$OD spectra of gaseous pyrazine (C\&D) and 5M aqueous pyrazine solution (G\&H), recorded over a $\sim$150~fs time window, with 10 fs steps. 

At the carbon K-edge in the gas phase (Fig.\,\ref{fig:2}C), a positive absorption band at 281.4 eV is observed in the first $\sim$50 fs, after which it disappears and is replaced by another positive band centered at 282.2 eV, which is broad in the initial $\sim$50-100 fs and then narrows. Concurrently, another positive band at 284.2 eV emerges, with a longer rise time than the previous two bands. At the pre-edge, there is a weak band at $\sim$285.7 eV. Above the pre-edge, a positive band at 287.2 eV can be observed rising earlier than the 284.2 eV band. Comparing the 284.2 eV and 287.2 eV bands, reveals that they weakly oscillate out of phase (Fig.~S1).

At the carbon K-edge in water solution (Fig.\,\ref{fig:2}G), a broad positive absorption band centered at $\sim$281.2 eV is also observed in the first $\sim$50 fs, after which it disappears. At the same time, another broad positive band centered at $\sim$282 eV is also emerging, such that the band shift from $\sim$281.2 to 282~eV is less clear compared to the gas-phase signal. Concurrently, another positive band at 284.4~eV emerges and peaks in intensity at $\sim$100~fs before gradually declining. At the pre-edge, there is a weak band at $\sim$285.6~eV. Above the pre-edge, a positive band at 286.8~eV can be observed to rise at nearly the same time as the 284.4~eV band, before peaking at $\sim$60~fs and subsequently declining in intensity. Comparing the 284.4~eV and 286.8~eV bands, no quantum beats can be identified in either of them. 

To assign the bands, both the carbon and nitrogen K-edge excited-state XA spectra were calculated at the RASPT2/RAS2(10e,8o) level at both the Franck-Condon (FC) geometry and the minimum of the first valence-excited singlet state (S$_1$), the latter obtained with and without polarizable continuum model (PCM) for gaseous and aqueous pyrazine, respectively. The relaxed geometries were obtained at the CASPT2 level. 
All RAS/CAS calculations were performed using OpenMOLCAS~\cite{li2023openmolcas}.
Additionally, coupled cluster calculations including perturbative triples (CC3)~\cite{CC3,CC3response} have been performed.
The carbon and nitrogen K-edge excited state XA spectra were calculated at the CC3 level of theory for the gas-phase FC and the S$_1$ minimum geometry using the eT program~\cite{folkestad2020,paul2020}.
The spectra and more details about the calculations can be found in the SI.
Note that, in the gas phase, the potential-energy minimum of the S$_1$ state has a mixed configuration character, with contributions from both of the n$\pi^*$ states. This is reflected in the simulated spectrum which shows peaks corresponding to the ones seen in both the $^1$B$_{\mathrm{3u}}$ and $^1$A$_{\mathrm{u}}$ states at the FC geometry. 

At the carbon K-edge in the gas phase,
based on the calculated spectra from Fig.\,\ref{fig:2}A and the schematics in Fig.~S9, we can assign the $\sim$281.4 eV experimental band to the 281.7 eV C 1s $\to \pi$
from the $^1$B$_{\mathrm{2u}}$($\pi\pi^*$) state at the FC geometry, 
the $\sim$282.2 eV band to the 282.1 eV C~1s$\,\to\,$n 
transitions from both
the $^1$B$_{\mathrm{3u}}$(n$\pi^*$) and $^1$A$_{\mathrm{u}}$(n$\pi^*$) states at {the} S$_1$ minimum geometry (highly overlapping in this region, see Figure~S12), 
the broad band around 
283~eV also to calculated C~1s$\,\to\,$n 
transitions at around 283.8 eV from both 
$^1$B$_{\mathrm{3u}}$(n$\pi^*$) and $^1$A$_{\mathrm{u}}$(n$\pi^*$),
the 284.2 eV band to the 284.2 eV C 1s $\to$ $\pi^\ast$ transition
{from} the $^1$A$_{\mathrm{u}}$(n$\pi^*$) state at the S$_1$ minimum geometry, the $\sim$285.7 eV band to the 285.9 eV C 1s $\rightarrow$ {$\pi^*$} transition from the $^1$B$_{\mathrm{3u}}$(n$\pi^*$) state at the S$_1$ minimum geometry, and the $\sim$287.2 eV band to the 287.2 eV C 1s $\rightarrow$ {$\pi^*$} transition from the $^1$B$_{\mathrm{3u}}$(n$\pi^*$) state at the S$_1$  minimum geometry.

At the carbon K-edge in solution phase, based on the calculated spectra from Fig.\,\ref{fig:2}E, we can assign the $\sim$281.2 eV band to the 281.8 eV C 1s $\to \pi$ transition from the $^1$B$_{\mathrm{2u}}$($\pi\pi^*$) state at the FC geometry, the $\sim$281.9 eV band to the 281.9 eV C 1s $\to$ n transition 
from both the $^1$B$_{\mathrm{3u}}$(n$\pi^*$) 
and $^1$A$_{\mathrm{u}}$(n$\pi^*$)
states at {the} S$_1$ minimum geometry (see Figure~S13), 
the weak broad band around 
283~eV to calculated 1s $\to$ n
transitions at around 283.6 eV from both 
$^1$B$_{\mathrm{3u}}$(n$\pi^*$) and $^1$A$_{\mathrm{u}}$(n$\pi^*$),
the $\sim$284.4 eV band mainly to the 284.7 eV  
{C 1s $\to$ $\pi^*$} 
transition from the $^1$A$_{\mathrm{u}}$(n$\pi^*$) at {the} S$_1$ minimum geometry, the $\sim$285.6 eV band to the 285.7~eV C 1s $\rightarrow$ $\pi^*$
transition from 
the $^1$B$_{\mathrm{3u}}$(n$\pi^*$) state at the S$_1$ minimum geometry, and the $\sim$286.8 eV band to the 286.8~eV C 1s $\rightarrow$ 
$\pi^*$
transition from the $^1$B$_{\mathrm{3u}}$(n$\pi^*$) state at {the} S$_1$ minimum geometry. 

Looking now at the nitrogen K-edge in the gas phase  (Fig.\,\ref{fig:2}D), where the pre-edge is at 398.8 eV, a broad positive absorption band from $\sim$393.6 to 395.2 eV is observed in the first $\sim$50 fs before disappearing. Concurrently, positive absorption bands centered at $\sim$395.3~eV, 396.4~eV, and 397.6~eV are emerging and show clear modulations in intensity. Looking above the pre-edge, there are two bands at 399.5 eV and 401 eV emerging at the same time as the others. These two above-pre-edge bands show clearer intensity modulations that are out of phase with each other.

At the nitrogen K-edge in solution phase (Fig.\,\ref{fig:2}H), where the pre-edge is at 398.9 eV, a weak positive absorption band at $\sim$396 eV appears initially before broadening into a broad band from $\sim$394 to 396.2 eV at $\sim$50 fs. After $\sim$50 fs, this broad band narrows into two bands at $\sim$395.2 eV and $\sim$396.3 eV. Meanwhile, a strong absorption band at 397.8 eV emerges from time zero and plateaus around 100 fs. Above the edge, a weak band at $\sim$400.3 eV starts emerging from time zero and merges into a broad peak from $\sim$400 to 402 eV at $\sim$70 fs.

At the nitrogen K-edge in the gas phase, based on the calculated spectra from Fig.\,\ref{fig:2}B 
(and the schematics in Fig.~S8), we can assign the broad $\sim$393.6 to 395.2 eV band to the 395.2 eV N~1s $\to \pi$ 
{[1b$_\mathrm{1u}$$\rightarrow$1b$_\mathrm{2g}$ in the $D_{2h}$ limit]} 
transition from the $^1$B$_{\mathrm{2u}}$($\pi\pi^*$) state at the FC geometry, 
the $\sim$395.3 eV band to the 395.5 eV 
N~1s~$\rightarrow$~n transition 
of the $^1$B$_{\mathrm{3u}}$(n$\pi^*$) state at the S$_1$ minimum geometry 
[$\mathrm{1b_{1u}} \to  6\mathrm{a_g}$ in $D_{2h}$, to a final core state of symmetry B$_\mathrm{2g}$],
the $\sim$396.4 eV band to the 396.4 eV N 1s $\rightarrow$ n transition of the $^1$A$_{\mathrm{u}}$(n$\pi^*$) state at the 
S$_1$ minimum geometry 
[also 1$\mathrm{b_{1u}} \to 6\mathrm{a_g}$ in the $D_{2h}$ limit, but to a final core state of symmetry B$_\mathrm{1g}$],
the $\sim$397.6 eV band to the 397.6 eV 
N 1s $\rightarrow$ 
$\pi^{\ast}$
{[1$\mathrm{a_g} \to 2\mathrm{b_{3u}(1\pi^*)}$ in the $D_{2h}$ limit, final core state 
A$_\mathrm{g}$]} 
transition 
from the $^1$B$_{\mathrm{3u}}$(n$\pi^*$) state at the S$_1$ minimum geometry, the $\sim$399.5 eV band to the 399.2 eV N~1s $\rightarrow$
$\pi^{\ast}$ transition 
from
the $^1$A$_{\mathrm{u}}$(n$\pi^*$) state 
{[also 1a$_\textrm{g}$ $\to$ 2b$_{\mathrm{3u}}(1\pi^*)$ in the $D_{2h}$ limit, but to a final $\mathrm{B_{3g}}$ core state]}
at the S$_1$ minimum geometry, and the $\sim$401 eV band to the 401 eV N 1s~$\rightarrow$~$\pi^*$ 
{[1b$_\mathrm{1u}$~$\to$~2b$_\mathrm{2g}(3\pi^*)$ 
in the $D_{2h}$ limit]}
transition from
the $^1$B$_{\mathrm{3u}}$(n$\pi^*$) state at the S$_1$ minimum geometry.

At the nitrogen K-edge in solution phase, based on the calculated spectra from Fig.\,\ref{fig:2}F, we can assign the initial 396 eV band to a mix of the 396.1 eV N 1s $\rightarrow$ $\pi$ transition 
from
the $^1$B$_{\mathrm{2u}}$($\pi\pi^*$) state at the FC geometry and 396.2 eV N 1s $\rightarrow$ {n} transition 
of the $^1$A$_{\mathrm{u}}$(n$\pi^*$) state at the S$_1$ minimum geometry. The $\sim$395.2 eV band is assigned 
{to} the 395.2 eV {N~1s~$\rightarrow$~{n}} transition 
from the $^1$B$_{\mathrm{3u}}$(n$\pi^*$) state, the $\sim$397.8 eV band to the 397.7 eV N 1s $\rightarrow$ 
{$\pi^*$} transition from the $^1$B$_{\mathrm{3u}}$(n$\pi^*$) state at the S$_1$ minimum geometry, the $\sim$400.3 eV band to the 399.4 eV N 1s $\rightarrow$ {$\pi^*$} transition 
from the $^1$A$_{\mathrm{u}}$(n$\pi^*$) state at {the} S$_1$ minimum geometry, and the $\sim$402 eV band to the 401.5 eV N 1s $\rightarrow$ {$\pi^*$} (3$\pi^ *$ in $D_{2h}$) transition from the $^1$B$_{\mathrm{3u}}$(n$\pi^*$) state at the S$_1$ minimum geometry.

\begin{figure}[h!]
\centering\includegraphics[width=\textwidth]{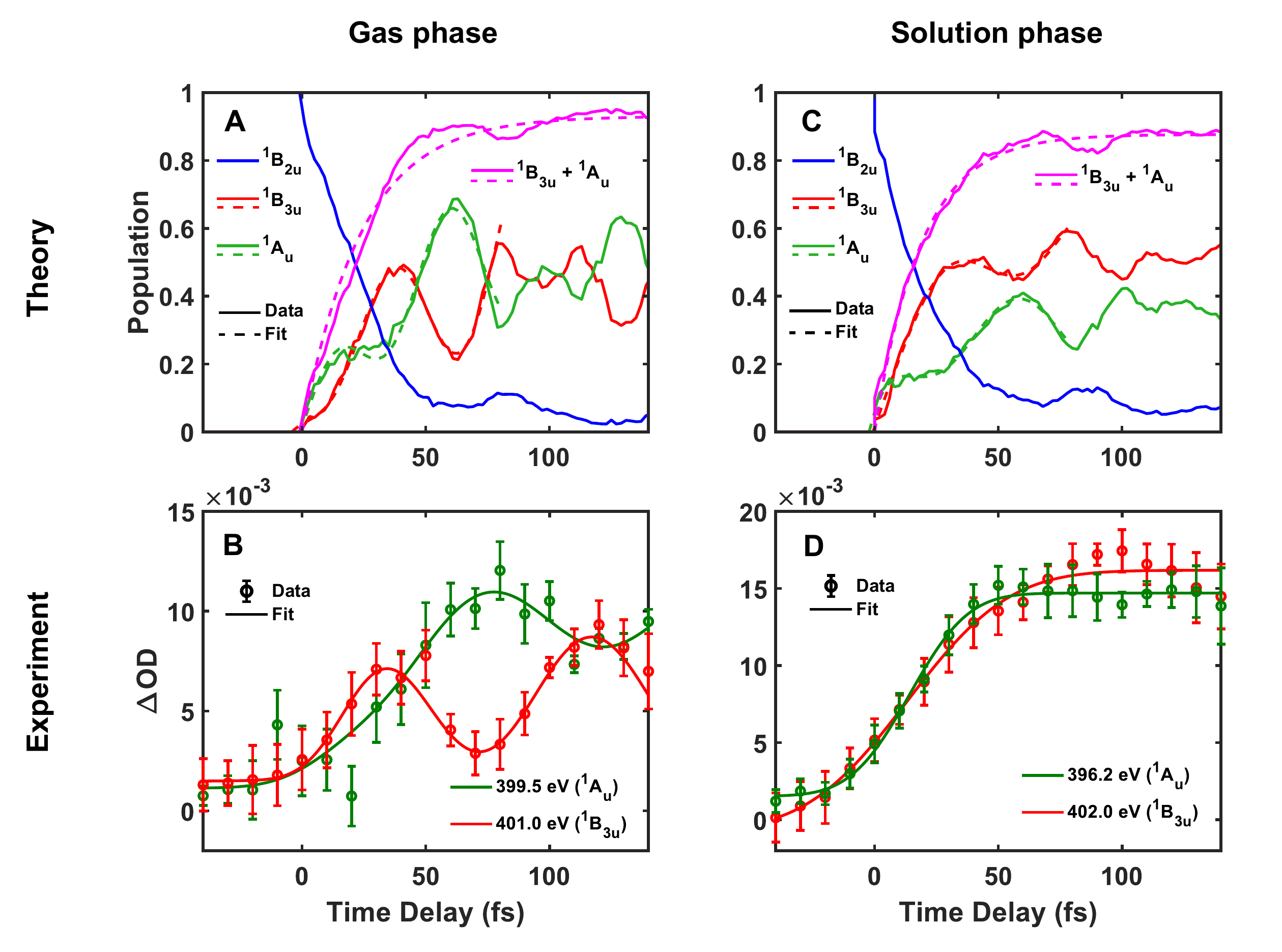}
\caption{\textbf{Observation of electronic and vibrational dynamics in gas-phase pyrazine and their dephasing in aqueous solution} \textbf{(A, C)} In gas phase and solution phase, respectively, populations of the diabatic $^1$B$_{\mathrm{2u}}$($\pi\pi^*$), $^1$B$_{\mathrm{3u}}$(n$\pi^*$) and $^1$A$_{\mathrm{u}}$(n$\pi^*$) states based on FSSH trajectories. \textbf{(B)} Time-dependent $\Delta$OD of gaseous pyrazine for the bands centered at 399.5~eV and 401.0~eV. According to calculations shown in Fig.\,\ref{fig:2}B, these bands are associated with the $^1$A$_{\mathrm{u}}$(n$\pi^*$) and $^1$B$_{\mathrm{3u}}$(n$\pi^*$) characters, respectively. \textbf{(D)} Time-dependent $\Delta$OD of 5M aqueous pyrazine for the bands centered at 396.2~eV and 402~eV. According to calculations shown in Fig.\,\ref{fig:2}F, these bands are associated with the $^1$A$_{\mathrm{u}}$(n$\pi^*$) character and $^1$B$_{\mathrm{3u}}$(n$\pi^*$) characters, respectively. The error bars indicate the measured standard deviations over 12 and 9 sets of scans for gas and solution phase measurements, respectively.}
\label{fig:nitrogen lineouts}
\end{figure}

The time dependence of these nitrogen K-edge differential absorbance bands over 150 fs is shown in Fig.\,\ref{fig:nitrogen lineouts}, together with the calculated time-dependent populations (in Fig.~3A). These diabatic populations were obtained by projecting wave functions calculated along trajectories using the fewest-switches surface hopping (FSSH) method onto diabatic electronic wave functions defined at the ground-state minimum geometry \cite{pitesa2021combined}. These deep, counter-phased quantum beats indicate an efficient population transfer between the $^1$A$_{\mathrm{u}}$ and $^1$B$_{\mathrm{3u}}$ states. The calculated spectral intensities corresponding to these FSSH calculations display the same trends as the population dynamics as shown in the SI (Fig.~S21), where additional details on these calculations are provided.
In gaseous pyrazine above the nitrogen pre-edge, the 399.5 and 401 eV bands (Fig.\,\ref{fig:nitrogen lineouts}B) are fitted with the product of a step function and a cosine convoluted with a Gaussian of 30 fs full width at half maximum (FWHM). From peak to peak, the 401 eV band rises earlier than the 399.5 eV band by $\sim$40 fs. It has two local maxima at $\sim$40 fs and $\sim$120 fs and a local minimum at $\sim$70 fs, with a period of $\sim$80 fs. The 399.5~eV band has local maxima at $\sim$80 fs and $\sim$150 fs and a local minimum at $\sim$120 fs, with a period of $\sim$70 fs. Given that the 399.5 and 401 eV bands are associated with dominant $^1$A$_{\mathrm{u}}$(n$\pi^*$) and $^1$B$_{\mathrm{3u}}$(n$\pi^*$) characters, respectively, the observed $\Delta$OD quantum beats in the two bands reflect the electronic dynamics involving these two states. When fitted with a sigmoidal function convoluted with a cosine, both the 399.5 and 401 eV bands have a rise time of $50 \pm 30$ fs with a time constant of $30 \pm 15$~fs, and are nearly $\pi$ out of phase with each other.

The experimentally observed quantum beat agrees well with the predictions of the gas-phase calculations shown in Fig.\,\ref{fig:nitrogen lineouts}A. The red and green curves for $^1$B$_{\mathrm{3u}}$(n$\pi^*$) and $^1$A$_{\mathrm{u}}$(n$\pi^*$), respectively, are fitted with the product of a rising exponential and a cosine function, revealing that they are close to $\pi$ out of phase with respect to each other. The brown curve for the summed populations of the $^1$B$_{\mathrm{3u}}$(n$\pi^*$) and $^1$A$_{\mathrm{u}}$(n$\pi^*$) states is fitted with a rising exponential function, yielding a time constant of $24.0 \pm 1.2$ fs.

At the nitrogen K-edge in aqueous pyrazine in Fig.\,\ref{fig:nitrogen lineouts}D, the 396.2 and 402 eV bands, which are associated with the $^1$A$_{\mathrm{u}}$(n$\pi^*$) and $^1$B$_{\mathrm{3u}}$(n$\pi^*$) characters, respectively, do not display any visible quantum beats. These data were therefore fitted with sigmoidal functions. The 396.2 and 402 eV bands have rise times of $50 \pm 20$ fs and $80 \pm 20$ fs, respectively. 
As a caveat, it is worth mentioning that observation of pure $^1$A$_{\mathrm{u}}$(n$\pi^*$)/$^1$B$_{\mathrm{3u}}$(n$\pi^*$) dynamics is prevented by overlap of three peaks at 396.2, 396.1 and 395.2 eV, which are assigned to transitions from the $^1$A$_{\mathrm{u}}$(n$\pi^*$), $^1$B$_{\mathrm{2u}}$($\pi\pi^*$) and $^1$B$_{\mathrm{3u}}$(n$\pi^*$) states, respectively. Nevertheless, the absence of quantum beats in the 402.0 eV band is a reliable indicator of the lack of quantum beats in the solution phase. The calculations shown in Fig.\,\ref{fig:nitrogen lineouts}C, which include two explicit water molecules and a conductor-like screening model (COSMO), do indeed predict a smaller (roughly by a factor of two) contrast of the quantum beat compared to the isolated-molecule calculations (Fig.\,\ref{fig:nitrogen lineouts}A). The complete suppression of the quantum beats in the experiment must therefore involve additional mechanisms, such as decoherence and dissipation, that go beyond those included in our simulations. The calculated rise time of the summed populations in solution phase is shorter than in the gas-phase calculations and amounts to $18.8 \pm 0.6$ fs.

\begin{figure}[h!]
\centering\includegraphics[width=\textwidth]{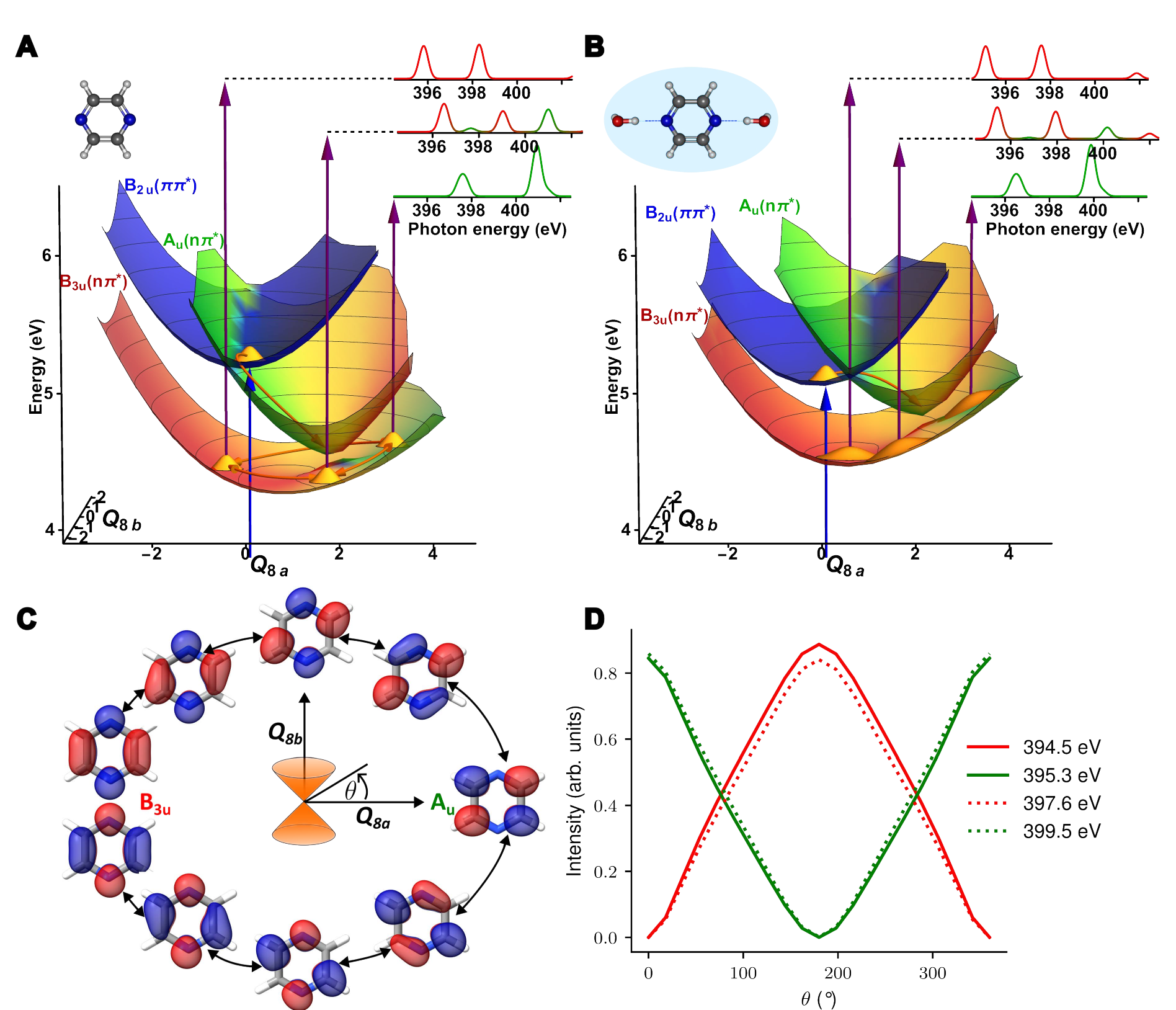}
\caption{{\bf Wavepacket dynamics in isolated and solvated pyrazine molecules.}
(\textbf{A}) Adiabatic potential-energy surfaces of the three lowest excited states of isolated pyrazine as a function of the $Q_{\rm 8a}$ and $Q_{\rm 8b}$ normal modes computed at the ADC(2)/aug-cc-pVDZ level of theory. Surfaces are colored according to contributions of $^1$B$_{\mathrm{3u}}$, $^1$A$_{\mathrm{u}}$ and $^1$B$_{\mathrm{2u}}$ electronic character in the wave functions of the excited states. (\textbf{B}) Same as (\textbf{A}) for the case of solvated pyrazine calculated with two explicit water molecules hydrogen-bonded to the nitrogen atoms and COSMO. Details of the calculations are given in the SI. (\textbf{C}) Particle NTOs for the S$_1$ state at geometries sampling a path around the conical intersection as indicated by arrows on the S$_1$ surfaces in (\textbf{A}). The colors of the isosurfaces encode the signs of the NTOs. (\textbf{D}) Relative intensities of absorption bands characteristic of the $^1$A$_{\mathrm{u}}$ (green) and $^1$B$_{\mathrm{3u}}$ (red) states along a 2D circular path spanned by the $Q_{\rm 8a}$ and $Q_{\rm 8b}$ normal modes encircling the $^1$B$_{\mathrm{3u}}$/$^1$A$_{\mathrm{u}}$ CI. 
}
\label{interpretation}
\end{figure}

We now discuss these new insights into the creation of electronic and vibrational dynamics in isolated and solvated pyrazine molecules in the light of the theoretical results. After photoexcitation to the $^1$B$_{\rm 2u}$ state, the molecular system moves away from the 
FC geometry in $<$30 fs following the slope of the excited-state potential-energy surface. This initial motion stabilizes all three excited states with respect to the ground state and shifts the corresponding XA peaks. After this, we find that the positions of the bands remain fairly constant and, especially at the nitrogen K-edge, they are most easily assigned based on those calculated at the minimum of the S$_1$ state. 
Our experimental results obtained at both edges and in both phases of matter clearly show that both the $^1$A$_{\mathrm{u}}$(n$\pi^*$) and the $^1$B$_{\mathrm{3u}}$(n$\pi^*$) states are populated within a few tens of femtoseconds, creating electronic and vibrational dynamics in the gas phase that are suppressed in solution phase. The spectra obtained in solution phase are analogous to those from the gas phase, but with some notable differences. Peaks corresponding to the $^1$A$_{\mathrm{u}}$(n$\pi^*$) state appear to rise earlier than in the gas phase, likely due to 
$^1$A$_{\mathrm{u}}$/$^1$B$_{2\mathrm{u}}$ CI lying directly in the FC region, as a consequence of different solvation shifts of the two states.

To rationalize these findings, Fig.~\ref{interpretation} displays the potential-energy surfaces of the relevant electronic states of isolated (A) and solvated (B) pyrazine. We chose the $Q_{8a}$ and $Q_{8b}$ modes since the $^1$A$_{\rm u}$/$^1$B$_{\rm 3u}$ CI lies along the $Q_{8a}$ mode, and the $Q_{8b}$ mode is responsible for the coupling between these two states \cite{sala2014role}. This is further confirmed by the fact that the population transfer is suppressed when the vibronic coupling mediated by the $Q_{8b}$ mode is set to zero (Fig. S21). The $Q_{8a}$ and $Q_{8b}$ modes thus represent the pair of tuning and coupling modes, respectively, that are most relevant to the $^1$A$_{\rm u}$/$^1$B$_{\rm 3u}$ dynamics.
In the gas phase, the $^1$B$_{\rm 2u}$/$^1$A$_{\rm u}$ CI is very close to the FC region, explaining the rapid population of both states. After crossing the $^1$A$_{\rm u}$/$^1$B$_{\rm 3u}$ CI, electronic dynamics is created that involves both states in the gas phase and manifests itself in the observed quantum beats. Motion of the FSSH trajectories along these modes is shown in Fig. S20 and further analyzed in Figs. S21 and S22 by running MCTDH calculations on the potential-energy surfaces of Ref. \cite{sala2014role}.

The electronic dynamics underlying the observed quantum beats are illustrated in Fig.~\ref{interpretation}C, which shows the natural transition orbitals (NTOs) for the S$_1$ state at geometries sampling a closed path around the $^1$B$_{\rm 3u}$/$^1\!$A$_{\rm u}$ CI. Following the evolution of the NTOs reveals their rotation around the aromatic ring of pyrazine reminiscent of a ring current. Ring currents in molecules have attracted considerable interest in recent years (Refs.\cite{barth06a,mineo13a,yuan17ringcurrent}, and references therein), but experimental evidence for such dynamics has been lacking. Although a uni-directional ring current cannot be generated in our experiment for symmetry reasons, the observed quantum beats can be interpreted as the consequence of a circular shift of the NTO around the aromatic ring. Our FSSH calculations indeed show that the excited-state trajectories can be grouped into two categories that encircle the $^1$B$_{\rm 3u}$/$^1\!$A$_{\rm u}$ CI either clockwise or counter-clockwise (Fig. S20). The corresponding electronic dynamics are in excellent agreement with MCTDH calculations run on the same potential-energy surfaces (Fig. S21). The MCTDH calculations themselves provide a consistent picture by showing that the excited-state wavepacket splits in two components that encircle the CI either clockwise or counter-clockwise (Fig. S22). Overall, these results show that the electronic dynamics observed in this work corresponds to a circular rearrangement of the electronic structure around the aromatic ring of the molecule. 

Our calculations additionally clarify why the observed electronic dynamics are directly accessible at the nitrogen K-edge.
Whereas the NTO of the $^1$B$_{\rm 3u}$ state has large amplitude on the nitrogen atoms, that of the $^1\!$A$_{\rm u}$ state has none, providing an intuitive interpretation why the observed quantum beats are out of phase and best visible at the nitrogen 
K-edge, as illustrated in Fig.~\ref{interpretation}D. The intensities in Fig.~\ref{interpretation}D are calculated along a tight circular path around the CI ($r=0.1$ in dimensionless normal-mode coordinates), showing that they are indeed primarily sensitive to the electronic character.
Finally, because of the geometric phase \cite{longuet58a,berry84a}, a closed loop around a CI is expected to result in a reversal of the signs of both the electronic and the nuclear wave functions, which is reflected in the phase of the NTOs shown in Fig.~\ref{interpretation}C, as well as in a node that appears in the nuclear wavepacket after its two components have encircled the CI in opposite directions (Fig. S22, \cite{duan18a}).

Turning to the role of solvation, our calculations show that solvation stabilizes the $\pi\pi^*$ state, whereas it destabilizes both of the n$\pi^*$ states. This finding has been qualitatively reproduced for a microsolvation model~\cite{tsuru2023time} (for details see section S3.3 of the SI) and is in agreement with solvation effects that were computationally found for nucleobases \cite{improta16a}. As Fig.~\ref{interpretation} shows, the solvation effects on pyrazine thus change both the location of the CIs and their slopes. Solvation indeed moves the location of the $^1$B$_{\rm 2u}$/$^1$A$_{\rm u}$ CI and causes a mixture of the two configuration characters near the FC point, explaining the earlier rise times of the corresponding bands in solution phase.
Comparing the solvation effects on the two n$\pi^*$ states, we find that the $^1$A$_{\rm u}$ state is more destabilized by solvation than the $^1$B$_{\rm 3u}$ state. This leads to an energetic skew of the path on which the wave packet encircles the CI, contributing to the dephasing of the quantum beats. However, as mentioned above, our solution-phase calculations (shown in Figs.~\ref{fig:nitrogen lineouts}B and \ref{interpretation}B) only partially account for solvation effects by including two explicit water molecules, hydrogen-bonded to the nitrogen atoms of pyrazine, and COSMO. Importantly, this partial account of solvation is nevertheless sufficient to reduce the contrast of the quantum beats by a factor of two as compared to the gas phase (Fig.~\ref{fig:nitrogen lineouts}A) and Fig.~\ref{fig:nitrogen lineouts}C identifies the differential solvation shifts as the likely origin of this partial dephasing. Since our experimental results show no discernible quantum beats, we conclude that additional mechanisms of decoherence and dissipation must be causing the observed complete dephasing on a time scale shorter than half of the 70-80~fs period of the observed quantum beats. 

Our work shows that CI dynamics can create electronic dynamics that correspond to large-amplitude rearrangement of the electronic structure and simultaneously resolves a decades-old controversy regarding the electronic-relaxation pathway and dynamics of pyrazine. 

Such electronic dynamics may be exploited for efficient long-range charge transfer \cite{falke14a} or for inducing very intense magnetic fields \cite{barth06a}, enabling novel molecular functionalities. Turning to the solution-phase results, our study shows that solvation can lead to dephasing of electronic dynamics and change the time constants of electronic relaxation. Our dynamical calculations moreover show that effects beyond the solvation shifts of the electronic states, such as dissipation and decoherence, are responsible for the observed complete dephasing of the electronic dynamics within 40 fs. Our work thus quantifies the effects of complex fluctuating environments under ambient conditions on electronic and vibrational dynamics. Our study demonstrates a general approach to unraveling the impact of solvation on conical-intersection dynamics \cite{zinchenko2021}, which also includes the predicted creation of electronic coherences at conical intersections \cite{kowalewski15a} and exploring the applicability of the concepts of charge-directed reactivity \cite{remacle98a,yin23a} on solution-phase dynamics. 

\section*{Acknowledgments}
We thank S. Gosh and A. Djorovic for supporting experiments on the ultraviolet absorbance of pyrazine solutions.
We thank M. Moret, A. Schneider, M. Seiler for technical support.
S.T. acknowledges discussion on solvent effects with Dr. B. Sharma, Prof. C. H\"{a}ttig, and Prof. D. Marx, Ruhr-Universit\"{a}t Bochum.

\bmhead{Funding:} We gratefully acknowledge funding from an ERC Consolidator Grant (Project No. 772797-ATTOLIQ), projects 200021\_204844 (USCOBIX), 200021\_172946 and 200020\_204928, as well as the NCCR-MUST, funding instruments of the Swiss National Science Foundation. Z.Y. acknowledges financial support from an ETH Career Seed Grant No SEED-12 19-1/1-004952-00. B.N.C.T. acknowledges  support from the European Union’s Horizon 2020 Research and Innovation Programme under the
Marie Sk{\l}odowska-Curie Individual Fellowship (Grant Agreement 101027796). S.C. acknowledges the Independent Research Fund Denmark – Natural Sciences, DFF-RP2 Grant No. 7014-00258B. S.T. acknowledges funding from the Alexander von Humboldt Foundation under the Humboldt Research Fellowship (JPN 1201668 HFST-P). M.S. acknowledges funding from the Croatian Science Foundation (project IP-2020-02-9932).
A.C.P, and H.K. acknowledge funding from the Research Council of Norway through FRINATEK (project No. 275506).
We acknowledge computing resources from DeIC -- Danish Infrastructure Cooperation (grant No. DeiC-DTU-N3-2023027),
through UNINETT Sigma2--the National Infrastructure for High Performance Computing
and Data Storage in Norway (project No. NN2962k) and through the Advanced computing service provided by University of Zagreb University Computing Centre - SRCE.

\section*{Author Contributions:} Y.-P.C., T.B., and Z.Y. carried out the experiments and analyzed the data. B.N.C.T., M.S., S.T., and S.C. carried out the CAS/RASPT2 calculations. A.C.P., S.C., and H.K. realized the CC calculations. M.S. realized the ADC(2) and the surface-hopping calculations. J.-P.W. and H.J.W. supervised the experimental work. S.C., H.K., and M.S. supervised the theoretical work. All authors discussed the data and contributed to the manuscript.

\section*{Competing interests}
The authors declare no competing interests.


\section*{Methods}
\subsection*{Experimental Setup}
The primary light source of this experiment is a Ti:Sapphire regenerative amplifier providing 4 mJ pulses that are subsequently amplified in a cryogenically--cooled two--pass amplifier. The two--stage Ti:Sapphire-based laser system provides 17~mJ, $\sim$30 fs pulses at 1 kHz repetition rate. 90\% of the output is used to pump a BBO--based type--II parametric amplifier seeded with a white--light supercontinuum originating from the same pump pulse. The parametric amplifier delivers $\sim$1.6~mJ, $\sim$30 fs idler pulses centered at $\lambda = 1.8$ $\mu\textrm{m}$ wavelength that are passively carrier-envelope-phase (CEP) stabilized. This mid-IR output is then compressed down to sub--three--cycle pulses in a dual-filamentation setup to $\sim$12 fs \cite{Schmidt2018}.
The compressed pulses is then focused by a $f=250$\,mm spherical mirror into a high-pressure He gas cell where the broadband soft X-ray probe is generated and focused via a toroidal mirror with a focal spot size of $\sim$62 $\mu \textrm{m}$ on the sub-$\mu$m liquid flat-jet.
The passing beam is then diffracted on a flat-field varied-line-spacing (VLS) grating on a 2D CCD soft X-ray camera from Andor. The energy resolution of the soft X-ray VLS spectrometer is around 0.3 eV and 0.4 eV on the carbon and nitrogen K-edges, respectively \cite{Yin2015}.
The other 10\% of the initial 800 nm beam is triple frequency doubled via BBO crystals to create 266 nm pump pulses. The pump pulses were characterized using self--diffraction frequency-resolved optical gating (FROG), which provided a measured pulse length of around $\sim$30\,fs, which is dominating the overall experimental time resolution. 
The pump pulse travel distance is matched with the beam path length of the probe and a high precision delay stage varies the time delays. 
Two 100 $\mu$m Ti filters were used to filter out the mid-IR beam from the HHG and also for differential pumping between the spectrometer and the experimental interaction chamber. Additional details on the experimental setup can be found in Ref. \cite{Smith2020}.
For both gaseous and aqueous phase measurements, the error bars are the measured standard deviations over 12 and 9 sets of scans, respectively, each having 20 spectra for each time delay and every spectra having an exposure time of 4s (given a laser repetition rate of 1 kHz).

\subsection*{Sample-Delivery Systems}
To generate the sub-$\mu$m liquid flat-jet, we utilized two cylindrical jets with an orifice of $\sim$20 $\mu$m and let them collide under a specific angle of 48 deg. The home-built sample-delivery device allows thickness within hundreds of nm thickness as measurements showed \cite{yin2020}. A more detailed description of the flat jet system and its properties can be found in Refs. \cite{Luu2018,yin2020}. Aqueous solutions of 5M pyrazine were used for the solution phase measurements and were freshly prepared on each day. For the gas-phase experiment, a heating device was designed to evaporate the crystalline pyrazine samples into the gas phase at $\sim$51 $^{\circ}$C, just below its melting point.
This was done in the experimental chamber to avoid condensation on the way. The gas sample was then guided into a metallic cuvette with ca. 1 mm thickness. 

\subsection*{Theoretical Methods}
To assign the peaks in the experimental spectra, electronic structure calculations were conducted at the level of regularized multi-state restricted-active-space second-order perturbation theory \cite{RMS-CASPT2} RASPT2/RAS2(10e,8o) for both the initial and final states of core excitations.  The population dynamics was investigated by nonadiabatic dynamics simulation in the fewest-switches surface hopping (FSSH) scheme based on ADC(2)/aug-cc-pVDZ. Time-resolved differential absorbance spectra were constructed using the nuclear ensemble approach by calculating XAS spectra for an ensemble of geometries sampled from FSSH trajectories.

In the RASPT2 calculations, RAS2 consisted of the two n$_{\mathrm{N}}$, three $\pi$ and three $\pi^{\ast}$ orbitals. The RAS1 space contained all the carbon and nitrogen 1s orbitals in the spectral simulations at the carbon and nitrogen K-edges, respectively. The RAS1 space was fully occupied in the initial states (ground state or valence excited states). The final core excited states were obtained by enforcing single electron occupation in RAS1 using the HEXS projection technique available in OpenMOLCAS~\cite{li2023openmolcas}, and thus the molecular orbitals in the core excited states were optimized under the presence of core hole in RAS1. In the solution-phase calculations, the solvent effects were considered by embedding the pyrazine molecule in PCM. The peak assignments based on RASPT2 were confirmed by calculating the core excitation energies and oscillator strengths with CC3 and CCSDT.

The 170 and 187 initial conditions of the FSSH simulation were randomly prepared in the 4.56-4.92 eV energy window, which corresponded to the low-energy part of the second absorption band in UV-Vis absorption, in the gas and aqueous-solution phases, respectively. Solvent effects were considered by including two explicit water molecules and COSMO in the aqueous solution phase. Each trajectory was propagated with a timestep of 0.5 fs up to 200 fs.

Further details concerning the calculations, including a more exhaustive list of references to the methods utilized, are given in the SI.


\section*{Data Availability}
The datasets generated during and/or analyzed during the current study are available from the corresponding author on reasonable request.

\end{document}


\title[SI]{\vspace{-1cm}\centering Supporting Information for\\ Electronic dynamics created at conical intersections and its dephasing in aqueous solution}

\author[1,2]{\fnm{Yi-Ping} \sur{Chang}}
\equalcont{These authors contributed equally to this work.}

\author[1,3]{\fnm{Tadas} \sur{Balciunas}}
\equalcont{These authors contributed equally to this work.}

\author*[3,4]{\fnm{Zhong} \sur{Yin}}\email{yinz@tohoku.ac.jp}
\equalcont{These authors contributed equally to this work.}

\author[5]{\fnm{Marin} \sur{Sapunar}}
\equalcont{These authors contributed equally to this work.}

\author[6,7]{\fnm{Bruno N. C.} \sur{Tenorio}}
\equalcont{These authors contributed equally to this work.}

\author[8]{\fnm{Alexander C.} \sur{Paul}}
\equalcont{These authors contributed equally to this work.}

\author[9,10]{\fnm{Shota} \sur{Tsuru}}

\author*[8]{\fnm{Henrik} \sur{Koch}}\email{henrik.koch@ntnu.no}

\author*[1]{\fnm{Jean--Pierre} \sur{Wolf}}\email{jean-pierre.wolf@unige.ch}

\author*[6]{\fnm{Sonia} \sur{Coriani}}\email{soco@kemi.dtu.dk}

\author*[3]{\fnm{Hans Jakob} \sur{W{\"o}rner}}\email{hwoerner@ethz.ch}

\affil[1]{\orgdiv{GAP--Biophotonics}, \orgname{Universit{\'e} de Gen{\`e}ve}, \postcode{1205} \city{Geneva}, \country{Switzerland}}

\affil[2]{\orgname{European XFEL}, \orgaddress{\postcode{22689} \city{Schenefeld}, \country{Germany}}}

\affil[3]{\orgdiv{Laboratory of Physical Chemistry}, \orgname{ETH Z{\"u}rich}, \postcode{8093} \city{Z{\"u}rich}, \country{Switzerland}}

\affil[4]{\orgdiv{International Center for SynchrotronRadiation Innovation Smart}, \orgname{Tohoku University}, \postcode{980-8577} \orgaddress{\city{Sendai}, \country{Japan}}}

\affil[5]{\orgdiv{Division of Physical Chemistry}, \orgname{Ruđer Bošković Institute},  \postcode{10000} \city{Zagreb}, \country{Croatia}}

\affil[6]{\orgdiv{Department of Chemistry}, \orgname{Technical University of Denmark}, \postcode{2800} \orgaddress{\city{Kongens Lyngby}, \country{Denmark}}}

\affil[7]{\orgdiv{Instituto Madrile\~{n}o de Estudios Avanzados en Nanociencia}, \orgname{IMDEA-Nanociencia}, \postcode{28049} \orgaddress{\city{Madrid}, \country{Spain}}}

\affil[8]{\orgdiv{Department of Chemistry}, \orgname{Norwegian University of Science and Technology}, \postcode{7034} \orgaddress{\city{Trondheim}, \country{Norway}}}

\affil[9]{\orgdiv{Lehrstuhl f\"{u}r Theoretische Chemie}, \orgname{Ruhr-Universit\"{a}t Bochum}, \postcode{44801} \orgaddress{\city{Bochum}, \country{Germany}}}

\affil[10]{\orgdiv{RIKEN Center for Computational Science}, \orgname{RIKEN}, \postcode{650-0047} \orgaddress{\city{Kobe}, \country{Japan}}}

\maketitle

\clearpage

\tableofcontents

\section{Additional Transient-Absorption Data}
\subsection*{Time-dependence of differential absorbance at the carbon K-edge}

\begin{figure}[htbp]
\centering\includegraphics[width=0.8\textwidth]{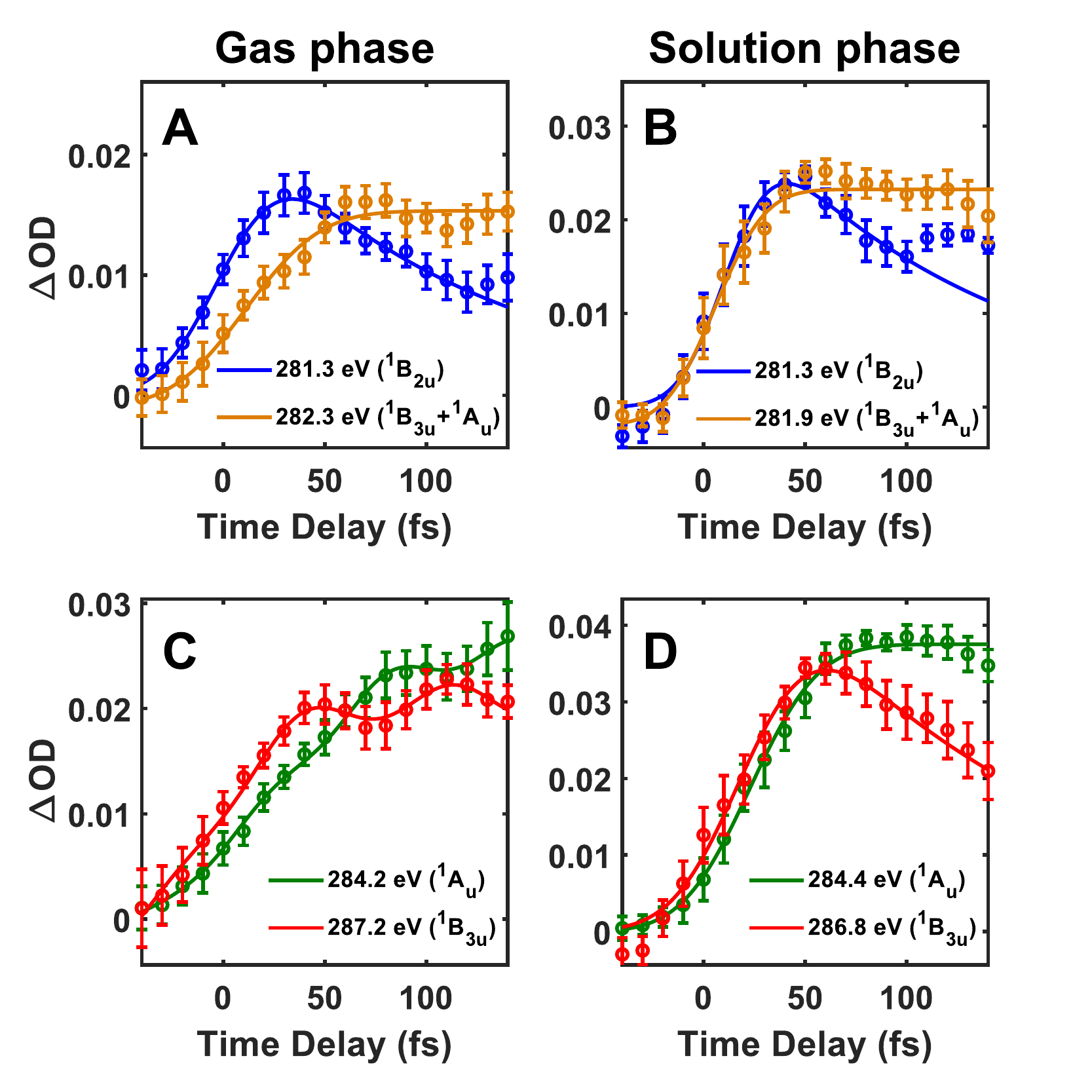}
\caption{Time dependence of the differential absorbance bands at {the} carbon K-edge. \textbf{(A)} Lineouts at the carbon K-edge of gaseous pyrazine for the 281.3 and 282.3 eV bands. According to calculations shown in Fig.\,\ref{figS:carbon dod}A, the 281.3 band is associated with the $^1$B$_{\mathrm{2u}}$ ($\pi\pi^*$) character, and the 282.3 eV band with mixed $^1$B$_{\mathrm{3u}}$(n$\pi^*$) and
$^1$A$_{\mathrm{u}}$(n$\pi^*$) characters, respectively. \textbf{(B)} Lineouts at the carbon K-edge of 5M aqueous pyrazine for 281.3 and 281.9 eV bands. According to calculations shown in Fig.\,\ref{figS:carbon dod}B, the 281.3 band is associated with the $^1$B$_{\mathrm{2u}}$($\pi\pi^*$) character, and the 281.9 eV band with mixed $^1$B$_{\mathrm{3u}}$(n$\pi^*$) and $^1$A$_{\mathrm{u}}$(n$\pi^*$) characters, respectively. \textbf{(C)} Lineouts at the 
carbon K-edge of gaseous pyrazine for 284.2 and 287.2 eV bands. According to calculations shown in Fig.\,\ref{figS:carbon dod}A, the 284.2 and 287.2 eV bands 
are associated with the
$^1$A$_{\mathrm{u}}$(n$\pi^*$) and $^1$B$_{\mathrm{3u}}$(n$\pi^*$) 
characters, respectively. The 284.2 eV band has an earlier rise-time than the 287.2 eV band and the two bands show weak out-of-phase oscillations.
\textbf{(D)} Lineouts at the carbon K-edge of 5M aqueous pyrazine for 284.4 and 286.8 eV bands. According to calculations shown in Fig.\,\ref{figS:carbon dod}B, the 284.4 eV band is associated with the $^1$A$_{\mathrm{u}}$(n$\pi^*$) character, while the 286.8 eV band is associated with the $^1$B$_{\mathrm{3u}}$(n$\pi^*$) character.}
\label{figS:carbon_lineouts}
\end{figure}

The time dependence of these carbon K-edge differential absorbance bands is shown in Fig.\,\ref{figS:carbon_lineouts}. 
In gaseous pyrazine, the 281.3-eV band is fitted with a convolution of the gaussian instrument-response function with an exponential decay function $e^{-t/\tau}$ and the 282.3-eV band with a sigmoidal function, as shown in Fig.\,\ref{figS:carbon_lineouts}A. The 281.3-eV band, associated with the $^1$B$_{\mathrm{2u}}$($\pi\pi^*$) character, has a time constant of $120 \pm 20$ fs. The 282.3 eV band, which is associated with mixed $^1$B$_{\mathrm{3u}}$(n$\pi^*$) and $^1$A$_{\mathrm{u}}$(n$\pi^*$) character, has a rise time of $70 \pm 20$ fs. 

The 284.2-eV and 287.2-eV bands (Fig.\,\ref{figS:carbon_lineouts}C) are both fitted with sigmoidal functions convoluted with cosines. The 284.2-eV band, which is associated with the $^1$A$_{\mathrm{u}}$(n$\pi^*$) character, has a rise-time of $130 \pm 40$ fs with a time constant of $70 \pm 20$ fs in the first 150 fs. The 287.2-eV band, which is associated with the $^1$B$_{\mathrm{3u}}$(n$\pi^*$) character, has a rise-time of $90 \pm 40$ fs with  a time constant of $50 \pm 20$ fs, and emerges earlier than the 284.2-eV band by $20 \pm 10$ fs. The two bands also show weak out-of-phase oscillations.

In aqueous pyrazine, again the 281.3~eV is fitted with a convolution of the gaussian instrument response function with an exponential decay function $e^{-t/\tau}$ and the 281.9 eV band with a sigmoid function, as shown in Fig.\,\ref{figS:carbon_lineouts}B. The 281.3 eV band, which is associated with the $^1$B$_{\mathrm{2u}}$($\pi\pi^*$) character, has a time constant of $130 \pm 40$ fs. The 281.9 eV band, which is associated with mixed $^1$B$_{\mathrm{3u}}$(n$\pi^*$) and $^1$A$_{\mathrm{u}}$(n$\pi^*$) character, has a rise time of $50 \pm 20$~fs. For the 284.4 and 286.8 eV bands (Fig.\,\ref{figS:carbon_lineouts}D), they are fitted with a sigmoidal function and an IRF convoluted with an exponential decay , respectively. The 284.4 eV band, which is associated with the $^1$A$_{\mathrm{u}}$(n$\pi^*$) character, has a rise-time of $70 \pm 10$ fs in the first 100 fs but starts declining in intensity afterward. The 286.8 eV band, which is associated with the $^1$B$_{\mathrm{3u}}$(n$\pi^*$) character, has a time constant of $140 \pm 40$ fs.

\subsection{Time-dependence of differential absorbance at the nitrogen K-edge}

\begin{figure}[htbp]
\centering\includegraphics[width=\textwidth]{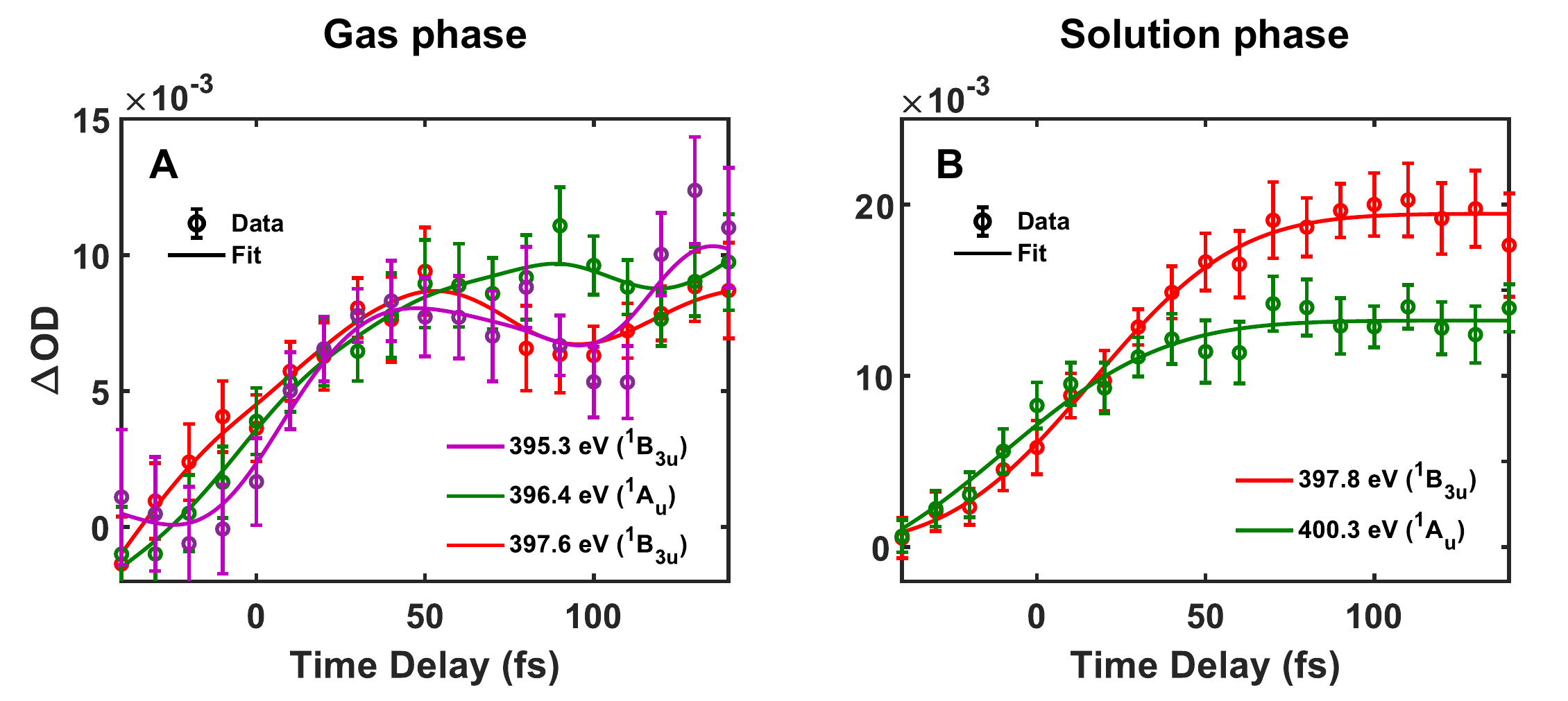}
\caption{ \textbf{(A,B)} Time dependence of differential absorbance bands at the nitrogen K-edge over 150 fs. \textbf{(A)} Lineouts at the nitrogen K-edge of gaseous pyrazine for 395.3, 396.4 and 397.6 eV bands. According to calculations shown in Fig.\,\ref{figS:nitrogen dod}A, the 395.3, 396.4 and 397.6 eV bands are associated with the $^1$B$_{\mathrm{3u}}$(n$\pi^*$), $^1$A$_{\mathrm{u}}$(n$\pi^*$) and $^1$B$_{\mathrm{3u}}$(n$\pi^*$) characters, respectively. \textbf{(B)} Lineouts at the nitrogen K-edge of 5M aqueous pyrazine for 397.8 eV and 400.3 eV bands. According to calculations shown in Fig.\,\ref{figS:nitrogen dod}B, the 397.8 eV and 400.3 eV bands are associated with the $^1$B$_{\mathrm{3u}}$(n$\pi^*$) and $^1$A$_{\mathrm{u}}$(n$\pi^*$) characters, respectively.}
\label{figS:nitrogen lineouts}
\end{figure}

Looking below the nitrogen pre-edge in gas phase, we have the polynomial fits of 395.3, 396.4 and 397.6 eV bands in Fig.\,\ref{figS:nitrogen lineouts}A, with character $^1$B$_{\mathrm{3u}}$(n$\pi^*$), $^1$A$_{\mathrm{u}}$(n$\pi^*$) and $^1$B$_{\mathrm{3u}}$(n$\pi^*$), respectively. These below-edge bands show weaker but still qualitatively comparable oscillations compared to the above-edge bands, showing a contrast in $\Delta$OD intensities near $\sim$90 fs. The weaker oscillations is likely due to a greater overlap of transition peaks below the pre-edge than above the pre-edge (Fig.~2B in the manuscript). Averaging all three of these bands together and fitting them with a sigmoid gives a rise time of $70 \pm 40$ fs. 

Above the nitrogen pre-edge in aqueous pyrazine in Fig.\,\ref{figS:nitrogen lineouts}B, the 397.8 eV and 400.3 eV bands, which are associated with the $^1$B$_{\mathrm{3u}}$(n$\pi^*$) and $^1$A$_{\mathrm{u}}$(n$\pi^*$) characters, respectively, do not display any visible oscillations. From sigmoidal function fits, the 397.8 and 400.3 eV bands have rise times of $90 \pm 20$ fs and $90 \pm 50$ fs, respectively.

\subsection{Depletion-corrected spectra}

\begin{figure}[htbp]
\centering\includegraphics[width=0.9\textwidth]{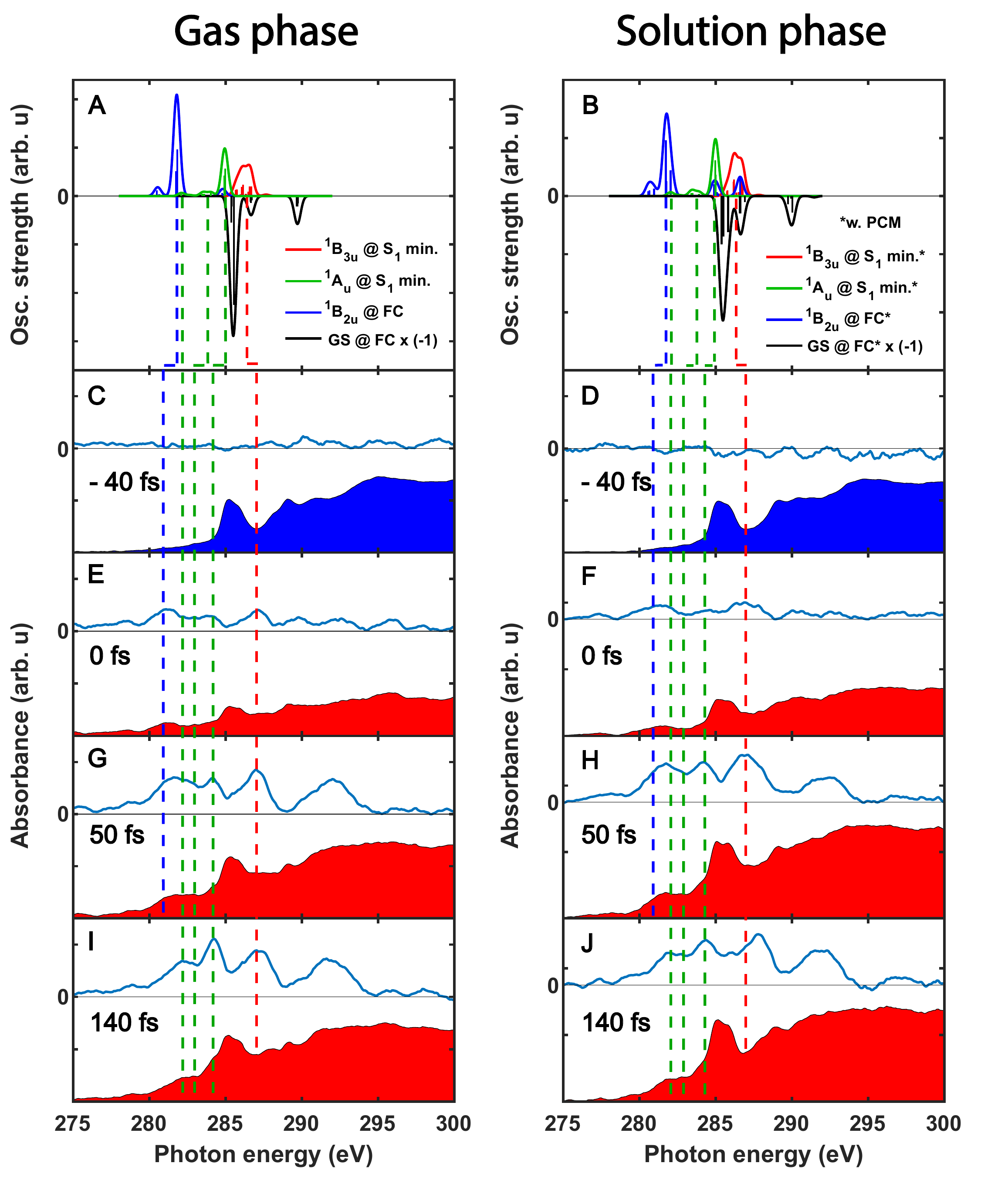} 
\caption{\textbf{(A)},\textbf{(B)} carbon K-edge excited-state XA spectra calculated at the RASPT2/RAS2(10e,8o) level at both the FC and relaxed geometries for the first valence-excited state (S$_1$), with and without PCM, respectively. Differential absorbance spectra (blue lines) and ground state depletion-corrected spectra (shaded areas) at the carbon K-edge of gaseous pyrazine (\textbf{(C)},\textbf{(E)},\textbf{(G)},\textbf{(I)}) and 5M aqueous pyrazine (\textbf{(D)},\textbf{(F)},\textbf{(H)},\textbf{(J)}) at different time delays of $-$40 fs, 0 fs, 50 fs, and 140 fs. The pure ground state spectra are shown by the blue-shaded areas at $-$40 fs and the depletion-corrected spectra by red-shaded areas at later delays.}
\label{figS:carbon dod}
\end{figure}

\begin{figure}[htbp]
\centering\includegraphics[width=0.9\textwidth]{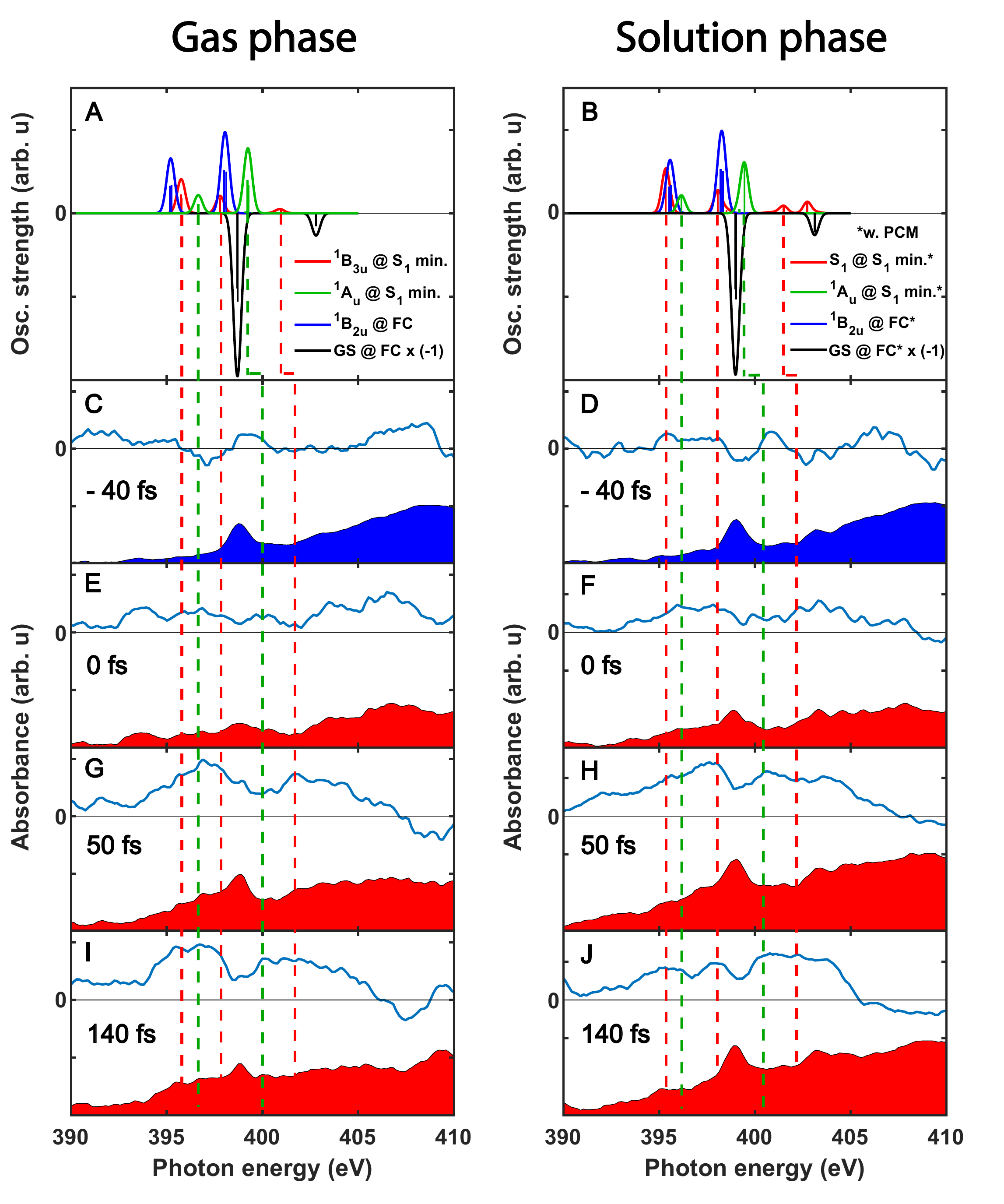} 
\caption{\textbf{(A)},\textbf{(B)} Nitrogen K-edge excited-state XA spectra calculated at the RASPT2/RAS2(10e,8o) level at both the FC and relaxed geometries for the first valence excited state (S$_1$), with and without PCM, respectively. Differential absorbance spectra (blue lines) and ground state depletion-corrected spectra (shaded areas) at the nitrogen K-edge of gaseous pyrazine (\textbf{(C)},\textbf{(E)},\textbf{(G)},\textbf{(I)}) and 5M aqueous pyrazine (\textbf{(D)},\textbf{(F)},\textbf{(H)},\textbf{(J)}) at different time delays of $-$40 fs, 0 fs, 50 fs and 140 fs. The pure ground state spectra are shown by the blue-shaded areas at $-$40 fs and the depletion-corrected spectra by red-shaded areas at later delays.}
\label{figS:nitrogen dod}
\end{figure}

This section presents depletion-corrected spectra at the carbon K-edge (Fig.~\ref{figS:carbon dod}) and the nitrogen K-edge (Fig.~\ref{figS:nitrogen dod}). These spectra were obtained by adding the ground-state spectra (blue shaded areas) scaled by the excitation fraction to the measured $\Delta$OD spectra at positive time delays. The excitation fraction was determined in each measurement by applying a positivity constraint to the depletion corrected spectra.

\subsection{Pump-probe scans extending to delays of 2 ps}

\begin{figure}[htbp]
\centering\includegraphics[width=\textwidth]{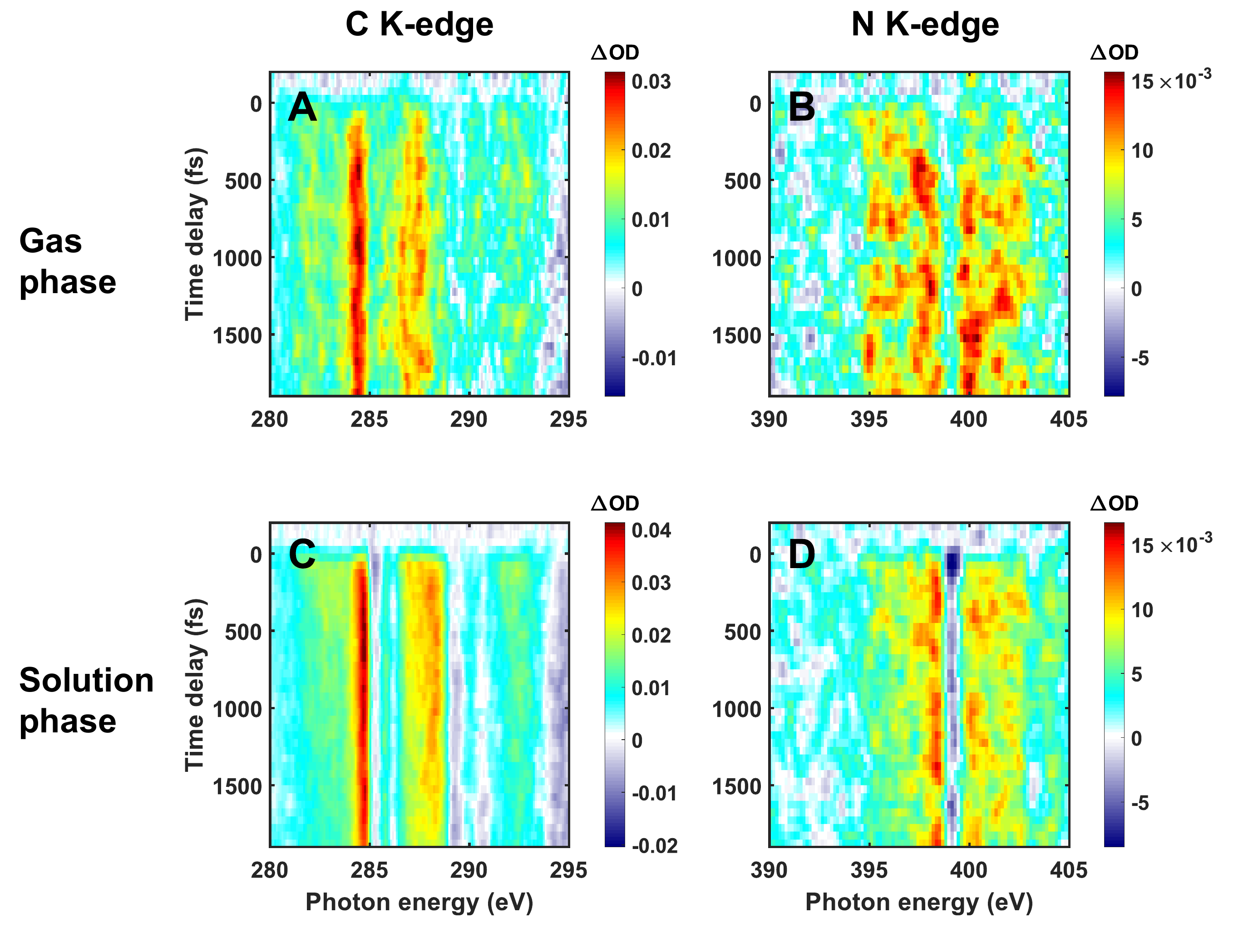}
\caption{Time-resolved differential absorbance spectra of pyrazine over the initial $\sim$2 ps. \textbf{(A,B)} Time-resolved differential absorbance spectra at the carbon and nitrogen 
K-edges of gaseous pyrazine, respectively. \textbf{(C,D)} Time-resolved differential absorbance spectra at the carbon and nitrogen K-edges of 5M aqueous pyrazine, respectively. These solution-phase data have been recorded with a $\sim$2 times higher pump intensity compared to the data shown in Fig. 2G,H of the main text.}
\label{figS:dod map 2ps}
\end{figure}

Measurements were also performed for the initial 2 ps delay range. These are shown in 
Fig.~\ref{figS:dod map 2ps} with mostly similar bands to the initial 150 fs measurements.

\clearpage

\section{Theoretical Methods}

\begin{figure}[hbtp!]
    \centering
\includegraphics[width=\textwidth]{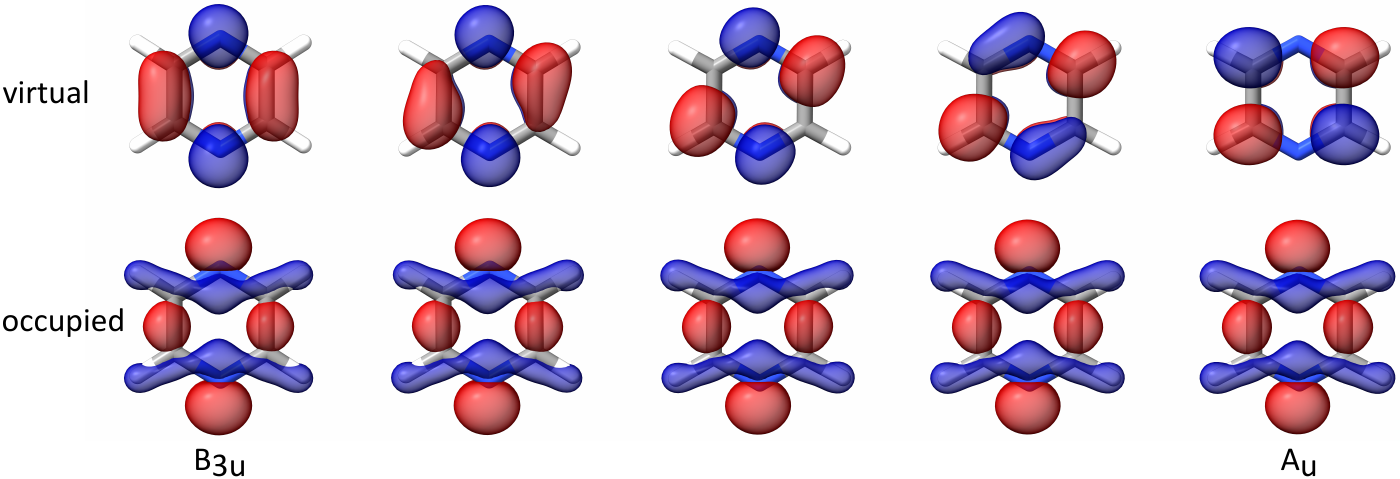}
    \caption{NTOs for the S$_1$ state at geometries along a path from a geometry where S$_1$ can be assigned to B$_{\mathrm{3u}}$
    to a geometry where the S$_1$ state can be assigned to A$_{\mathrm{u}}$ symmetry.
    The geometries used are part of a PES scan utilized to prepare Fig.~4.
The scan is a path around the conical intersection as indicated by the arrows on the S$_1$ surface in Fig.~4. The potential is symmetric and therefore completing the cycle is just mirroring the plot, as shown in Fig.~4 of the main text.}
    \label{fig:pyrazine_ntos_path}
\end{figure}

\clearpage

\begin{figure}[hbtp!]
    \centering
\includegraphics[width=0.85\textwidth]{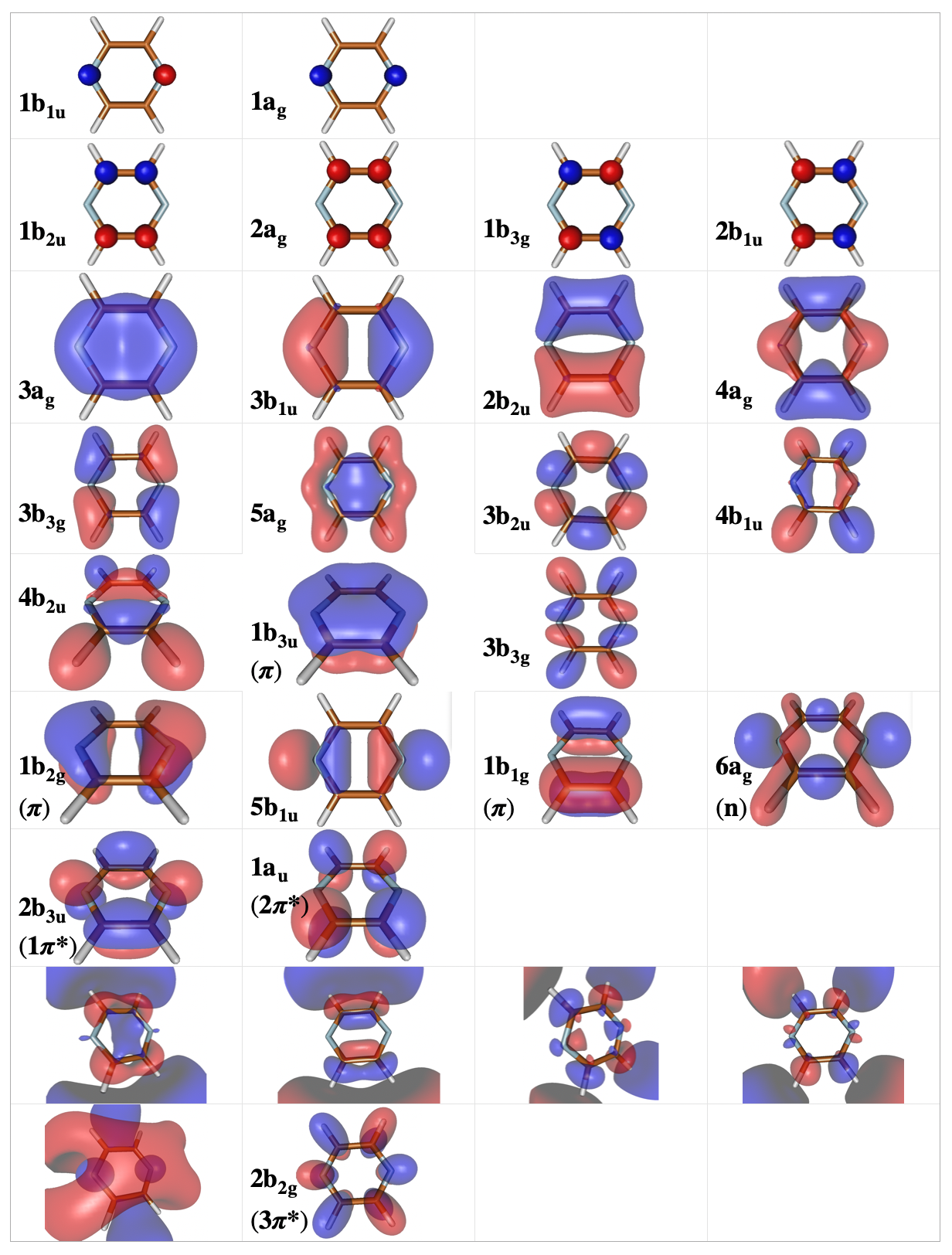}
    \caption{Occupied molecular orbitals and first virtual orbitals 
    of pyrazine (CAMB3LYP/cc-pVTZ). Mulliken symmetry notation.}
    \label{fig:MOs_pyrazine_Mulliken}
\end{figure}

\clearpage

\subsection{Symmetry analysis at the FC geometry ($D_{2h}$ symmetry)}
\begin{figure}[hbtp!]
    \centering
\includegraphics[width=\textwidth]{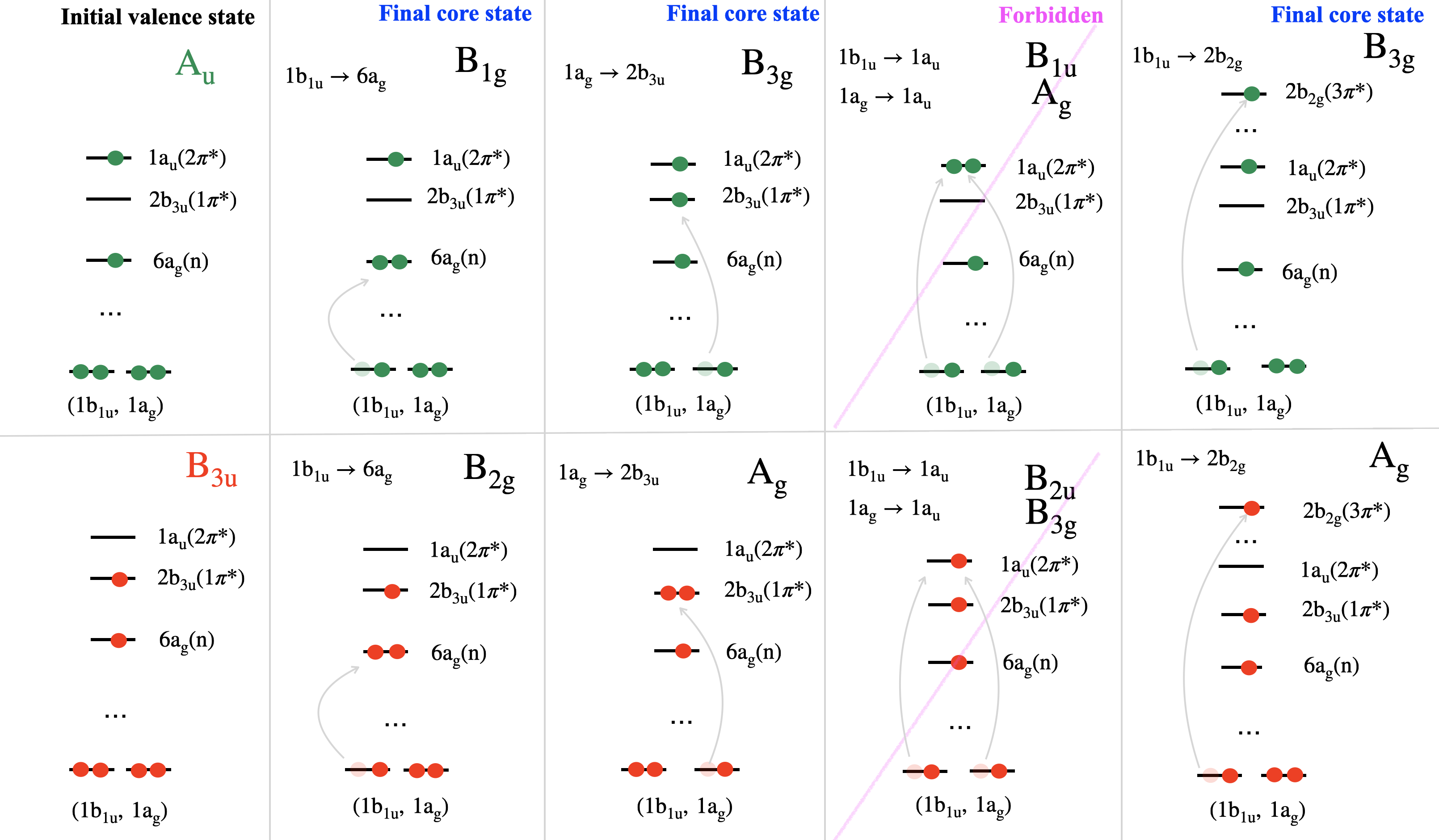}
\caption{Schematics of the main (dipole allowed) transitions in the XA spectra of the ${}^1$B$_{\mathrm{3u}}$(n$\pi^*$) and 
${}^1\mathrm{A_u}$(n$\pi^*$) 
states at $D_{2h}$ symmetry at the nitrogen K-edge. As indicated, the core excitations of the two states involve the same orbital transitions, but  the final symmetry of the core state is different.
From ${}^1\mathrm{A_u}$, final  dipole-allowed core state are $\mathrm{B_{1g}}$,
$\mathrm{B_{2g}}$ and 
$\mathrm{B_{3g}}$.
From ${}^1\mathrm{B_{3u}}$, final  dipole-allowed core state are $\mathrm{B_{2g}}$,
$\mathrm{B_{1g}}$ and 
$\mathrm{A_{g}}$.
Hence, transitions from the N 1s orbitals to $\mathrm{1a_u}$(2$\pi^*$) are dipole forbidden in $D_{2h}$.}
\label{fig:Symm_N-Kedge}
\end{figure}

\begin{figure}[hbtp!]
    \centering
\includegraphics[width=\textwidth]{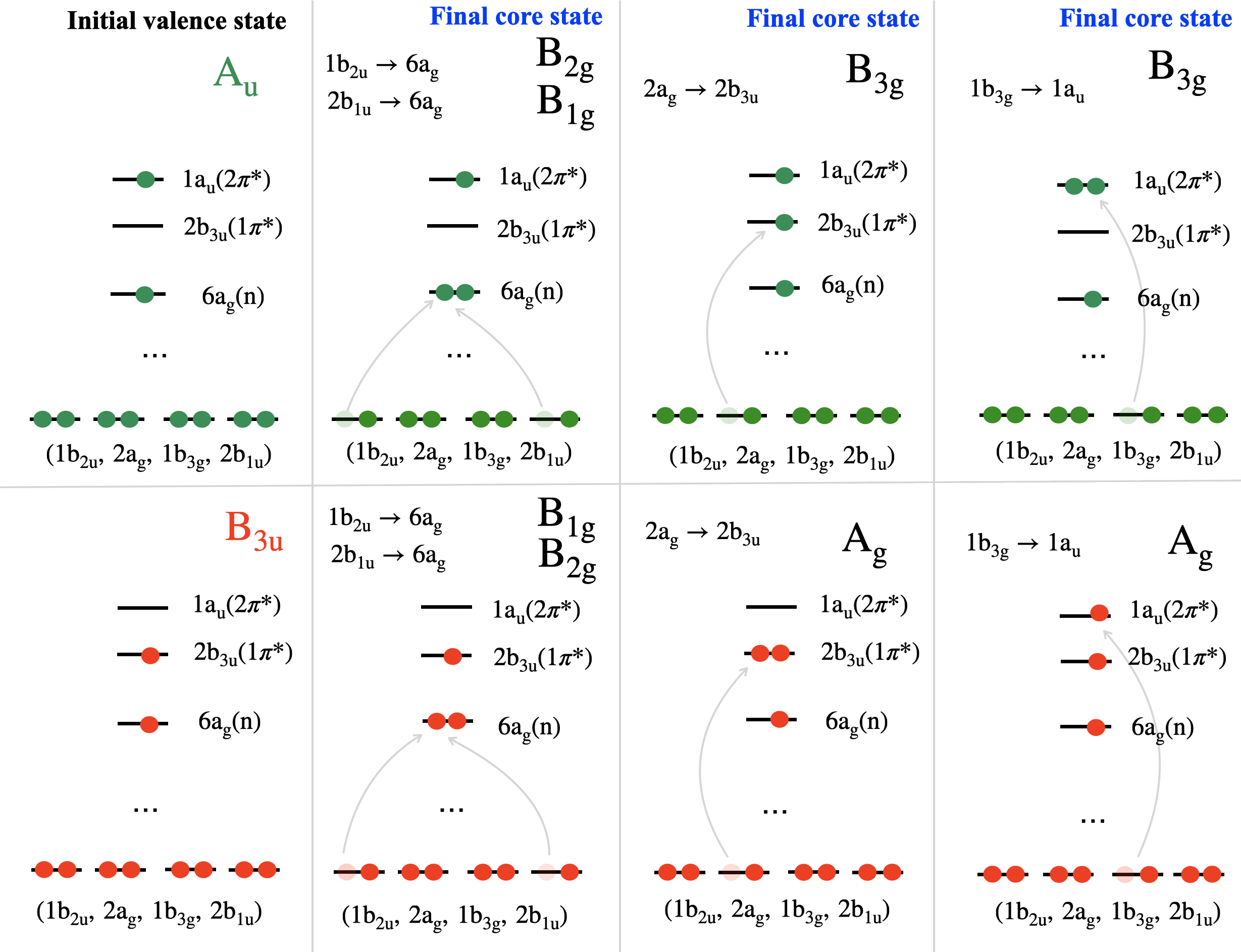}
    \caption{Schematics of the main transitions in the XA spectra of the $^1$B$_{\mathrm{3u}}$(n$\pi^*$) and $^1$A$_{\mathrm{u}}$(n$\pi^*$) states at $D_{2h}$ symmetry at the carbon K-edge.
    As indicated, the core excitations of the two states involve the same orbital transitions, but  the final symmetry of the core state is different.
    From ${}^1\mathrm{A_u}$, final  dipole-allowed core state are $\mathrm{B_{1g}}$,
$\mathrm{B_{2g}}$ and 
$\mathrm{B_{3g}}$.
From ${}^1\mathrm{B_{3u}}$, final  dipole-allowed core state are $\mathrm{B_{2g}}$,
$\mathrm{B_{1g}}$ and 
$\mathrm{A_{g}}$.}
    \label{fig:Symm_C-Kedge}
\end{figure}

\clearpage

\clearpage

\subsection{RASPT2 calculations}
RASPT2 calculations of the X-ray absorption spectra (XAS) of valence-excited states of pyrazine in combination with trajectory-based surface hopping (SH) dynamics performed at the ADC(2) level of theory
have been used to simulate the time-resolved X-ray absorption spectra (TR-XAS) at the nitrogen K-edge.

Moreover, we also provide an interpretation of the TR-XAS experiments measured at the nitrogen and carbon K-edges for longer time-delays 
with relaxed geometries of the S$_1$ state provided by CASPT2 geometry optimization. 
The influence of the solvent in the spectra is also investigated with a polarizable continuum model (PCM). 
For calculations at the nitrogen K-edge, we used the Roos Augmented Double Zeta ANO basis set \cite{Widmark1990} with the contraction [\textit{4s3p2d}] for \ce{N} and the ANO-L-DZVP basis set for \ce{C} and \ce{H}.
For the carbon K-edge calculations, we used the ANO-L-DZVP basis set on all atoms.

Core-excited states were computed by
placing the pertinent core orbitals in the RAS1 space and enforcing single electron occupation in it using the HEXS projection technique \cite{HEXS-RAS} available in {\sc OpenMolcas} ({\it 30\/}),
which corresponds to applying the core-valence separation~\cite{Cederbaum1980}.
RAS2 was used for complete electron distribution, i.e., to define the complete active space for the valence electrons while RAS3 was kept empty. 

\begin{figure}
    \centering
    \includegraphics[width=\textwidth]{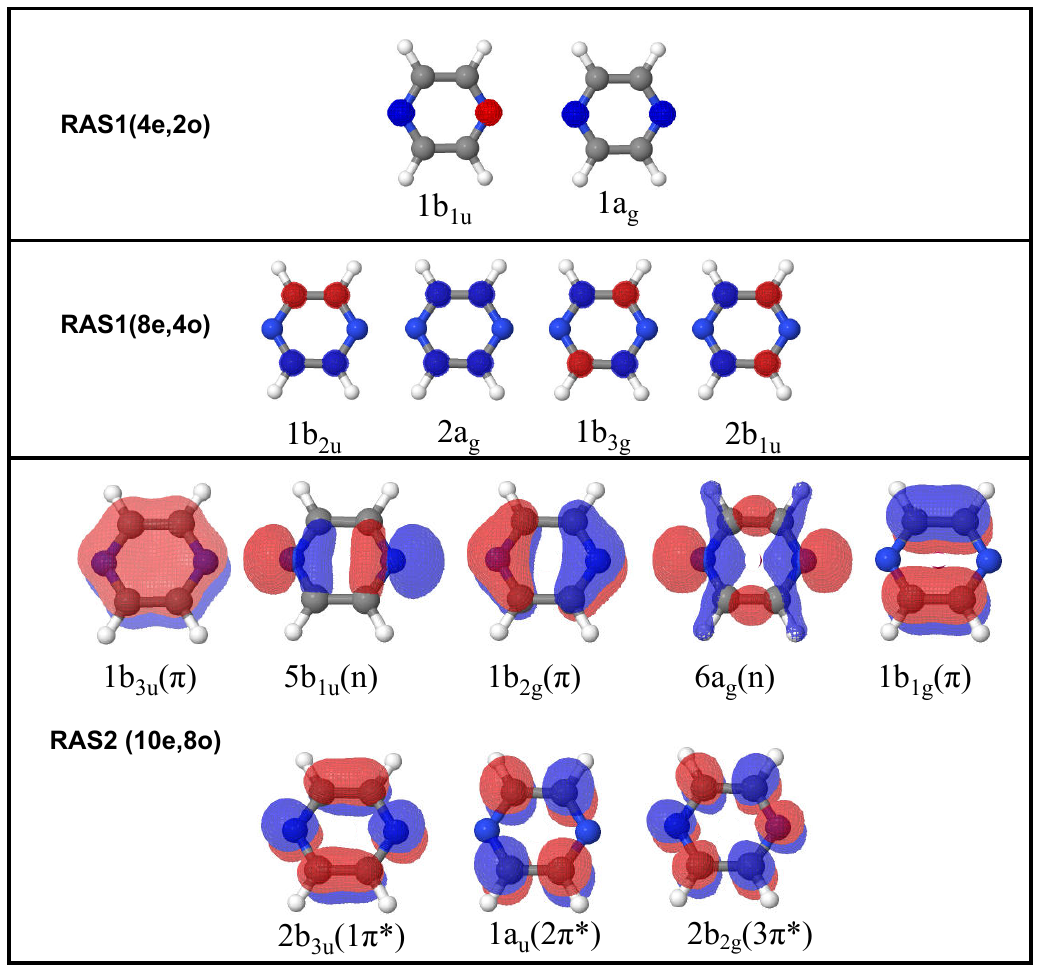}
    \caption{
        RASSCF active spaces. RAS1(4e,2o) + RAS2(10e,8o) are used for the nitrogen K-edge calculations, while RAS1(8e,4o) + RAS2(10e,8o) are used for the carbon K-edge calculations.}
    \label{rasscf}
\end{figure}

We used the same active space described previously by Northey et al.~\cite{C9CP03019K}: the  1s orbitals of interest in the RAS1 subspace (two orbitals in case of \ce{N} and four orbitals in case of \ce{C}) 
and eight valence orbitals and ten electrons in the RAS2 subspace (two n$_{\mathrm{N}}$ lone-pairs, three $\pi$ and three $\pi^*$). Throughout, we will refer to this active space with the short notation RASPT2/RAS2(10e,8o). 
The active orbitals computed at the Franck–Condon (FC) geometry are schematically represented in Fig. \ref{rasscf}.

Dynamical correlation effects are further included in the reference space using the regularized multi-state restricted-active-space perturbation theory of the second-order (RMS-RASPT2) approach \cite{RMS-CASPT2}. 
An imaginary level shift of 0.25 Hartree was applied to avoid intruder-state singularities. IPEA shift has not been used.
For the nitrogen K-edge calculations, initial valence and final core-excited states were obtained by state averaging over 5 and 15 states, respectively. 
For the carbon K-edge calculations, we state-averaged the same number of initial and 25 final core-excited states.

\begin{figure}[hbpt!]
    \centering
\includegraphics[width=0.7\textwidth]{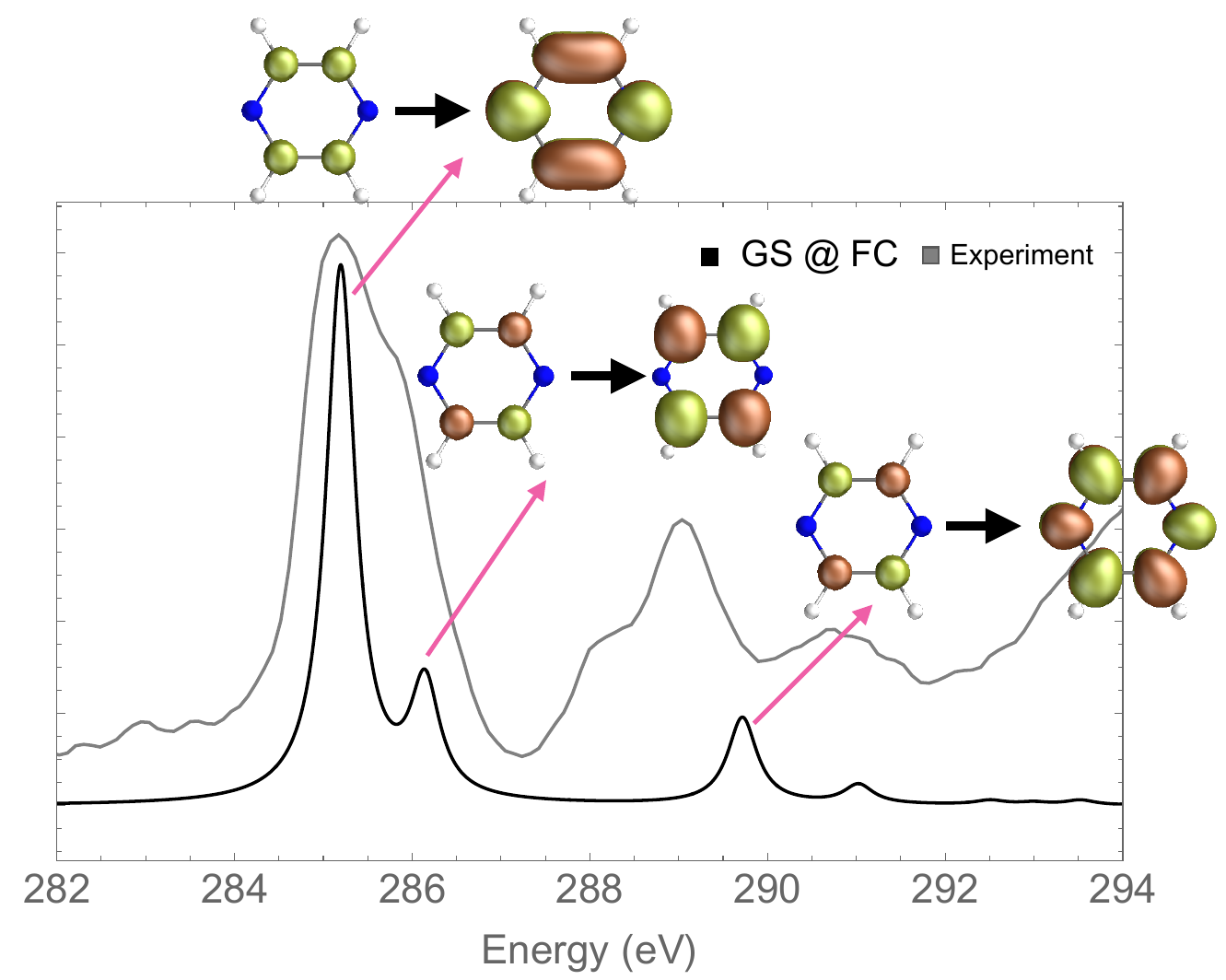}\\
\includegraphics[width=0.7\textwidth]{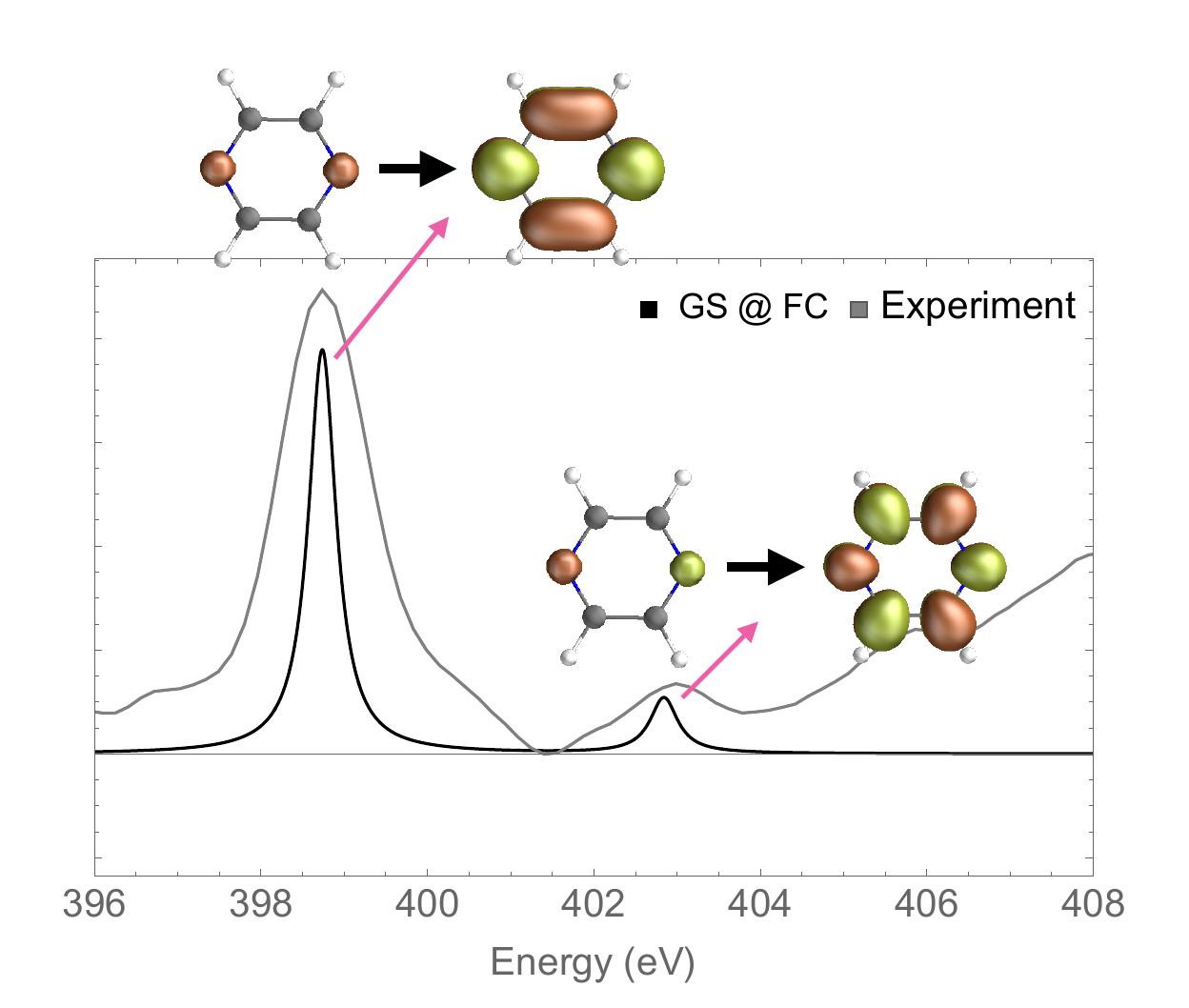}
    \caption{Ground-state XA spectra (with NTO-based peak-assignments) at the carbon (upper panel) and nitrogen (lower panel) K-edges computed at the RASPT2/RAS2(10e,8o) level versus experiment. {Note that no transitions to $\sigma^*$ are present in the RASPT2 simulated spectra, since these orbitals were not included in the active space.}}
    \label{fig:GS_both-K_edges_RASPT2}
\end{figure}

The ground-state XA spectra at the carbon and nitrogen K-edges computed at the RAS\-PT2/RAS2(10e,8o) level along with the natural transition orbitals (NTOs) of the main transitions are shown in Fig.~\ref{fig:GS_both-K_edges_RASPT2}. The choice of active space containing three $\pi^*$ orbitals and no $\sigma^*$ or Rydberg orbitals in the active space means that transitions to the latter orbitals cannot be described at the current level of theory. Therefore, some peaks are missing in the simulated spectrum, but the peaks that are present match very well with the experiment.

\begin{figure}[h]
\centering
    \begin{subfigure}[b]{0.45\textwidth}
    \centering
    \footnotesize
    \includegraphics[width=\textwidth]{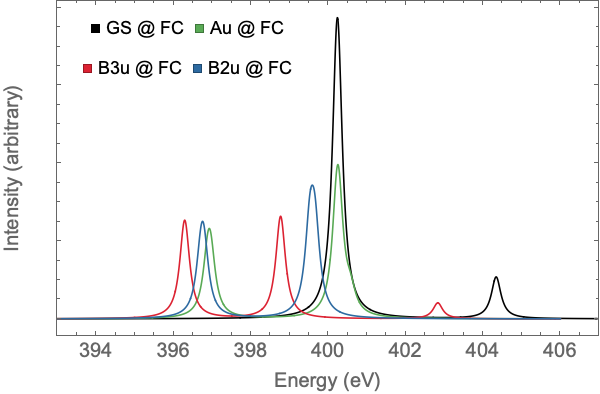}
    \caption{FC nitrogen K-edge}
    \label{4a}
    \end{subfigure}
    \hspace{0.5cm}
    \begin{subfigure}[b]{0.45\textwidth}
    \centering
    \footnotesize
    \includegraphics[width=\textwidth]{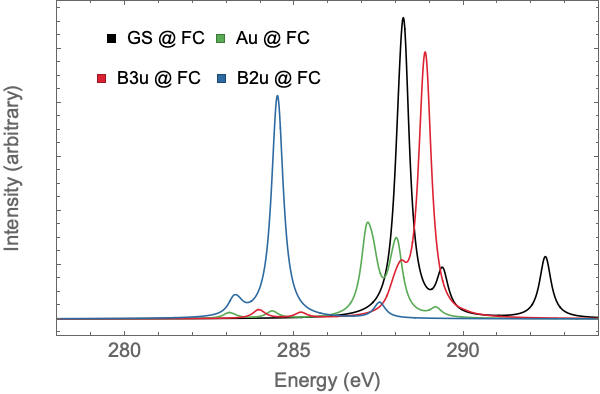}
    \caption{FC carbon K-edge}
    \label{4b}
    \end{subfigure} 
    \\
    \begin{subfigure}[b]{0.45\textwidth}
    \centering
    \footnotesize
    \includegraphics[width=\textwidth]{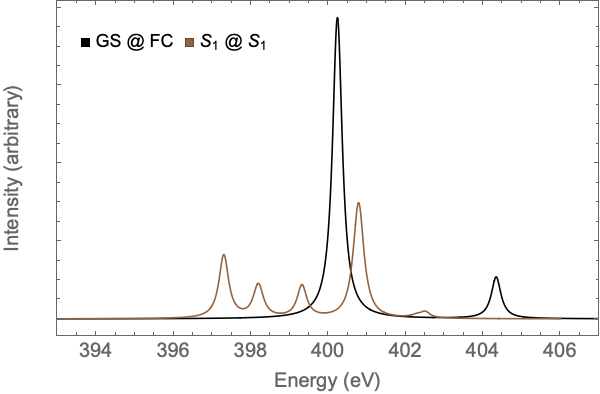}
    \caption{S$_1$ nitrogen K-edge}
    \label{4c}
    \end{subfigure}
    \hspace{0.5cm}
    \begin{subfigure}[b]{0.45\textwidth}
    \centering
    \footnotesize
    \includegraphics[width=\textwidth]{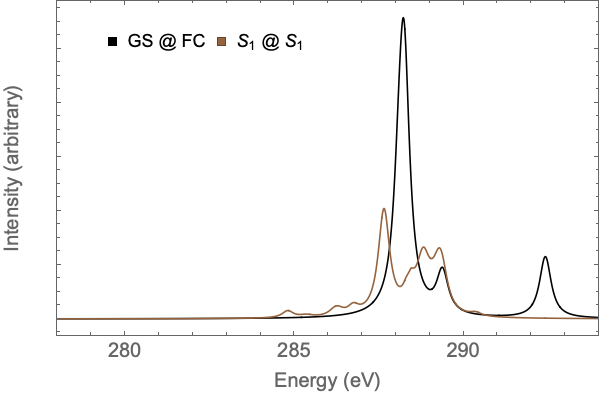}
    \caption{S$_1$ carbon K-edge}
    \label{4d}
    \end{subfigure}
\caption{RASPT2 spectra computed with the equilibrium geometry of the ground state (FC) and the equilibrium geometry of the first excited state (S$_1$) in vacuum.
The S$_1$ equilibrium geometry was optimized at the CASPT2 level. 
The computed spectra have been convoluted with Lorentzian functions using HWHM = 0.2 eV. No energy shift has been applied to the calculated energies.}
\label{raspt2_spec_vac}
\end{figure}

\begin{figure}[h]
\centering
    \begin{subfigure}[b]{0.45\textwidth}
    \centering
    \footnotesize
    \includegraphics[width=\textwidth]{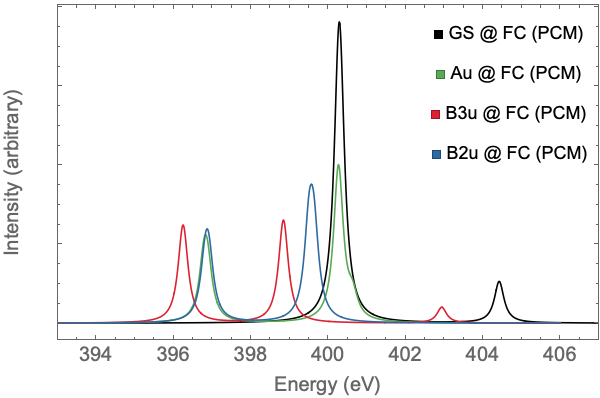}
    \caption{FC nitrogen K-edge}
    \label{4ap}
    \end{subfigure}
    \hspace{0.5cm}
    \begin{subfigure}[b]{0.45\textwidth}
    \centering
    \footnotesize
    \includegraphics[width=\textwidth]{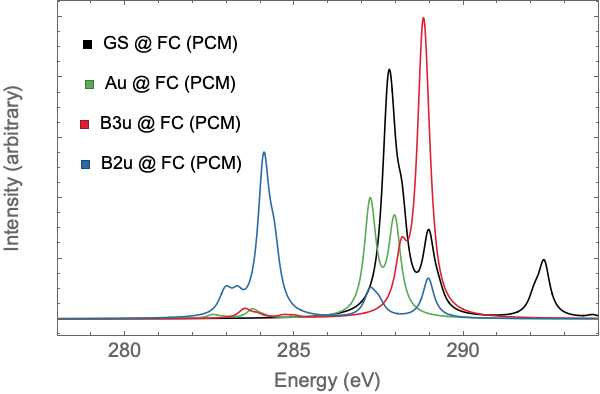}
    \caption{FC carbon K-edge}
    \label{4bp}
    \end{subfigure} 
    \\
    \begin{subfigure}[b]{0.45\textwidth}
    \centering
    \footnotesize
    \includegraphics[width=\textwidth]{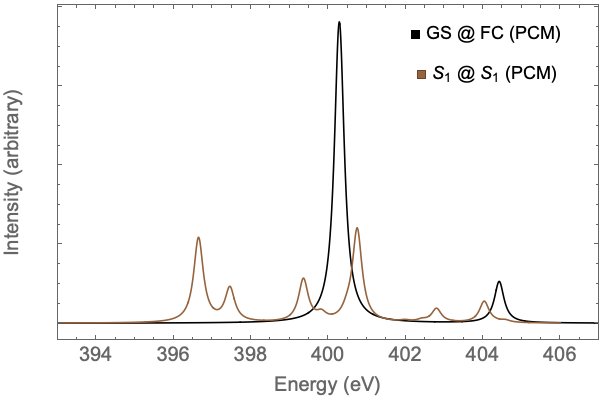}
    \caption{S$_1$ nitrogen K-edge}
    \label{4cp}
    \end{subfigure}
    \hspace{0.5cm}
    \begin{subfigure}[b]{0.45\textwidth}
    \centering
    \footnotesize
    \includegraphics[width=\textwidth]{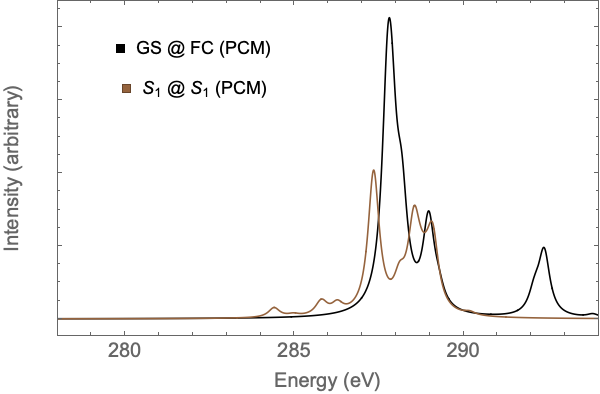}
    \caption{S$_1$ carbon K-edge}
    \label{4dp}
    \end{subfigure}
\caption{RASPT2 spectra computed with the equilibrium geometry of the ground state (FC) and the equilibrium geometry of the first excited state (S$_1$) using a polarizable continuum model (PCM) to mimic solvent effects. 
The computed spectra have been convoluted with Lorentzian functions using HWHM = 0.2 eV. No energy shift has been applied to the calculated energies.}
\label{raspt2_spec_pcm}
\end{figure}

The spectra computed with the equilibrium geometry of the ground state (FC) and the equilibrium geometry of the first excited state (S$_1$), both in vacuum, are shown in Fig.~\ref{raspt2_spec_vac}.
The same but using the PCM to mimic solvent effects on the liquid phase are shown in Fig.~\ref{raspt2_spec_pcm}.

One advantage of using the RASPT2 approach for TR-XAS is that it can reproduce all orbital configurations, such as $1h1p$, $2h2p$, $3h3p$, etc, within the RAS2 subspace. As long as the selected active space is large enough to account for all important orbital configurations, which is barely guaranteed.
On the other hand, the RASPT2 approach has some disadvantages when compared to other single reference methods. It tends to be computationally expensive and time-consuming. Additionally, the accuracy of the calculations strongly relies on the choice of the active space and the number of states included in the state average.

\subsection{CCSD and CC3 calculations} \label{sec:CC3}

Calculations on pyrazine in gas-phase using CCSD and CC3 were performed with the eT program ({\it 33, 34\/}).
Core-valence separation~\cite{CVS_CC} is employed to obtain core-excited states.
To ensure that the valence-excited states are orthogonal to the core-excited states the inverse projection
(i.e. removing the contributions of core orbitals) 
is used for the valence excited states.

\begin{figure}[hbpt!]
    \centering
    \includegraphics[width=0.65\textwidth]{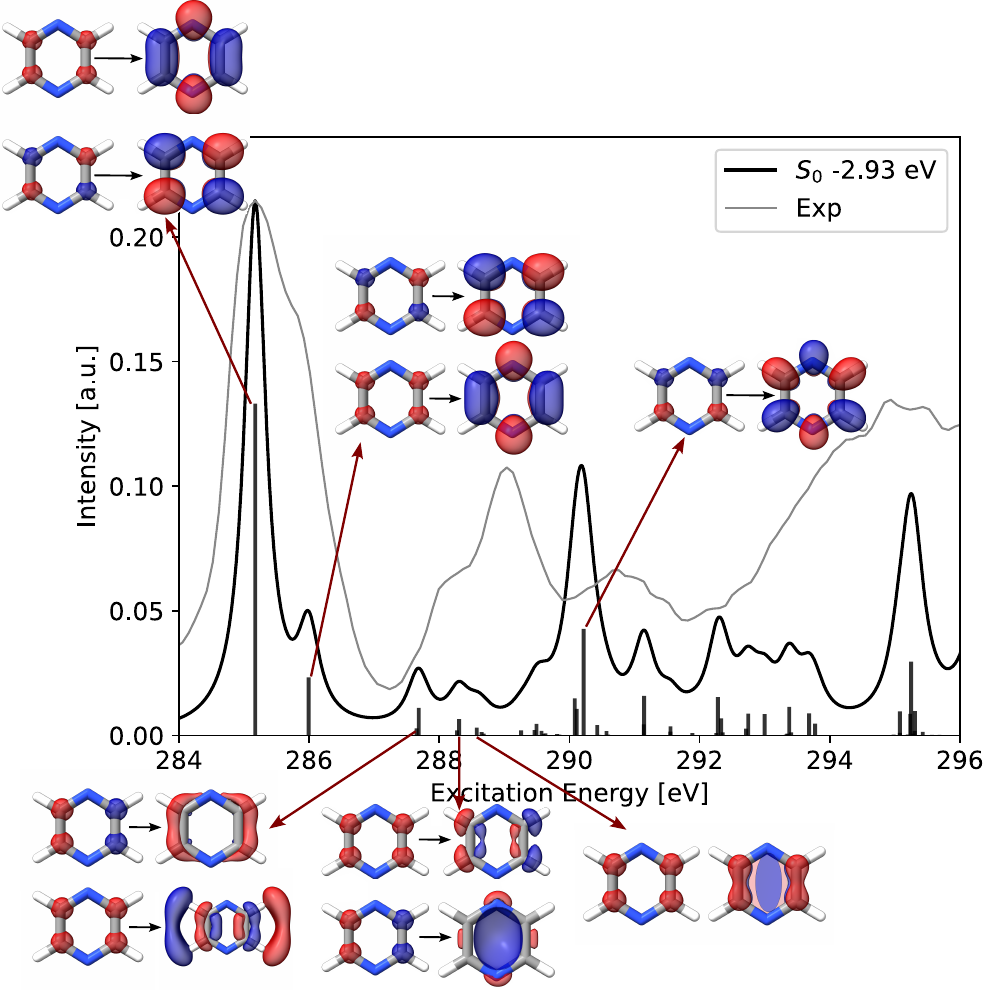}\\[.5cm]
    \includegraphics[width=0.66\textwidth]{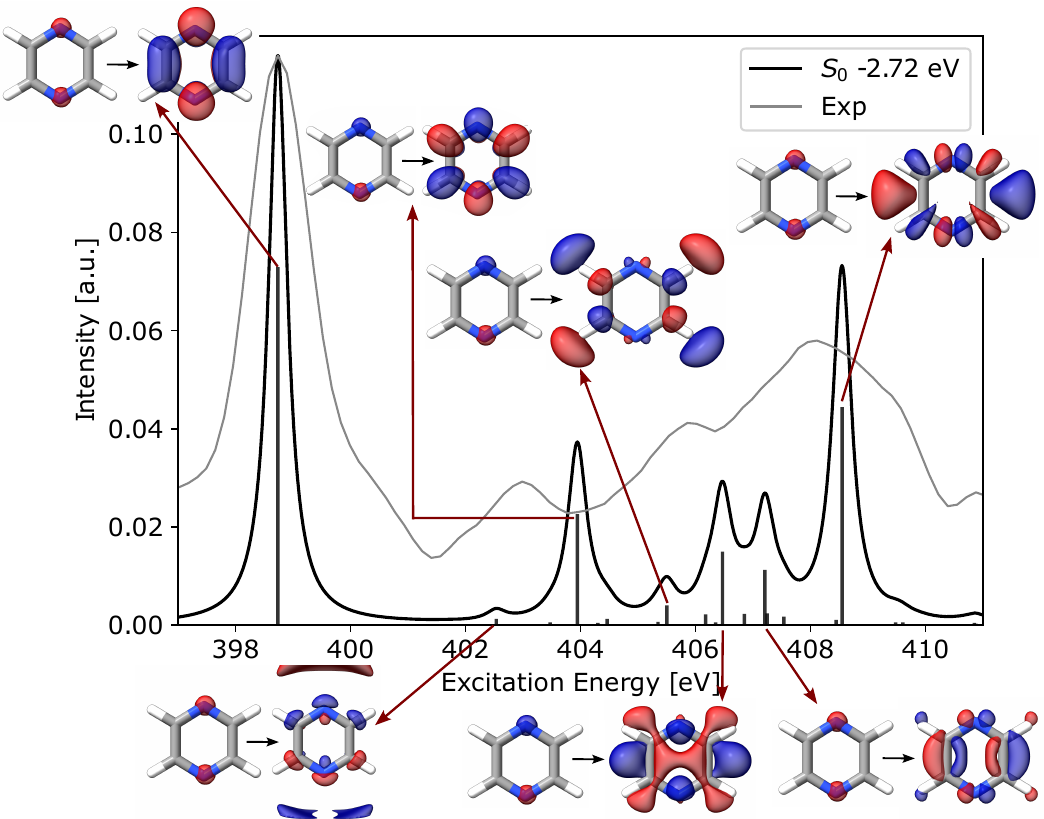}
    \caption{
    Ground state XA spectra at the carbon and nitrogen K-edges computed at the CC3 level of theory using the cc-pVDZ basis 
    plus additional Rydberg type functions generated according to Kaufmann, Baumeister and  Jungen's prescription~\cite{KBJ_bases}
    with quantum numbers $n$~=~2, 2.5, 3 and angular momentum $s$ and $p$. NTOs of the main transitions are reported for assignment.
    }
\label{fig:cc3_C_and_N_gs_dz_Ry}
\end{figure}

\begin{figure}[h]
\centering
    \begin{subfigure}[b]{0.45\textwidth}
        \centering
        \footnotesize
        \includegraphics[width=\textwidth]{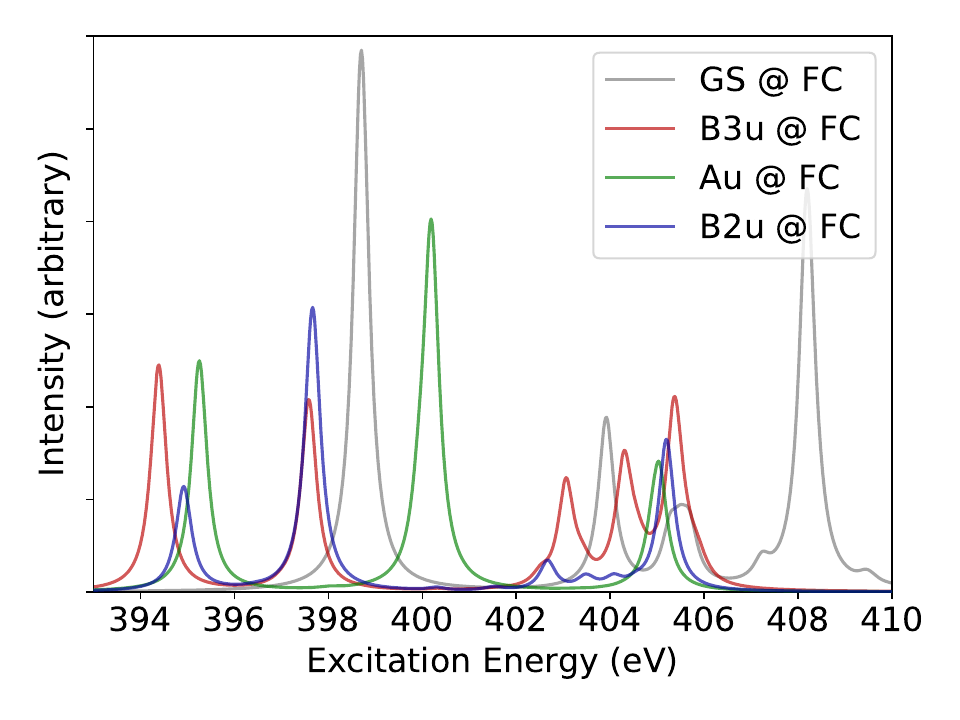}
        \caption{FC nitrogen K-edge}
    \end{subfigure}
    \begin{subfigure}[b]{0.45\textwidth}
        \centering
        \footnotesize
        \includegraphics[width=\textwidth]{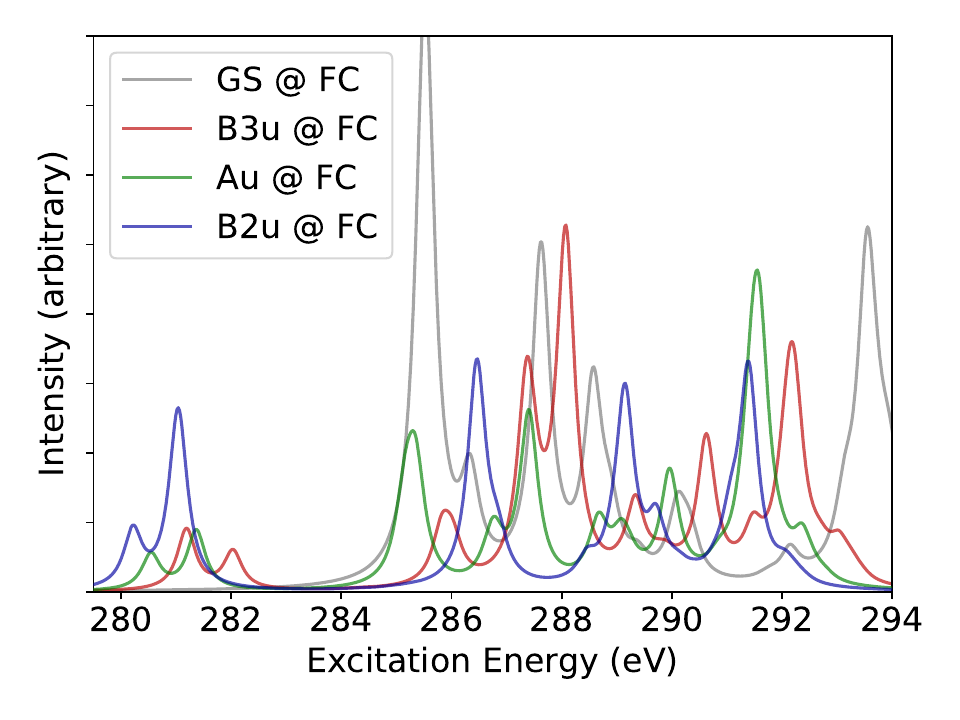}
        \caption{FC carbon K-edge}
    \end{subfigure}
    \\
    \begin{subfigure}[b]{0.45\textwidth}
        \centering
        \footnotesize
        \includegraphics[width=\textwidth]{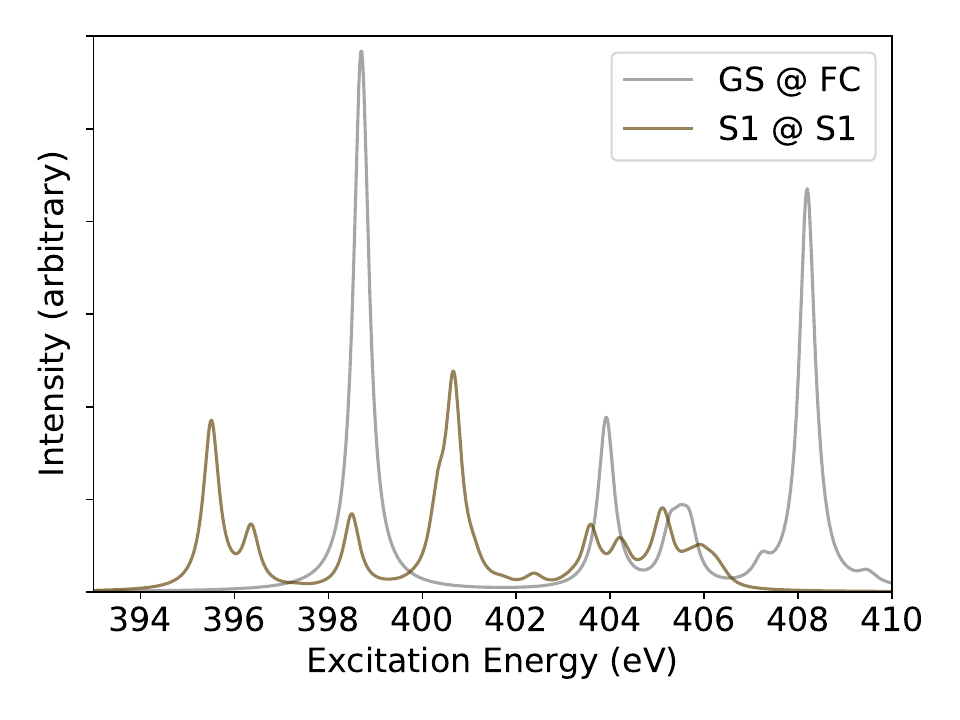}
        \caption{S$_1$ nitrogen K-edge}
    \end{subfigure}
    \begin{subfigure}[b]{0.45\textwidth}
        \centering
        \footnotesize
        \includegraphics[width=\textwidth]{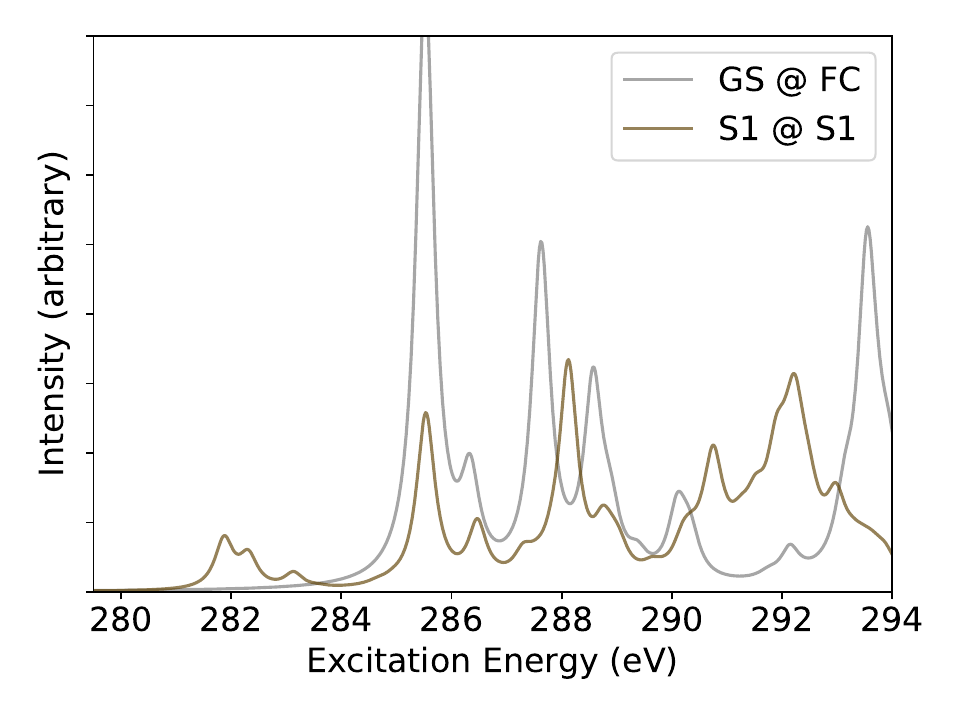}
        \caption{S$_1$  carbon K-edge}
    \end{subfigure}
    \caption{
        CC3 spectra computed at the equilibrium geometry of the ground state (FC) 
        and the equilibrium geometry of the first excited state (S$_1$) in vacuum.
        The S$_1$ equilibrium geometry was optimized at the CASPT2 level. 
        The computed spectra have been convoluted with Lorentzian functions using $\text{HWHM}=0.2$\,eV.
    }
    \label{fig:cc3_FC_and_S1_spectra}
\end{figure}

The X-ray absorption spectrum of the ground state at the carbon and nitrogen K-edges along with the NTOs of the main transitions are shown in Fig.~\ref{fig:cc3_C_and_N_gs_dz_Ry}.
For these calculations, the cc-pVDZ basis set plus additional Rydberg-type functions generated according to Kaufmann, Baumeister, and Jungen's prescription~\cite{KBJ_bases} was used.
The X-ray absorption of the ground and excited states at the Franck-Condon and S$_1$ minimum geometry was calculated using cc-pVDZ only, and the spectra are depicted in 
Fig.~\ref{fig:cc3_FC_and_S1_spectra}.
Additionally, X-ray absorption spectra were calculated for geometries sampled 
every $\SI{10}{\fs}$ from 34 ADC(2)/aug-cc-pVDZ SH trajectories using the cc-pVDZ basis set.
The resulting false color plot is shown in Fig.~\ref{fig:trxas_cc3_n}.
In the nitrogen spectra, 
all peaks have been shifted by $-2.72$~eV to match the ground state bleach. 
Peaks predicted to lay above the ground state bleach involve significant 
double excitation character with respect to the ground state and are thus not described as accurately as the pre-edge peaks at the CC3 level. 
For this reason, coupled cluster calculations using full triples (CCSDT) and the STO-3G basis have been performed using MRCC~\cite{MRCC,MRCC_website,CC_MRCC} as a benchmark of the CC3 results.
Based on these calculations, an additional shift of $-2.6$~eV was applied to the core excitations from the valence excited states with an excitation energy larger than the ground state bleach. 

In the carbon spectra, all peaks have been shifted by $-2.6$~eV to match the ground state bleach. 
Based on CCSDT/STO-3G calculations performed with the eT program ({\it 33\/}),
the excitations from the valence excited states with an excitation energy 
larger than the ground state bleach have been additionally shifted by $-2.0$~eV.

\clearpage

\subsection{Comparison of static CC3 and RASPT2 spectra with experiment at the Carbon and Nitrogen K-edges}

\begin{figure}[htbp!]
\centering\includegraphics[width=\textwidth]{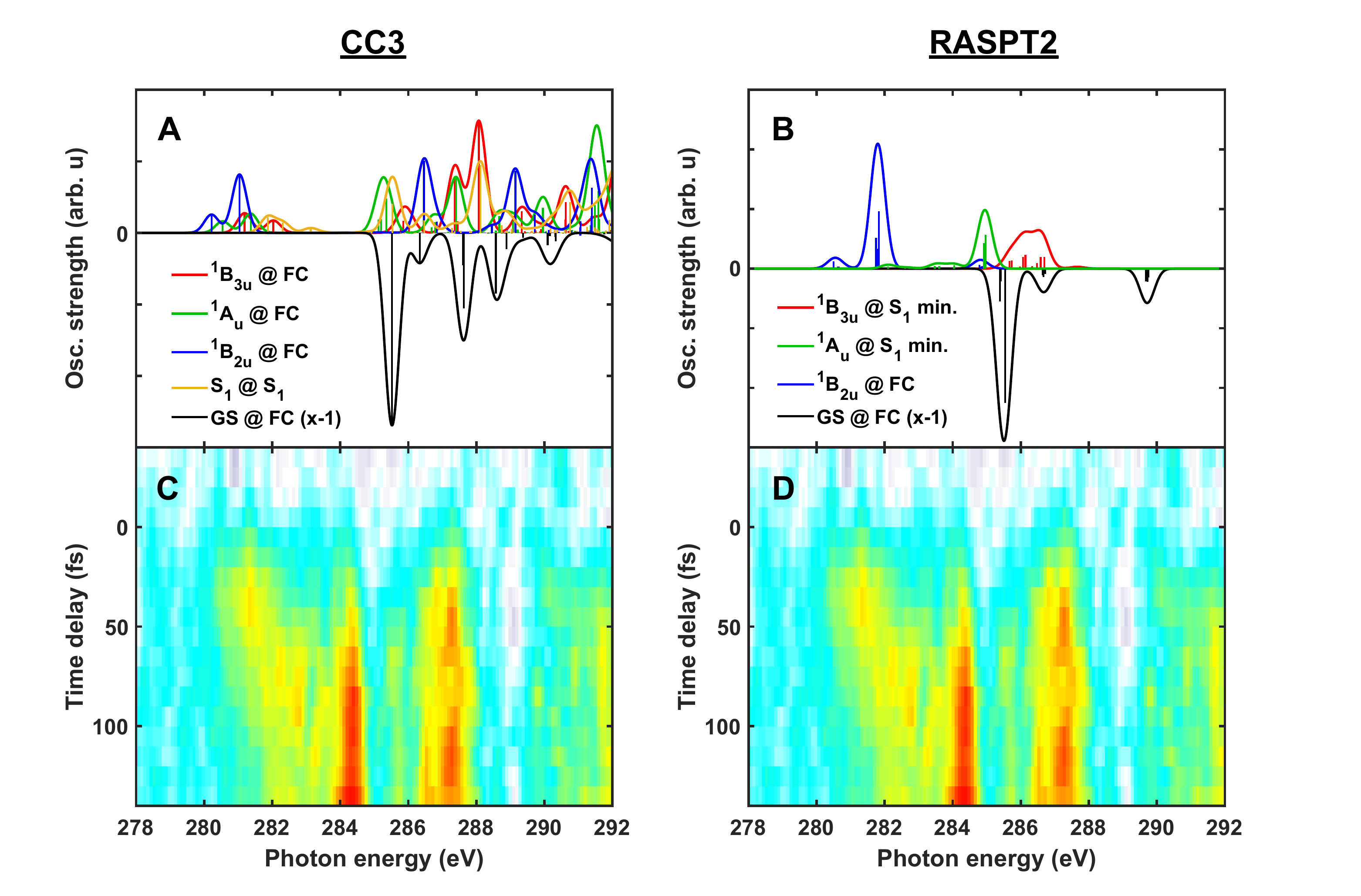} 
\caption{Carbon K-edge excited-state XA spectra calculated using CC3/cc-pVDZ \textbf{(A)} and RASPT2/RAS2(10e,8o) \textbf{(B)} at both the FC and relaxed geometries for the first valence excited state (S$_1$). \textbf{(C,D)} Experimental differential absorbance spectra at the carbon K-edge.
Note that some transitions are missing in the RASPT2 simulated spectra,
since $\sigma^\ast$ orbitals were not included in the active space.}
\label{figS:CC3_vs_RASPT2_carbon}
\end{figure}

\begin{figure}[htbp!]
\centering\includegraphics[width=\textwidth]{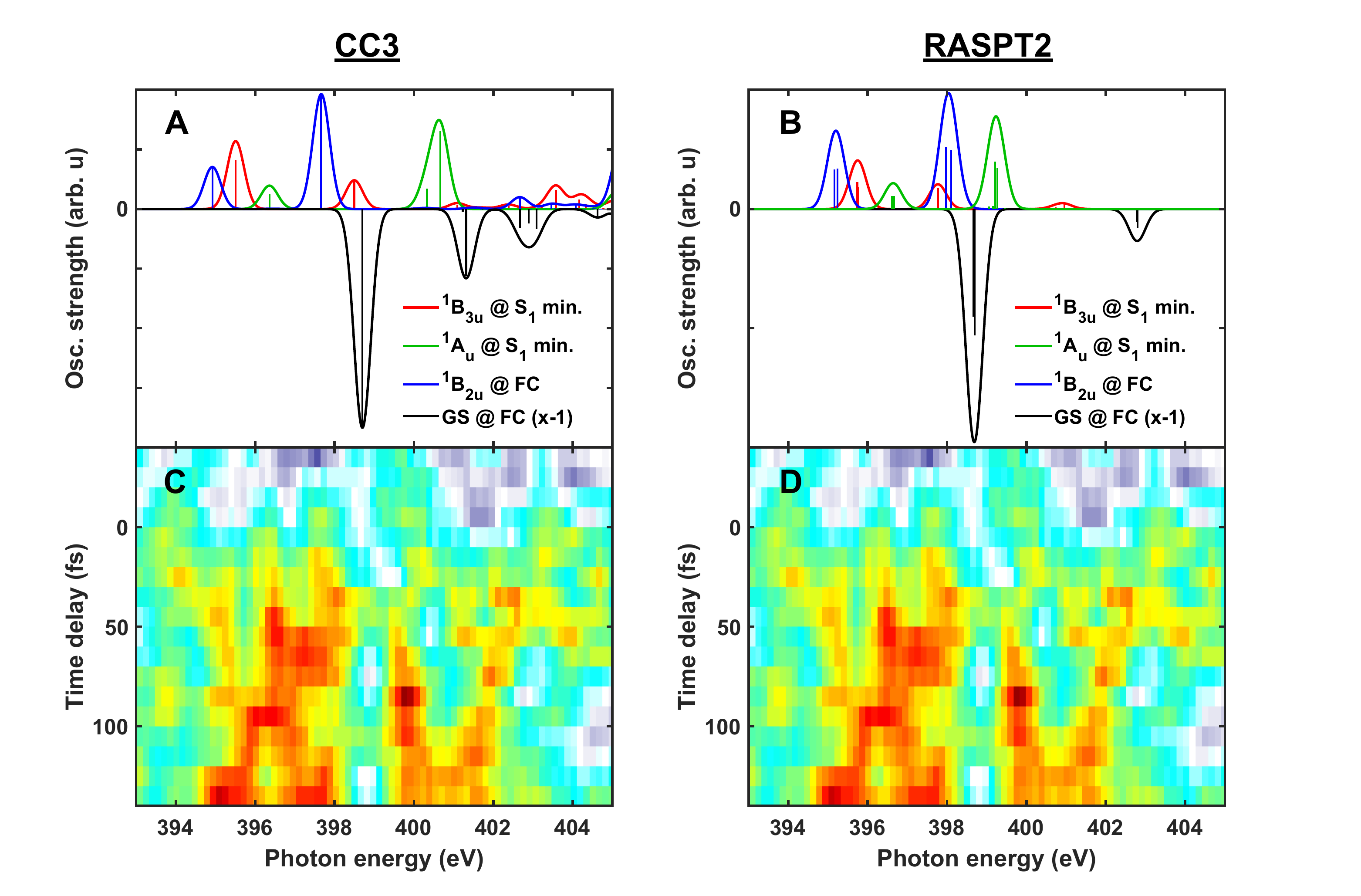} 
\caption{Nitrogen K-edge excited-state XA spectra calculated using CC3/cc-pVDZ \textbf{(A)} and RASPT2/RAS2(10e,8o) \textbf{(B)} at both the FC and relaxed geometries for the first valence excited state (S$_1$). \textbf{(C,D)} Experimental differential absorbance spectra at the nitrogen K-edge.
Note that some transitions are missing in the RASPT2 simulated spectra,
since $\sigma^\ast$ orbitals were not included in the active space.}
\label{figS:CC3_vs_RASPT2_nitrogen}
\end{figure}

\clearpage

\subsection{Nuclear dynamics}

\subsubsection{Simulated UV spectra}

\begin{figure}[hb]
    \centering
    \includegraphics[]{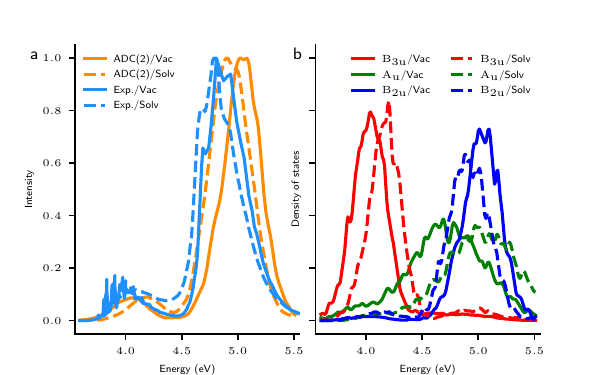}
    \caption{a) Simulated and experimental UV absorption spectrum of pyrazine in vacuum and in solution. b) Density of states of pyrazine calculated at the ADC(2)/aug-cc-pVDZ level for the ensemble in vacuum and in solution.}
    \label{fig:uv_gas_solv}
\end{figure}

The gas phase and solution UV absorption spectra are shown in Fig.~\ref{fig:uv_gas_solv}a. The spectra are calculated at the ADC(2)/aug-cc-pVDZ level (as implemented in the Turbomole program package~\cite{TURBOMOLE, TURBOMOLE_review}) based on an ensemble generated from the harmonic ground state Wigner distribution. For the vacuum spectrum a total of 4000 geometries of pyrazine was sampled. To model the effect of solvation, two water molecules (hydrogen bonded to the nitrogen atoms) were added while the rest of the effect of the environment was modeled implicitly using the conductor like screening model (COSMO)~\cite{cosmo_original, cosmo_ri_adc2}. For this system 2000 initial conditions were sampled from the Wigner distribution with the six lowest frequency normal modes (corresponding to motion of the water molecules relative to pyrazine) frozen as such large amplitude motions are not well described by this approximation.

The first two peaks of the experimental spectrum are reproduced at the ADC(2) level, with a shift of approx. 0.16 eV with respect to the experiment. We see that the solvent shift of the spectrum is correctly reproduced by a combination of two explicit water molecules and COSMO for the bulk environment. Adding either of these effects by itself also resulted in a shift of the spectrum, but by a smaller amount. In Fig.~\ref{fig:uv_gas_solv}b the density of states in the nuclear ensemble is shown, decomposed into contributions from the states of different character. We see that the solvent shifts the two n$\pi^*$ states and the $^1$B$_{\mathrm{2u}}$($\pi\pi^*$) state in opposite directions, with the greatest effect on the $^1$A$_{\mathrm{u}}$(n$\pi^*$) which is shifted above the $^1$B$_{\mathrm{2u}}$($\pi\pi^*$) state at most geometries.

\subsubsection{Fewest-Switches Surface Hopping calculations}

Nonadiabatic dynamics simulations were based on the locally diabatic variant of the fewest-switches surface hopping algorithm (LD-FSSH) with potential energy surfaces (PESs) calculated at the ADC(2)/aug-cc-pVDZ level, following a previous benchmark of the method~\cite{xie2019assessing}. This procedure was also successfully used to simulate the time-resolved photoelectron spectrum~\cite{pitesa2021combined}. A total of 170 initial conditions for nonadiabatic dynamics were randomly (weighted by oscillator strengths) selected from the ground state ensemble among all states falling in an energy window between 4.56 eV and 4.92 eV (excitation window A). This energy window captures the low-energy part of the peak in Fig.~\ref{fig:uv_gas_solv}. Trajectories were propagated for 200 fs using a time step of 0.5 fs. Geometries were sampled from time slices of these trajectories and XA spectra from the currently populated state $L_i(t)$ to a manifold of final core excited states were calculated at the sampled geometries. The TR-XAS spectrum was calculated using the nuclear ensemble approach
\begin{equation} \label{eqn:nea_spectrum}
    \sigma(E,t) \propto \sum_i \sum_F \Delta E_{L_i(t)F} \abs{\mu_{L_i(t)F}}^2 k\qty(E - E_{L_i(t)f}, \delta) \,,
\end{equation}
where the sums are over all trajectories $i$ and final states $F$, $\Delta E$ and $\mu$ are the (shifted) energy difference and cross sections between the currently populated state $L_i(t)$ and final state $F$ at geometry $R_i(t)$. 

For the solution model, an identical computational procedure was employed as in the gas phase to obtain results which are directly comparable. A total of 187 initial conditions were selected for propagation. Hydrogen transfer from the water molecule to the nitrogen atom of pyrazine occurred in one single trajectory. The population dynamics of the remaining trajectories was similar to the one in vacuum. To test the impact of the excitation window on the dynamics, an additional set of (54 in vacuum, 94 in solution) trajectories was calculated with excitation energies between 4.92 eV and 5.5 eV encompassing the higher energy part of the second peak of the UV spectrum of pyrazine (excitation window B).

\begin{figure}[hbt!]
    \centering
    \includegraphics[width=\textwidth]{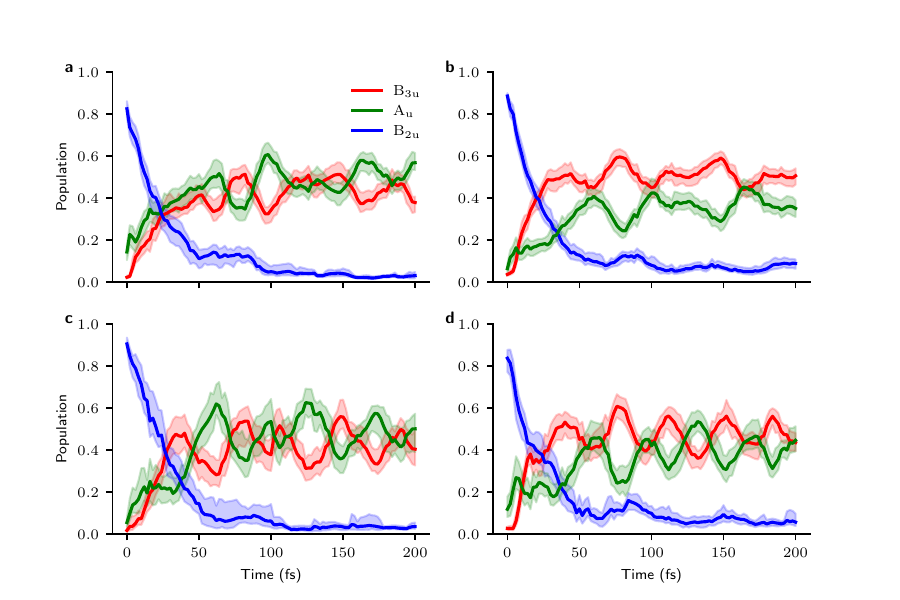}
    \caption{Diabatic populations along FSSH trajectories of (\textbf{a}, \textbf{c}) pyrazine excited in excitation windows A and B and (\textbf{b}, \textbf{d}) pyrazine in solution excited in excitation windows A and B. Shaded areas represent 95\% confidence intervals for the population of each state estimated using the bootstrap method}.
    \label{fig:dyn_pop}
\end{figure}

Analysis of the trajectories and of the potential energy scans was based on projecting the electronic wave functions $\qty{\ket{\Phi(\vb{R})}}$ calculated at each geometry onto the wave functions of the ground state minimum geometry $\qty{\ket{\Phi(\vb{R}_0)}}$. The squares of the elements of an orthogonalized matrix of coefficients obtained in this way~\cite{sapunar2019highly, pitesa2021combined} are then used as a quantitative measure of the electronic character of each state at each geometry described in terms of $^1$B$_{\mathrm{3u}}$(n$\pi^*$), $^1$A$_{\mathrm{u}}$(n$\pi^*$), $^1$B$_{\mathrm{2u}}$($\pi\pi^*$) state contributions. Figure \ref{fig:dyn_pop} shows the diabatic populations in each set of trajectories. The three sets of trajectories share the same qualitative picture of population dynamics. However, we also see some differences in the population oscillations between the $^1$A$_{\mathrm{u}}$ and $^1$B$_{\mathrm{3u}}$ states depending on the excitation window. This is especially true at very early times where the first maximum of the $^1$B$_{\mathrm{3u}}$ population seen at higher energies is missing in the lower excitation window. We also see that the population of the $^1$A$_{\mathrm{u}}$ is lower in solution than in vacuum.

\begin{figure}[hbt!]
    \centering
    \includegraphics[width=\textwidth]{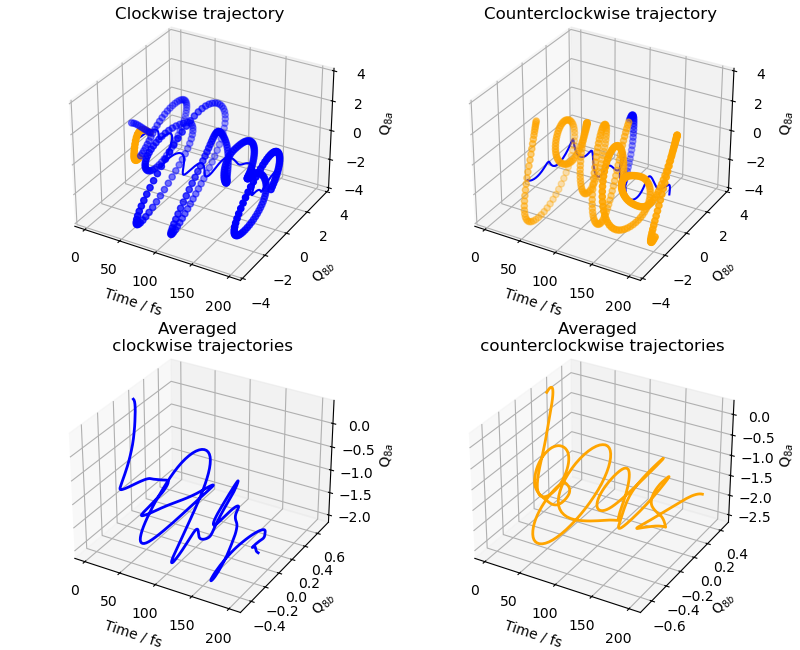}
    \caption{Dynamics of the FSSH trajectories along the 8a and 8b modes over time. The upper panels show an example trajectory moving primarily clockwise (blue) and counterclockwise (orange) around an average (blue full line). The lower panels show the average dynamics of all trajectories in each group.}
    \label{fig:fssh_8a_8b}
\end{figure}

To support the qualitative description of the dynamics shown in Fig. 4a of the manuscript, we analyze the motion of FSSH trajectories along the 8a and 8b modes in Fig.~\ref{fig:fssh_8a_8b}. The trajectories have been grouped into two sets depending on whether their motion around the CI can primarily be described as clockwise or counterclockwise and the counterclockwise by looking at their motion at each time step with respect to a rolling average representing the central axis at the given time. As expected from symmetry arguments, the number of trajectories falling into each category is approximately the same.

\subsubsection{Comparing FSSH and MCTDH calculations}

\begin{figure}[hbt!]
    \centering
    \includegraphics[width=\textwidth]{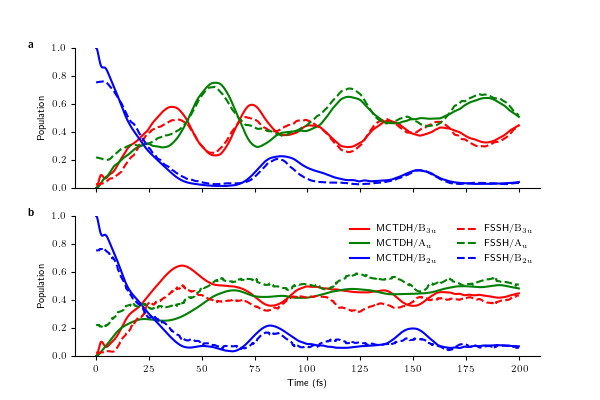}
    \caption{Comparison of diabatic populations from MCTDH and FSSH calculations performed using (a) the full three-state model from Ref.~\cite{sala2014role} and (b) the same model, but with coupling between the $^1$A$_{\rm u}$ and $^1$B$_{\rm 3u}$ states along the 8b mode set to zero.}
    \label{fig:fssh_mctdh_dpop}
\end{figure}

Additional dynamics simulations were performed for the three-state diabatic model Hamiltonian developed by Sala \textit{et al.} \cite{sala2014role}. This model includes four electronic states (the ground state and the three lowest excited states) and nine vibrational modes ($\nu_{6a}$,  $\nu_{1}$, $\nu_{9a}$, $\nu_{8a}$, $\nu_{10a}$, $\nu_{4}$, $\nu_{5}$, $\nu_{3}$, $\nu_{8b}$). The nuclear dynamics for this model systems are propagated using FSSH and also the MCTDH method \cite{beck2000, meyer1990} implemented in the Quantics package \cite{quantics_code}. For details on the model system and the MCTDH propagation, the reader is referred to Ref.~\cite{sala2014role}. Populations obtained for the chosen model using FSSH and MCTDH were previously shown to be in good agreement between methods and also with full dimensional ADC(2) calculations \cite{xie2019assessing}. In Fig.~\ref{fig:fssh_mctdh_dpop} we confirm this agreement and also show that the population oscillations disappear when coupling between the $^1$A$_{\rm u}$ and $^1$B$_{\rm 3u}$ states along the $Q_{8b}$ mode is set to zero.

\begin{figure}[hbt!]
    \centering
    \includegraphics[width=\textwidth]{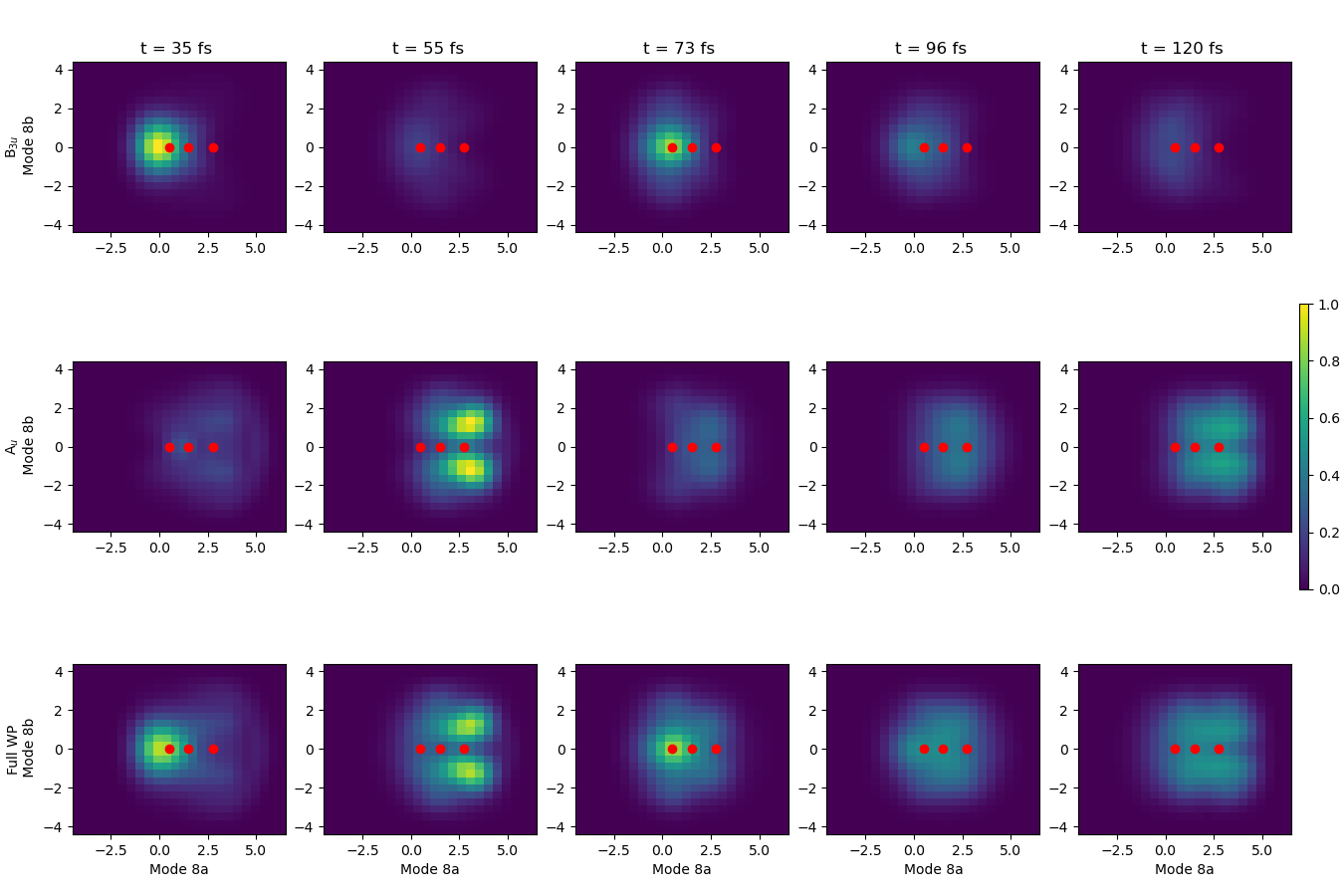}
    \caption{Motion of the wave packet in the space spanned by the 8a and 8b modes on the potential-energy surfaces from Ref. \cite{sala2014role}. Components of the wave packet on the $^1$B$_{\rm 3u}$ (top row) and $^1$A$_{\rm u}$ (middle row) and the two components together (bottom row) are shown. The red dots mark the positions of the $^1$B$_{\rm 3u}$ minimum, the CI and the $^1$A$_{\rm u}$} minimum (from left to right).
    \label{fig:mctdh_8a_8b}
\end{figure}

In Fig.~\ref{fig:mctdh_8a_8b}, we show the 2D diabatic reduced density of the wave packet along the 8a and 8b modes. Upon passing through the CI, amplitude is equally sent towards both sides around $Q_{8b} = 0$ on the $^1$A$_{\rm u}$ state. This coincides with a reflection along the 8a mode sending the system back towards the CI. The two components of the wave packet then recombine on the $^1$B$_{\rm 3u}$ surface on the other side of the CI and the process is repeated. Before 100 fs, a recurrence of the $^1$B$_{\rm 2u}$ state occurs after which the wave packet is spread out further, but oscillations along the 8a mode along with the spreading + recombining of the wave packet along the 8b mode are still visible.

\clearpage

\subsection{Simulated TR-XAS}

Carbon and nitrogen K-edge TR-XAS up to 200 fs were calculated at the RASPT2/RAS2(10e,8o) levels of theory based on time slices taken every 2 fs from 34 of the gas phase trajectories. 

Calculating the spectra in this way introduces a discontinuity when sampling geometries from trajectories where the currently populated state is, obviously, calculated at the ADC(2)/aug-cc-pVDZ level and now needs to be re-calculated at a different level of theory. To ensure the consistency of the calculated results, we checked the distribution of excitation energies for all geometries sampled from the dynamics calculated at the ADC(2)/aug-cc-pVDZ, CC3/cc-pVDZ and RASPT2/RAS2(10e,8o) levels of theory.

The overall agreement between the three methods is very good, with only a slight difference in the distribution of energies of the S$_2$ state at the RASPT2 level. However, taking only the subset of geometries at $t=0$ we see that the agreement between RASPT2 and the other methods is significantly worse, with a significant shift of the S$_2$ state which is the initially populated state in many trajectories. Despite this difference in the region of the ground state minimum, the energies become much closer at later times, indicating that the shapes of the PES are similar at all three levels. To fix the issue with the different ordering of the states at some geometries at early times, we apply a reordering based on the overlap of the excited state wave functions at the ADC(2) and RASPT2 levels. For the simulated XAS we always choose the RASPT2 state with the largest overlap with the ADC(2) excited state that is populated at the given geometry during the dynamics simulations.

\begin{figure}[h!] 
    \centering \includegraphics[]{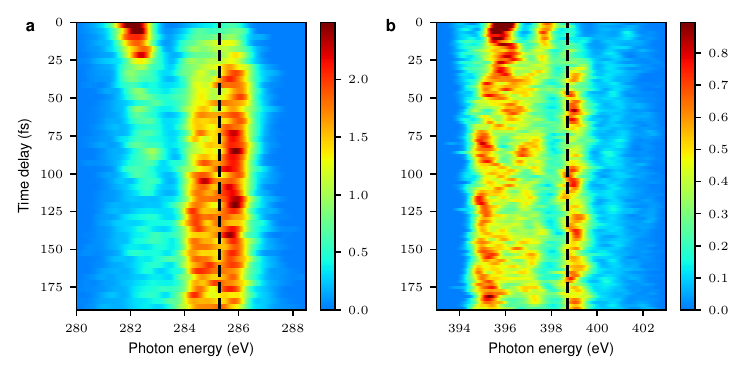}
    \caption{(\textbf{a}) Carbon and (\textbf{b}) nitrogen K-edge excited-state XA spectra calculated at the RMS-RASPT2/RAS2(10e,8o) level based on 34 FSSH trajectories calculated at the ADC(2)/aug-cc-pVDZ level. The simulated spectrum has been shifted by $-2.9$ eV (carbon K-edge) and $-1.5$ eV (nitrogen K-edge). The position of the ground state bleach is marked by the dashed black line.}
    \label{fig:trxas_raspt2}
\end{figure}

Fig.~\ref{fig:trxas_raspt2}a shows the simulated spectrum at the RASPT2/RAS2(10e,8o) level. The positions of the peaks are accurately represented by the simulated spectrum. Interaction between the two n$\pi^*$ states is clearly visible from the oscillations in the intensity of the peaks. However, these oscillations still have a significantly different time scale and amplitude to the ones seen in the experiment.

As with the nitrogen K-edge, we see a good qualitative agreement with the experimental spectrum. The main difference is the shift of the peaks just below and above the edge in opposite directions, bringing them significantly closer to each other than they are in the experiment.

\begin{figure}[h!]
    \centering
    \includegraphics[]{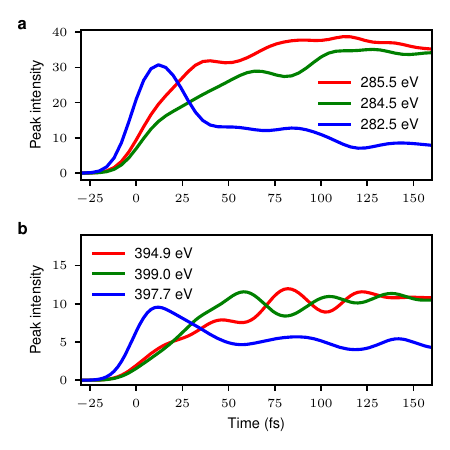}
    \caption{Integrated intensities of selected peaks in the simulated (\textbf{a}) carbon K-edge and (\textbf{b}) nitrogen K-edge XA spectra calculated at the RMS-RASPT2/RAS2(10e,8o) level. The red, green and blue lines represent peaks primarily originating from the $^1$B$_{\mathrm{3u}}$(n$\pi^*$), $^1$A$_{\mathrm{u}}$(n$\pi^*$) and $^1$B$_{\mathrm{2u}}$($\pi\pi^*$) states, respectively.  The full spectrum is shown in Fig.~\ref{fig:trxas_raspt2}.}
    \label{fig:lineouts_raspt2}
\end{figure}

Fig.~\ref{fig:lineouts_raspt2} shows the integrated intensities of the peaks in the simulated spectra. We see the oscillations and relative intensities of the peaks. These lineouts are highly correlated, but not directly proportional, with the populations of the diabatic states to which the peaks are assigned. 

The nitrogen K-edge TR-XAS was also simulated at the CC3/cc-pVDZ level using time slices taken every 10 fs from the same trajectories used for the RASPT2/RAS2(10e,8o) spectra and the same procedure for handling the discontinuity between electronic structure methods for dynamics and spectrum simulations. The state ordering between the ADC(2) and CC3 calculations is significantly more consistent, even at early times close to the CI, than when comparing with RASPT2 calculations.

\begin{figure}[h!]
    \centering
    \includegraphics[width=0.5\textwidth]{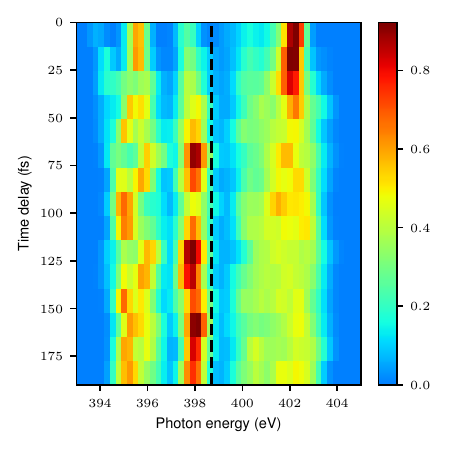}
    \caption{Nitrogen K-edge excited-state XAS spectra calculated at the CC3/cc-pVDZ level based on 34 FSSH trajectories calculated at the ADC(2)/aug-cc-pVDZ level. The position of the ground state bleach is marked by the dashed black line. The simulated spectrum has been shifted by $-2.75$ eV with an additional $-2.6$ eV shift for the above-edge peaks (see Sec~\ref{sec:CC3}).
    }
    \label{fig:trxas_cc3_n}
\end{figure}

\clearpage
\section{Interpretation of the solvent effects}

\subsection{Concentration-dependent absorption spectra \label{Experiment:LB}}

Fig.~\ref{fig:trxas_cc3_n} shows the 266 nm UV pump absorbance of pyrazine solutions across a broad range of concentrations, namely ranging from a 1 mM to 5 M. This was achieved by using various liquid jet thicknesses, down to the sub-$\mu$m range \cite{yin2020few, Ekimova2015, fuledevelopment}, in order to cover the entire dynamic range. The linear increase in absorbance with pyrazine concentration indicates that self-association does not play a major role on the absorption properties of pyrazine at the studied concentrations. The $\sim$18-$\mu$m nozzle orifice size, which was used for 1-5 M linear absorption measurements, was utilized for the time-resolved experiments. The 266 nm source was the same as the one used for the time-resolved experiments.
\begin{figure}[h!]
    \centering
    \includegraphics[width=\textwidth]{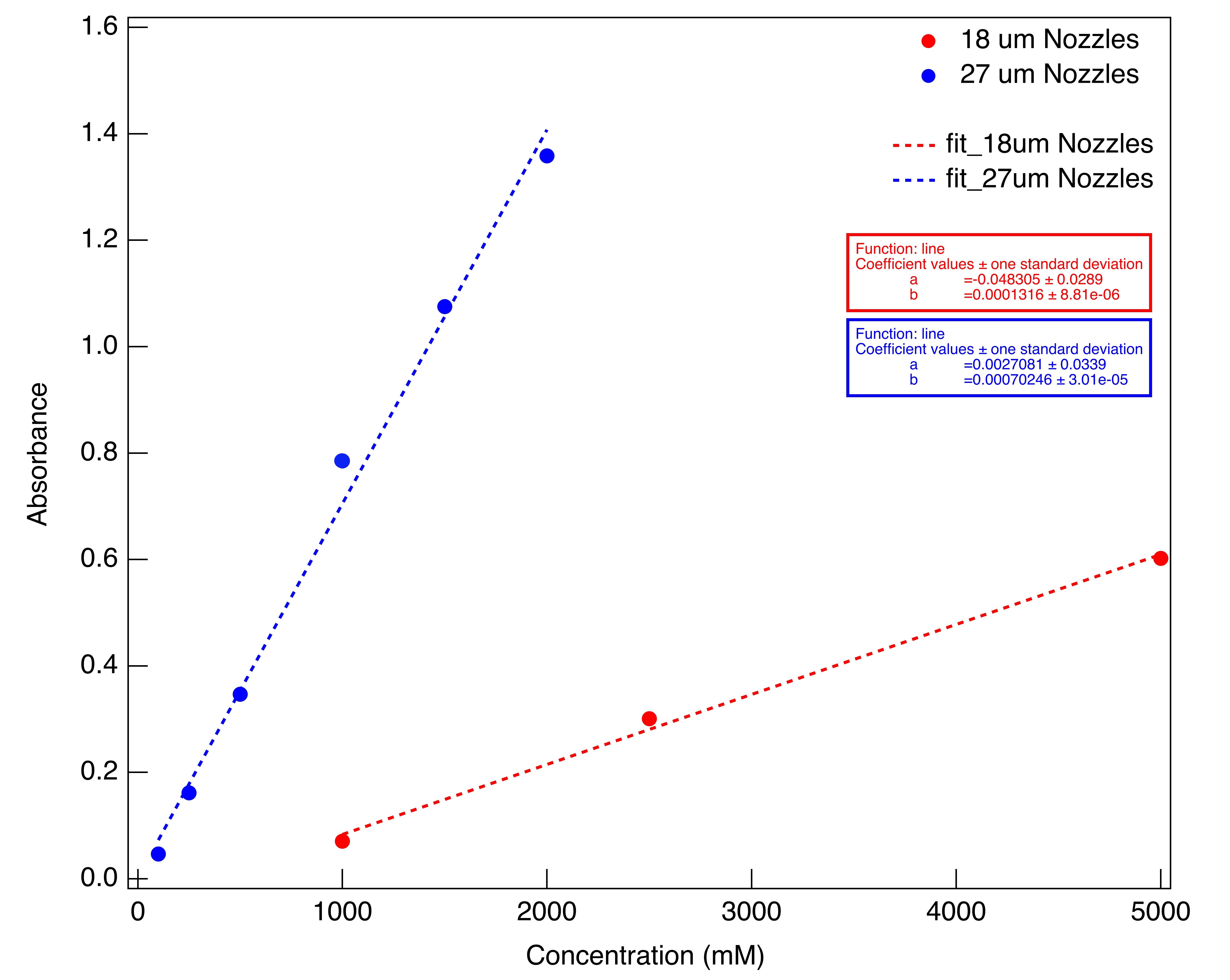}
    \caption{Absorbance at the pump wavelength (266 nm) by liquid flat jets with different inner diameters of nozzles running aqueous solutions of pyrazine at various concentrations. The linear relationship across the entire concentration range indicates that self-association does not play a major role in the absorption properties of pyrazine at the studied concentrations. The parameters $a$ and $b$ of the linear regressions stand for the intercept and slope, respectively.
    }
    \label{fig:trxas_cc3_n}
\end{figure}
\\Using a standard UV-Vis instrument (Thermo Scientific GENESYS 50), and pyrazine concentrations in the 0.1-2.5 mM range, the extinction coefficient of aqueous pyrazine solutions was found to be $\varepsilon$ = 10580 L mol$^{-1}$ cm$^{-1}$. Using the jet thickness previously measured for $\sim$27 $\mu$m nozzle orifice jets \cite{yin2020few}, the aqueous pyrazine extinction coefficient for jet measurements with these nozzles was extracted via the Beer-Lambert relation and found to be $\varepsilon_{27}$ = 8570 L mol$^{-1}$ cm$^{-1}$, demonstrating reasonable agreement with the value measured from a traditional UV-Vis system. This agreement further suggests that there are no major self-association effects on the absorption properties of aqueous pyrazine even at concentrations above 1 M.

\subsection{Self-association effect}

The experimental results shown in the previous subsection indicated no major effects of self-association on the UV absorption spectra. 

A previous study~\cite{peral} reported small deviations from the Beer-Lambert law over a very broad range (5.5 orders of magnitude) of concentration.

\begin{figure}[hbt!]
    \centering
\includegraphics[width=0.5\textwidth]{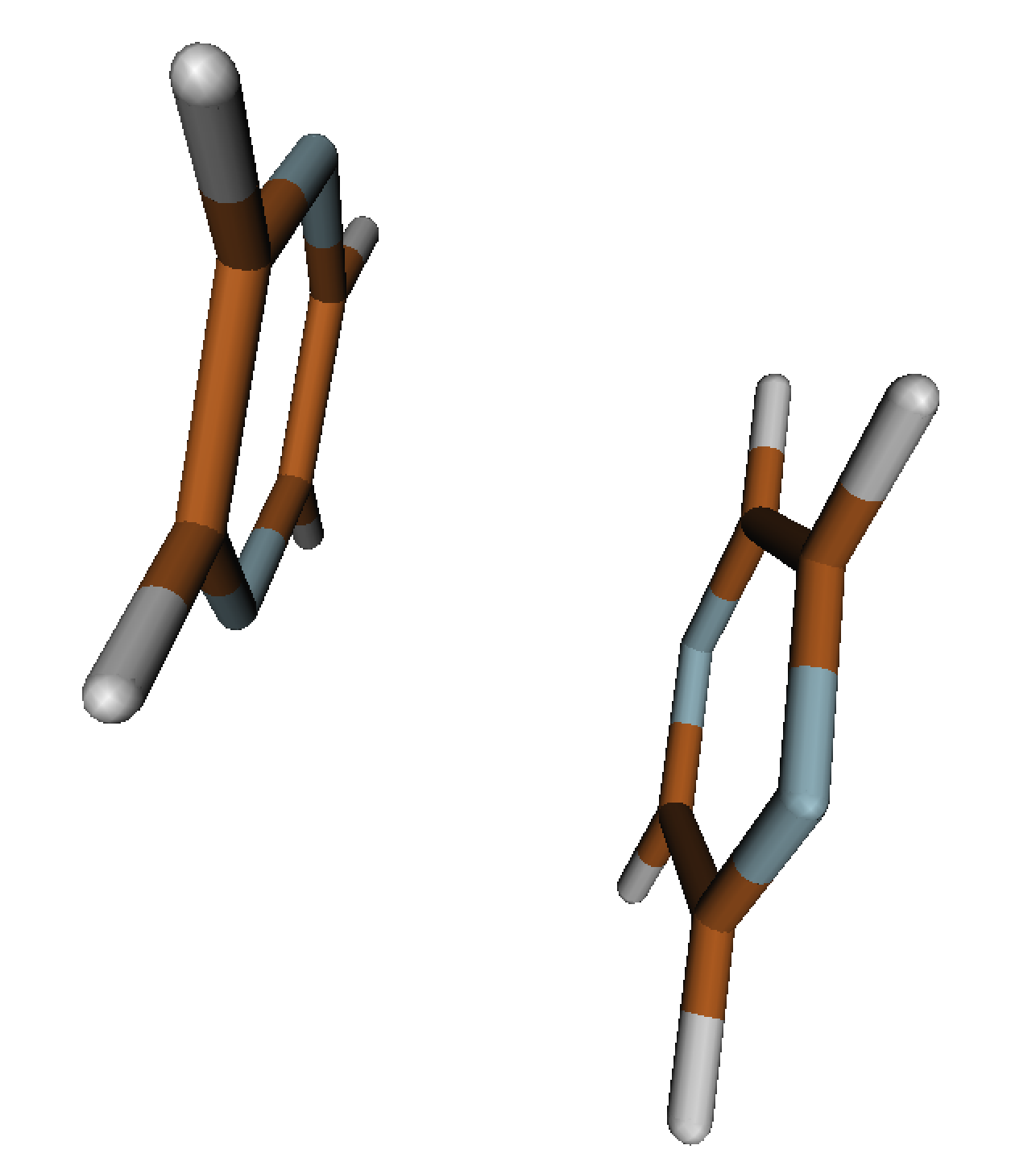}

    \caption{Visualized geometry of the pyrazine dimer optimized at the B3LYP-D3/aug-cc-pVDZ level.}
    \label{fig:dimer}
\end{figure}

In this section, we computationally explored whether 
insight into
self-association of the pyrazine molecules 

could be obtained from
calculations of the UV excitation energies and oscillator strengths
of a pyrazine dimer. 
To this end, we assumed a dimer with a sandwich 
configuration, where the intermolecular interaction should be the largest, and optimized the geometry at the B3LYP-D3/aug-cc-pVDZ level as in Table~\ref{tab:dimer_geom} and Fig.~\ref{fig:dimer}.
The optimization yielded as most stable structure  a cross-displaced $\pi$–$\pi$ stacked structure,
consistent with the CCSD(T) results in Ref.~\cite{Feng:dimer:2023}.
We then calculated the excitation energies of the dimer at the optimized geometry at the ADC(2)/aug-cc-pVDZ level of theory.

\begin{table}[h]
  \centering
  \caption{
  Geometry of pyrazine dimer optimized at the B3LYP-D3/aug-cc-pVDZ level of theory; xyz format, coordinates in Angstrom (\AA).}
  
  \begin{tabular}{cccc}
     \hline

C    &      2.127257  &  1.134516   & -0.323795 \\
C    &      1.409320  &  1.133535    & 0.877978 \\
C    &      2.127895 &  -1.133551   & -0.325302 \\
C    &      1.409930 &  -1.134621   & 0.876444 \\
H    &      2.408516  &  2.074124  &  -0.803604 \\
H    &       1.111302  &   2.071613   &  1.349250 \\
H     &     2.409713  & -2.072344  & -0.806377 \\
H     &     1.112321 &  -2.073489  &  1.346403 \\
N     &     2.493008  &  0.001000  & -0.933839\\
N     &     1.048120  & -0.001038   & 1.487955\\
C     &    -2.334762 &   0.698987  &  0.706912 \\
C     &    -2.333932  & -0.701496   & 0.705570 \\
C     &    -1.201511 &   0.700790   & -1.257781 \\
C     &    -1.200694  & -0.698182  & -1.259106 \\
H     &    -2.799606 &   1.254518 &   1.523908 \\
H     &    -2.798103 &  -1.259158  &  1.521499 \\
H     &    -0.727873 &   1.258310 &  -2.067611 \\
H     &    -0.726384 &  -1.253564  & -2.070013 \\
N     &    -1.772184  &  1.409789  & -0.275635 \\
N     &    -1.770503 &  -1.409732   & -0.278336 
\\
     \hline
      \end{tabular}
\label{tab:dimer_geom}
\end{table}

\begin{figure}[hbpt!]
\centering
\includegraphics[width=0.8\textwidth]{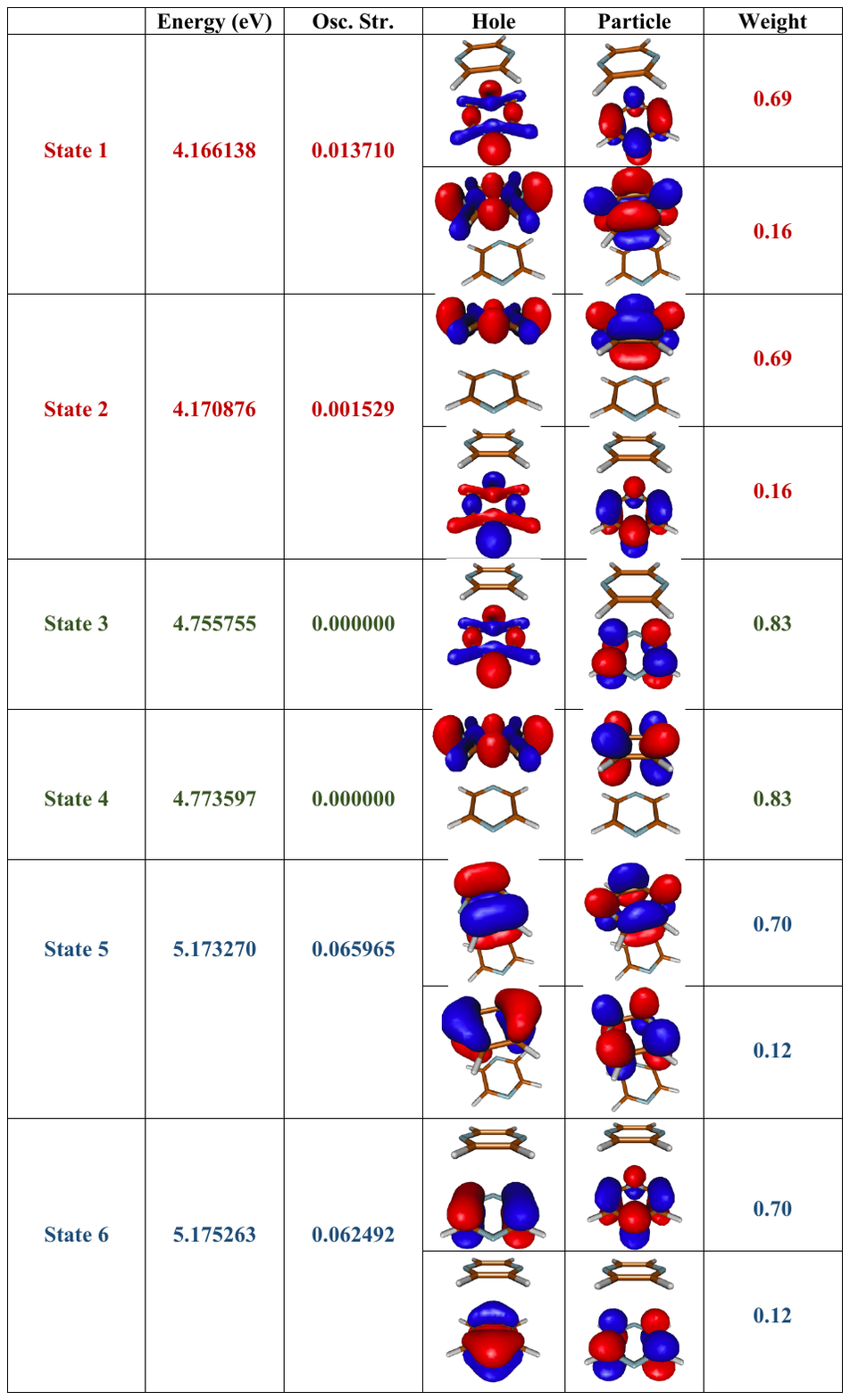}
\caption{Natural transition orbitals of the first 6 valence excited states of pyrazine dimer. ADC(2)/aug-cc-pVDZ results.
\label{fig:NTOs-dimer}}
\end{figure}

The energies and oscillator strengths of the excitations of each pyrazine monomer in the dimer and of the dimer as a whole are given in Table~\ref{tab:dimer_exci}. 
NTOs of the dimer excitations are shown in Figure~\ref{fig:NTOs-dimer}.

\begin{table}[hbt!]
 \centering
 \caption{Excited states of each pyrazine monomer in the B3LYP-D3 optimized dimer and of the dimer as a whole, calculated 
 using ADC(2) in conjunction with the aug-cc-pVDZ basis set.}
 \begin{tabular}{c|cc|cc}
\hline
State &  \multicolumn{2}{c|}{Monomer 1 / Monomer 2)} &  \multicolumn{2}{c}{Dimer}  
     \\
&  Exc. energy (eV) & Osc. strength & Exc. energy (eV) & Osc. strength 
     \\
     \hline
     $^1$B$_{\mathrm{3u}}$(n$\pi^*$)   & 4.18153  & 0.00606  &  4.16614  & 0.01371
     \\
    &  4.18085   & 0.00606 &  4.17088  &  0.00153
     \\
     \hline
     $^1$A$_{\mathrm{u}}$(n$\pi^*$)   & 4.78947  & 0.00000 &  4.75575  & 0.00000
     \\
     & 4.78909  & 0.00000 &  4.77360  &  0.00000
     \\
     \hline
     $^1$B$_{\mathrm{2u}}$($\pi\pi^*$)   & 5.20376  & 0.09022 &  5.17327  & 0.06597
     \\
   &  5.20554     & 0.09024 &  5.17526  &  0.06249
     \\
     \hline
     \end{tabular}
 \label{tab:dimer_exci}
 \end{table}

In the dimer, 
the $^1$B$_{\mathrm{3u}}$(n$\pi^*$) and $^1$A$_{\mathrm{u}}$(n$\pi^*$) 
and  $^1$B$_{\mathrm{2u}}$($\pi\pi^*$)
pairs of states 
experience a slight red shift from those of the monomers, although by less than 0.1 eV. 
While the pair of $^1$A$_{\mathrm{u}}$(n$\pi^*$) states remains dark,
the oscillator strengths  of the two $^1$B$_{\mathrm{3u}}$(n$\pi^*$) states
sum up to a value that is 
slightly larger than
twice the strength of
the monomer (hyperchromism).

For the $^1$B$_{\mathrm{2u}}$($\pi\pi^*$) state,
on the other hand, the oscillator strength in the dimer is lower than twice 
that of the monomer (hypochromism). 
These results are consistent with the experimental observations.

\subsection{Solvent effects in each excited state}

In this section, we study the effects of solvation on pyrazine by building on the spectral simulations based on microsolvation models from Ref.~\citenum{tsuru2023time}. In Ref.~\citenum{tsuru2023time}, 100 snapshots based on an ab initio molecular dynamics (AIMD) simulation for a system where a pyrazine molecule was dissolved in 112 water molecules were obtained. This system corresponds to 0.5 M pyrazine aqueous solution. In each snapshot, the pyrazine molecule and the water molecules in the first and second solvation shells were extracted. The number of extracted water molecules was 24 on average, and the microsolvation model consisting of one pyrazine molecule and 24 water molecules may qualitatively describe the solvent effects in the present 5M pyrazine aqueous solution, where the unit cell contains one pyrazine and 12 water molecules. For each snapshot of the microsolvation model consisting of the extracted pyrazine and water molecules, the valence excitation energies were calculated at the ADC(2) level, adopting the aug-cc-pVTZ and aug-cc-pVDZ basis sets to the pyrazine and water molecules, respectively. The oscillator strengths for the 100 snapshots were collected and plotted in histogram style in Fig. 5(f) of Ref.~\citenum{tsuru2023time}. Such UV-Vis spectral simulation based on 200 AIMD snapshots was also conducted for an isolated pyrazine molecule and the result was plotted in Fig. 4(a) of Ref.~\citenum{tsuru2023time}.

For one of the 100 snapshots, where the NTOs of the transitions from the ground state to the three lowest adiabatic excited states ($\tilde{\mathrm{A}}$, $\tilde{\mathrm{B}}$ and $\tilde{\mathrm{C}}$) were similar to those at the equilibrium geometry of pyrazine, the solvent effects from the water molecules in the first and second solvation shells were investigated by comparing the excitation energies, the oscillator strengths, and the NTOs for the three models: (i) only the pyrazine molecule, (ii) the pyrazine molecule with the water molecules in the first solvation shell, and (iii) the pyrazine molecule with the water molecules in the first and second solvation shells. The excitation energies and the oscillator strengths were given in the “no-COSMO” columns of Table 3 of Ref.~\citenum{tsuru2023time}. The NTOs were given in Table 4 of Ref.~\citenum{tsuru2023time}. As shown in Table 4 of Ref.~\citenum{tsuru2023time}, without water molecules, $\tilde{\mathrm{A}}$, $\tilde{\mathrm{B}}$ and $\tilde{\mathrm{C}}$ states had almost the pure configuration characters of $^1$B$_{\mathrm{3u}}$(n$\pi^*$), $^1$A$_{\mathrm{u}}$(n$\pi^*$) and $^1$B$_{\mathrm{2u}}$($\pi\pi^*$), respectively. Energy ordering of the configuration characters of $^1$A$_{\mathrm{u}}$(n$\pi^*$) and $^1$B$_{\mathrm{2u}}$($\pi\pi^*$) switched when the water molecules in the second solvation shell were added. Considering the configuration characters in Table 4 of Ref.~\citenum{tsuru2023time} and the excitation energies given in Table 3 of Ref.~\citenum{tsuru2023time}, it is estimated that the diabatic $^1$B$_{\mathrm{3u}}$(n$\pi^*$) and $^1$A$_{\mathrm{u}}$(n$\pi^*$) states are raised by 0.3 and 0.5 eV, respectively, whereas the diabatic $^1$B$_{\mathrm{2u}}$($\pi\pi^*$) state is lowered by 0.2 eV, due to the solute-solvent interactions. 

\begin{figure}
    \centering
    \includegraphics[width=\textwidth]{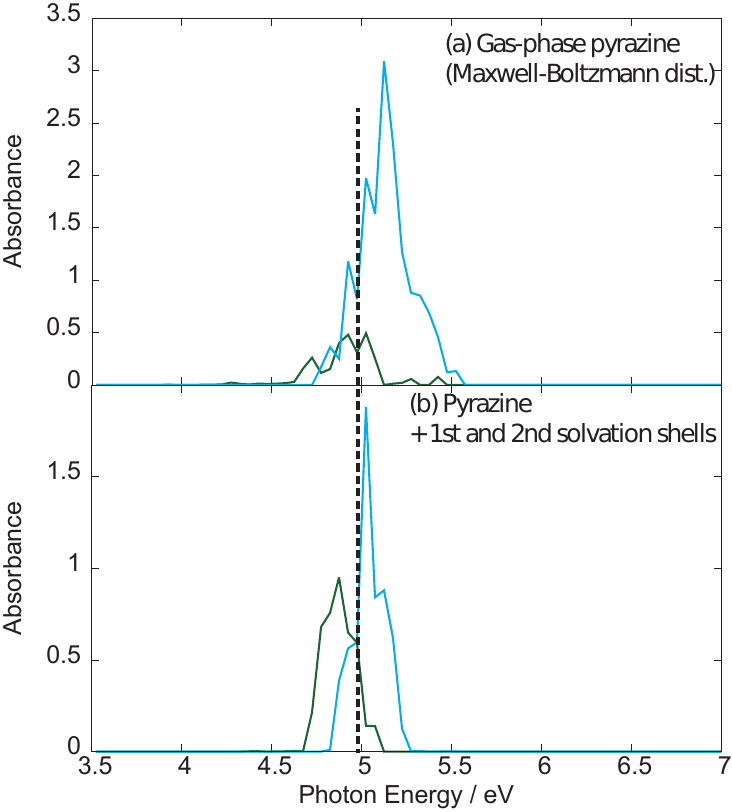}
    \caption{Absorbances of the electronic transitions from the ground state to the (deep green) $\tilde{\mathrm{B}}$ and (cyan) $\tilde{\mathrm{C}}$ adiabatic states plotted based on the Maxwell-Boltzmann distributions for (a) pyrazine in the gas phase and (b) pyrazine embedded in the first and second solvation shells of water molecules. The excitation energies were calculated with ADC(2) adopting the aug-cc-pVTZ and aug-cc-pVDZ basis sets to the pyrazine and water molecules, respectively. The dashed bar indicates the energy corresponding to the central wavelength of the pump pulse (4.66 + 0.31 eV). These plots are based on computational data for Fig. 4(a) and Fig. 5(f) of Ref.~\citenum{tsuru2023time}.}
    \label{fig:BC_separation}
\end{figure}

The second absorption band in Fig. 4(a) and Fig. 5(f) of Ref.~\citenum{tsuru2023time} are separated for the adiabatic $\tilde{\mathrm{B}}$ and $\tilde{\mathrm{C}}$ states and shown in panels (a) and (b) of Fig.~\ref{fig:BC_separation}, respectively. The dashed bar corresponds to the central wavelength of the pump pulse of the present experiment, considering the difference between the band maximum of the simulated and experimental spectra shown in Fig. 4(a) of Ref.~\citenum{tsuru2023time}. Since the adiabatic $\tilde{\mathrm{B}}$ state has the configuration character of $^1$A$_{\mathrm{u}}$(n$\pi^*$) at the equilibrium geometry and the excitation from the ground state is dipole forbidden there, contribution from the $\tilde{\mathrm{B}}$ state to the second absorption band in the UV-Vis absorption spectrum of pyrazine in the gas phase is minor, as seen in panel (a) of Fig.~\ref{fig:BC_separation}. Meanwhile as seen in panel (b) of Fig.~\ref{fig:BC_separation}, contributions from the $\tilde{\mathrm{B}}$ and $\tilde{\mathrm{C}}$ states to the second absorption band in the UV-Vis spectrum simulated for the microsolvation model are comparable. The two contributions are almost in equal amount, especially at the energy corresponding to the central wavelength of the experimental pump laser. This behavior is consistent with the switching of the energy ordering observed in Table 4 of Ref.~\citenum{tsuru2023time} and indicates that a conical intersection between the diabatic $^1$A$_{\mathrm{u}}$(n$\pi^*$) and $^1$B$_{\mathrm{2u}}$($\pi\pi^*$) states lies in the FC region. 

\begin{figure}
    \centering
    \includegraphics[width=\textwidth]{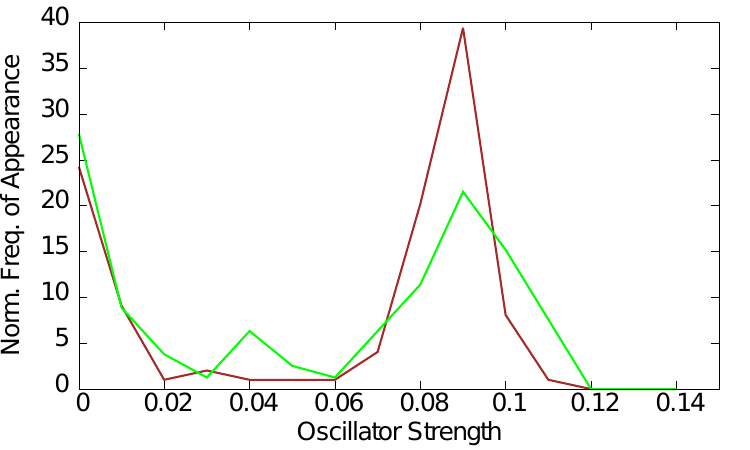}
    \caption{Normalized frequency distribution of the oscillator strengths in excitation energy of 4.9-5.1 eV calculated for pyrazine (brown) in the gas phase and (green) embedded in the first and second solvation shells of water molecules. The oscillator strengths of the transitions from the ground state to the second and third excited states at the optimized geometry are 0.00 and 0.10, respectively. These plots are based on computational data from Fig. 4(a) and Fig. 5(f) of Ref.~\citenum{tsuru2023time}.}
    \label{fig:BC_os}
\end{figure}

Normalized frequency distribution of the oscillator strength in the energy window of 4.9--5.1 eV of Fig.~\ref{fig:BC_separation} (a) and (b) are plotted in Fig.~\ref{fig:BC_os} (a) and (b), respectively, to examine whether the $^1$A$_{\mathrm{u}}$(n$\pi^*$) and $^1$B$_{\mathrm{2u}}$($\pi\pi^*$) configurations are noticeably mixed in the FC region. When the target state has the pure $^1$A$_{\mathrm{u}}$(n$\pi^*$) or $^1$B$_{\mathrm{2u}}$($\pi\pi^*$) configuration character, the oscillator strength of the transition from the ground state is 0.00 and 0.10, respectively, at the ADC(2)/aug-cc-pVTZ level. Panel (a) thus indicates that the two configurations are not mixed when a pyrazine molecule is isolated. Meanwhile, panel (b) exhibits evidence of the configuration mixing in the FC region by distribution of the oscillator strength around 0.04 due to the solute-solvent interactions. This configuration mixing may enable faster nonadiabatic transition to $^1$A$_{\mathrm{u}}$(n$\pi^*$) or almost direct optical transition with the pump laser to the diabatic $^1$A$_{\mathrm{u}}$(n$\pi^*$) state when pyrazine is dissolved in water. This interpretation is consistent with the early appearance of the core-excitation peaks which are assigned to the $^1$A$_{\mathrm{u}}$(n$\pi^*$) state in the solution phase (see Figs.~2 and~3 of the main manuscript and Fig.~\ref{figS:carbon_lineouts} of this SI).

The solvent effects observed in the simulation of Ref.~\citenum{tsuru2023time} also gives possible interpretation of suppression of the oscillatory flow of population between the $^1$B$_{\mathrm{3u}}$(n$\pi^*$) and $^1$A$_{\mathrm{u}}$(n$\pi^*$) states. The potential energy surface of $^1$A$_{\mathrm{u}}$(n$\pi^*$) is raised by ~0.5 eV whereas that of $^1$B$_{\mathrm{3u}}$(n$\pi^*$) is raised by only ~0.3 eV in the solution phase (compare $\tilde{\mathrm{A}}$-$\tilde{\mathrm{B}}$ of bare/no-COSMO and $\tilde{\mathrm{A}}$-$\tilde{\mathrm{C}}$ of +second shell/no-COSMO in Table 3 of Ref.~\citenum{tsuru2023time}. The higher the $^1$A$_{\mathrm{u}}$(n$\pi^*$) state surface with respect to the $^1$B$_{\mathrm{3u}}$(n$\pi^*$) state surface, the smaller the area of the $^1$B$_{\mathrm{3u}}$(n$\pi^*$)/$^1$A$_{\mathrm{u}}$(n$\pi^*$) conical intersection becomes. Moreover, energy dissipates from the nuclear wave packet of pyrazine dissolved into the water bath. As a result, the transitions between the $^1$B$_{\mathrm{3u}}$(n$\pi^*$) and $^1$A$_{\mathrm{u}}$(n$\pi^*$) states might be prevented, especially in the direction from $^1$B$_{\mathrm{3u}}$(n$\pi^*$) to $^1$A$_{\mathrm{u}}$(n$\pi^*$).

\clearpage

\bibliography{refs, attobib, shota}